\documentclass{aa}

\usepackage{graphicx}
\usepackage[round]{natbib}
\usepackage{amssymb}

\bibpunct{(}{)}{;}{a}{}{,}
\def\ion#1#2{{\rm #1}\,{\sc #2}}

\begin{document}
   \title{A comprehensive set of elemental abundances in damped Ly$\alpha$ systems\,: 
          revealing the nature of these high-redshift galaxies
   \thanks{Based on the UVES observations collected during the ESO programmes ID No. 65.O-0296 
           and 67.A-0022 at the VLT/Kueyen telescope, Paranal, Chile
   }}
   

   \author{M. Dessauges-Zavadsky
          \inst{1},
	  F. Calura
	  \inst{2},
	  J. X. Prochaska
	  \inst{3},
          S. D'Odorico
          \inst{4}
	  \and
	  F. Matteucci
	  \inst{2,5}
          }

   \offprints{M. Dessauges-Zavadsky}

   \institute{Observatoire de Gen\`eve, CH-1290 Sauverny, Switzerland 
              \and
	      Dipartimento di Astronomia-Universit\'a di Trieste, Via G. B. Tiepolo 11, I-34131
	      Trieste 
              \and
	      UCO/Lick Observatory, University of California, Santa Cruz, Santa Cruz, CA 95064 
	      \and
	      European Southern Observatory, Karl-Schwarzschildstr. 2, D-85748 Garching 
              bei M\" unchen, Germany 
	      \and
	      INAF, Osservatorio Astronomico di Trieste, Via G. B. Tiepolo 11, I-34131 Trieste 
             }
   
   \date{Received ; accepted}

   \authorrunning{M. Dessauges-Zavadsky et al.}

   \titlerunning{A comprehensive set of elemental abundances in DLAs}

   \abstract{
By combining our UVES-VLT spectra of a sample of four damped Ly$\alpha$ systems (DLAs) toward the 
quasars Q0100+13, Q1331+17, Q2231$-$00 and Q2343+12 with the existing HIRES-Keck spectra, we 
covered the total optical spectral range from 3150 to 10\,000 \AA\ for the four quasars. This large 
wavelength coverage and the high quality of the spectra allowed us to measure the column densities 
of up to 21 ions, namely of 15 elements $-$ N, O, Mg, Al, Si, P, S, Cl, Ar, Ti, Cr, Mn, Fe, Ni, 
Zn. This comprehensive set of ionic column densities and elemental abundances severely contrasts 
with the majority of DLAs for which only a handful of ions and elements is typically observed. Such 
a large amount of information is necessary to constrain the photoionization and dust depletion 
effects, two important steps in order to derive the intrinsic chemical abundance patterns of DLAs. 
We evaluated the photoionization effects with the help of the Al$^{+}$/Al$^{++}$, 
Fe$^{+}$/Fe$^{++}$, N$^0$/N$^+$ and Ar/Si,S ratios, and computed dust corrections. Our analysis 
revealed that the DLA toward Q2343+12 requires important ionization corrections. This makes the
abundance determinations in this object uncertain. The access to the complete series of relatively
robust intrinsic elemental abundances in the other three DLAs allowed us to constrain their star 
formation history, their age and their star formation rate by a detailed comparison with a grid of 
chemical evolution models for spiral and dwarf irregular galaxies. Our results show that the 
galaxies associated with these three DLAs in the redshift interval $z_{\rm abs} = 1.7-2.5$ are 
either outer regions of spiral disks (radius $\geq 8$ kpc) or dwarf irregular galaxies (showing a 
bursting or continuous star formation history) with ages varying from some 50 Myr only to $\gtrsim 
3.5$ Gyr and with moderate star formation rates per unit area of $-2.1 < \log \psi < -1.5$ 
M$_{\odot}$ yr$^{-1}$ kpc$^{-2}$.

   }

   \maketitle
%

\section{Introduction}

The study of high redshift galaxies and the access to their physical properties can be done using 
either emission or absorption line spectroscopy. The absorption line spectroscopy is a very 
powerful technique, and it presents several advantages compared to the information which can be 
derived from emission. First, it allows to detect objects up to very high redshift. Indeed, the 
quasars (QSOs) detected up to redshifts of 6.2 \citep{fan01} can be used as background searchlights 
for useful probes of the intervening Universe up to lookback times of 95\% of the age of the 
Universe. Second, through the analysis of QSO absorption lines we can study the spatial 
distribution, motion, chemical enrichment and ionization histories of gaseous structures on a 
variety of scales, ranging from the intergalactic medium to high column density absorption systems 
associated with galaxies. Third, since the detection of material intercepting a line of sight to a 
given QSO is dependent only on the column density of the gas and the luminosity of the QSO, this is 
a unique technique for probing the chemical composition and physical conditions in the interstellar 
medium of various types of galaxies over a large range of lookback times, detected independently of 
their distance, luminosity, star formation history and morphology. 

While the major part of the baryon content of the present-day galaxies is concentrated in stars, it 
must exist an epoch at which the essential of the mass of galaxies still lay in the gas. The 
damped Ly$\alpha$ systems (DLAs) observed in QSO spectra are characterized by large column densities 
of neutral hydrogen, $>2\times 10^{20}$ cm$^{-2}$, and dominate the cosmic mass density of neutral 
hydrogen gas \citep[e.g.][]{storrie96,storrie00}. The general view is that the DLA systems probably 
represent some early stages in the evolution of the galaxies we see around us today, perhaps at a 
time shortly after they had condensed out of the intergalactic medium, but before they had time to 
form many stars, so that most of their mass still resided in the interstellar medium. These objects 
thus are by far our best laboratory for studying the galaxies at high redshift, in their early 
stages of evolution, and for tracking the galactic chemical evolution through the cosmic ages. 

However, the reconstruction of the star formation histories of DLAs from the abundance pattern 
measurements is not straightforward and led to contradictory results 
\citep{lu96,prochaska99,centurion00,molaro00,vladilo02a}. The access to abundance ratios involving 
two elements formed on different timescales, in particular the $\alpha$/Fe ratios, examined 
together with [Fe/H], or any other metallicity tracer such as [Zn/H], is crucial, since the star 
formation history of a galaxy is completely determined by the [$\alpha$/Fe] versus [Fe/H] 
distributions \citep[e.g.][]{matteucci01}. The principal difficulty, however, is to disentangle the 
nucleosynthetic contributions from dust depletion effects. Because we are studying gas-phase 
elemental abundances in DLAs, the observed abundances may not represent the intrinsic composition 
of the system if part of the elements is removed from the gas to the solid phase \citep{savage96}. 
Several pieces of evidence show that some dust is indeed present in DLAs 
\citep[e.g.][]{pei91,prochaska02c}. Another effect which has to be carefully examined, although
usually assumed negligible is the photoionization effect \citep[e.g.][]{viegas95,vladilo01}, since 
in gas-phase studies a fraction of the gas may also be ionized.

The difficulty to correctly evaluate the photoionization effects and the degeneracy between dust 
depletion and nucleosynthesis in DLAs is accentuated by the limited number of ions and elements 
typically detected in these galaxies. To determine the ``intrinsic'' chemical abundance patterns, 
free from ionization and dust depletion effects and to be then able to study the DLA galaxies {\it 
individually}, we need to examine several column density ratios of adjacent ions of the same 
element and the relative abundances of as many elements as possible. We thus aimed at obtaining for 
a few damped Ly$\alpha$ systems the column density measurements of many ions of different 
ionization levels and the abundance measurements of the complete series of accessible elements in 
DLAs. Until now the DLA galaxy population has been analyzed as a whole and chemical evolution models 
were constructed in order to interpret the abundance patterns observed in DLAs as an ensemble, 
considering them as an evolutionary sequence \citep[e.g.][]{matteucci97,jimenez99,hou01,mathlin01}, 
while several pieces of evidence $-$ the low redshift deep imaging revealing a variety of 
morphological types belonging to the DLA population \citep[e.g.][]{lebrun97,nestor02}, the large 
scatter in the $\alpha$ over Fe-peak element abundance ratios at a given metallicity and the large 
scatter observed in the metallicities~$-$ show that the DLAs trace galaxies with different 
evolutionary histories. Some may have formed stars on timescales similar to that of the early Milky 
Way, while others apparently did so more slowly or intermittently, so that the Fe-peak elements 
could catch up with the $\alpha$-elements. The DLA systems thus likely sample a wide range of 
galaxy types, and consequently a variety of star formation histories. It is very important to 
determine the star formation history of each of these high redshift galaxies individually to better 
understand the galaxy formation and evolution.

We thus attempted a new approach for studying the DLA galaxy population, focusing on individual 
systems. By combining our UVES-VLT spectra with the existing HIRES-Keck spectra we 
obtained the column density measurements of some 21 ions, namely of 15 elements $-$ N, O, Mg, Al, 
Si, P, S, Cl, Ar, Ti, Cr, Mn, Fe, Ni, Zn~$-$, in four DLA systems in the redshift interval $z_{\rm 
abs} = 1.7-2.5$. The complete analysis of the four DLA systems is described in 
Sects.~\ref{reduction} and \ref{data-analysis}. In Sect.~\ref{ionization}, we address the question 
of photoionization effects, and in Sect.~\ref{abundances} we analyze the dust content and discuss 
for each system individually the intrinsic chemical abundances. Finally, the relative abundances as 
a function of the metallicity and the redshift are systematically compared with a grid of chemical 
evolution models for spiral and dwarf irregular galaxies in Sect.~\ref{SFH}. 
%

\begin{table*}[t]
\begin{center}
\caption{Journal of observations} 
\label{journal}
\begin{tabular}{l c c c l c c c}
\hline \hline
\\[-0.3cm]
Quasar  & V     & $z_{\rm em}$ & $z_{\rm abs}$ & Observing date & Mode & Wavelength range & Exposure time \\
        & [mag] &              &               &                &      & [nm]             & [s]     
\smallskip 
\\     
\hline  
Q0100+13   & 16.6 & 2.69 & 2.309 & September 2001 & dic2 (390+860) & 330$-$452/670$-$1000 & 7200 \\
Q1331+17   & 16.7 & 2.08 & 1.776 & June-July 2001 & dic1 (346+860) & 312$-$387/670$-$1000 & 13500 \\
Q2231$-$00 & 17.5 & 3.02 & 2.066 & September 2000 & dic2 (390+800) & 335$-$452/610$-$990  & 15700 \\
Q2343+12   & 17.0 & 2.55 & 2.431 & August 2001    & dic1 (346+860) & 315$-$387/670$-$1000 & 14400 \\
\hline
\end{tabular}
\end{center}
\end{table*}
%

\section{Observations and data reduction}\label{reduction}

The selected quasars Q0100+13, Q1331+17, Q2231$-$00 and Q2343+12 with four intervening DLAs in the
redshift interval $z_{\rm abs} = 1.7-2.5$ are relatively bright with V=16.5$-$17.5 and already known 
from the literature. Indeed, the DLAs toward Q0100+13, Q1331+17 and Q2231$-$00 were carefully and 
accurately analyzed by \citet{prochaska99} and \citet{prochaska01} thanks to high-resolution spectra 
obtained with the HIRES echelle spectrograph on the Keck I telescope at Mauna Kea, in Hawaii, and 
some results also obtained from HIRES-Keck spectra on the DLA toward Q2343+12 were presented by e.g. 
\citet{lu98}.

To complete the wavelength coverage of the HIRES-Keck spectra of these quasars, we used the unique 
capability of the Ultraviolet-Visual Echelle Spectrograph UVES \citep{dodorico00} on the VLT 
8.2m Kueyen ESO telescope at Paranal, Chile, and we obtained high resolution, high signal-to-noise
ratio spectra for each quasar in the blue $\lambda$ 3150$-$4500 \AA\ and in the far-red $\lambda$ 
6700$-$10000 \AA. The observations were performed in visitor mode in September 2000 for one object 
and in service mode in summer 2001 for the three other objects under good seeing conditions 
(between 0.5'' and 1.0''). For each observation, slit widths of 1'' in the blue and of 0.9'' in the 
red were chosen with a CCD binning of $2\times 2$ resulting in a resolution of FWHM $\simeq 6.9$ 
km~s$^{-1}$ and 6.4 km~s$^{-1}$, respectively. Relevant details of the observations are collected 
in Table~\ref{journal}. The total exposure times of each quasar were split in multiple exposures 
of 3600 or 4500~s.

The spectra were reduced using the ESO data reduction package {\tt MIDAS} and the UVES pipeline in 
an interactive mode available as a {\tt MIDAS} context. A detailed description of the pipeline can 
be found in \citet{ballester00}. To make sure of the best result, we made a systematic check of 
each step of the pipeline reduction. The wavelengths of the reduced one-dimensional spectra were 
converted to a vacuum-heliocentric scale, and the individual spectra of each object were co-added 
using their signal-to-noise ratio as weights in order to get the maximum signal-to-noise ratios. 
The final step was the normalization of the resulting spectra obtained by dividing them by a spline 
function fitted to smoothly connect the regions free from absorption features. The continuum in the 
Ly$\alpha$ forest was fitted by using small regions deemed to be free of absorptions and by 
interpolating between these regions with a spline. An average signal-to-noise ratio per pixel of 
$\sim 30$, 55 and 45 was achieved in the final spectra at $\lambda$ $\sim 3700$ \AA, 7500 \AA\ and 
9000 \AA, respectively.
%

\section{Data analysis and ionic column densities}\label{data-analysis}

By combining our UVES-VLT spectra with the existing HIRES-Keck spectra obtained by 
\citet{prochaska99} we cover the total spectral range from 3150 to 10000 \AA\ for the four observed 
quasars, and hence have access for the first time to up to 21 ions and 15 elements for each of 
their intervening DLAs.

The column densities of the metal species were derived with the Voigt profile fitting technique.
This technique consists in fitting theoretical Voigt profiles to the observed DLA absorption metal 
lines well described as a complex of components, each defined by a redshift $z$, a Doppler 
parameter $b$, a column density $N$ and the corresponding errors. The fits were performed using a 
$\chi^2$ minimization routine {\tt fitlyman} in {\tt MIDAS} \citep{fontana95}. We assumed that 
metal species with {\em similar ionization potentials} can be fitted using identical component 
fitting parameters, i.e. the same $b$ (which means that macroturbulent motions dominate over 
thermal broadening) and the same $z$ in the same component, and allowing for variations from metal 
species to metal species in $N$ only. We distinguish three categories of metal species with similar 
ionization potentials: the low-ion transitions (i.e. the neutral and singly ionized species), the 
intermediate-ion transitions (e.g. \ion{Fe}{iii}, \ion{Al}{iii}), and the high-ion transitions 
(e.g. \ion{C}{iv}, \ion{Si}{iv}). We used relatively strong (but not saturated) lines to fix the 
component fitting parameters ($b$ and $z$), and excellent profile fits could then be achieved for 
weak metal lines and for metal lines located in the Ly$\alpha$ forest where the probability of 
blending is high by allowing only the column density to vary. We had a sufficient number of 
relatively strong metal-line profiles to well constrain the fitting parameters in the four studied 
DLAs exhibiting multicomponent velocity structures. 
%

\begin{table*}[t]
\begin{center}
\caption{Atomic data} 
\label{atomic-data}
\begin{tabular}{l c c c | l c c c}
\hline \hline
\\[-0.3cm]
Transition & $\lambda$ & $f$-value & Reference & Transition & $\lambda$ & $f$-value & Reference 
\smallskip 
\\     
\hline  
\ion{O}{i}$\lambda$950    &  950.8846 & 0.00157100 & 1 & \ion{C}{i}$\lambda$1328   & 1328.8333 & 0.05804000 & 1 \\
\ion{P}{ii}$\lambda$961   &  961.0410 & 0.34890000 & 1 & \ion{C}{ii}$\lambda$1334  & 1334.5323 & 0.12780000 & 1 \\
\ion{P}{ii}$\lambda$963   &  963.8010 & 1.45800000 & 1 & \ion{C}{ii}$^*$$\lambda$1335&1335.7077& 0.11490000 & 1 \\
\ion{N}{i}$\lambda$963.9  &  963.9903 & 0.01837000 & 1 & \ion{Cl}{i}$\lambda$1347  & 1347.2400 & 0.11860000 & 1 \\
\ion{N}{i}$\lambda$964.6  &  964.6256 & 0.01180000 & 1 & \ion{O}{i}$\lambda$1355   & 1355.5977 & 0.00000124 & 1 \\
\ion{N}{i}$\lambda$965.0  &  965.0413 & 0.00580100 & 1 & \ion{Ni}{ii}$\lambda$1370 & 1370.1310 & 0.07690000 & 5 \\
\ion{O}{i}$\lambda$971    &  971.7380 & 0.01480000 & 1 & \ion{Si}{ii}$\lambda$1526 & 1526.7066 & 0.12700000 & 6 \\
\ion{O}{i}$\lambda$976    &  976.4481 & 0.00330000 & 1 & \ion{C}{i}$\lambda$1560   & 1560.3092 & 0.08041000 & 1 \\
\ion{C}{iii}$\lambda$977  &  977.0200 & 0.76200000 & 1 & \ion{Fe}{ii}$\lambda$1608 & 1608.4511 & 0.05800000 & 7 \\
\ion{O}{i}$\lambda$988    &  988.7734 & 0.04318000 & 1 & \ion{Fe}{ii}$\lambda$1611 & 1611.2005 & 0.00136000 & 2 \\
\ion{Si}{ii}$\lambda$989  &  989.8731 & 0.13300000 & 1 & \ion{C}{i}$\lambda$1656   & 1656.9283 & 0.14050000 & 1 \\
\ion{Cl}{i}$\lambda$1002  & 1002.3464 & 0.07046000 & 1 & \ion{Al}{ii}$\lambda$1670 & 1670.7874 & 1.88000000 & 1 \\
\ion{S}{iii}$\lambda$1012 & 1012.5020 & 0.03550000 & 1 & \ion{Ni}{ii}$\lambda$1709 & 1709.6000 & 0.03240000 & 8 \\
\ion{Si}{ii}$\lambda$1020 & 1020.6989 & 0.02828000 & 1 & \ion{Ni}{ii}$\lambda$1741 & 1741.5490 & 0.04270000 & 8 \\
Ly$\beta$                 & 1025.7223 & 0.07912000 & 1 & \ion{Ni}{ii}$\lambda$1751 & 1751.9100 & 0.02770000 & 8 \\
\ion{Cl}{i}$\lambda$1028  & 1028.1740 & 0.02228000 & 1 & \ion{Si}{ii}$\lambda$1808 & 1808.0126 & 0.00218000 & 9 \\
\ion{Cl}{i}$\lambda$1030  & 1030.8845 & 0.02870000 & 1 & \ion{Al}{iii}$\lambda$1854& 1854.7164 & 0.53900000 & 1 \\
\ion{Cl}{i}$\lambda$1031  & 1031.5070 & 0.15060000 & 1 & \ion{Al}{iii}$\lambda$1862& 1862.7895 & 0.26800000 & 1 \\
\ion{C}{ii}$\lambda$1036  & 1036.3367 & 0.12310000 & 1 & \ion{Ti}{ii}$\lambda$1910.60&1910.6000& 0.20200000 & 10 \\
\ion{O}{i}$\lambda$1039   & 1039.2304 & 0.00919700 & 1 & \ion{Ti}{ii}$\lambda$1910.97&1910.9700& 0.09800000 & 10 \\
\ion{Ar}{i}$\lambda$1048  & 1048.2199 & 0.24410000 & 1 & \ion{Zn}{ii}$\lambda$2026 & 2026.1360 & 0.48900000 & 11 \\
\ion{Ar}{i}$\lambda$1066  & 1066.6600 & 0.06652000 & 1 & \ion{Cr}{ii}$\lambda$2026 & 2026.2690 & 0.00471000 & 12 \\
\ion{N}{ii}$\lambda$1083  & 1083.9900 & 0.10310000 & 1 & \ion{Mg}{i}$\lambda$2026  & 2026.4768 & 0.11200000 & 1 \\
\ion{Fe}{iii}$\lambda$1122& 1122.5260 & 0.05390000 & 2 & \ion{Cr}{ii}$\lambda$2056 & 2056.2539 & 0.10500000 & 11 \\
\ion{N}{i}$\lambda$1134.165&1134.1653 & 0.01342000 & 1 & \ion{Cr}{ii}$\lambda$2062 & 2062.2340 & 0.07800000 & 11 \\
\ion{N}{i}$\lambda$1134.415&1134.4149 & 0.02683000 & 1 & \ion{Zn}{ii}$\lambda$2062 & 2062.6640 & 0.25600000 & 11 \\
\ion{N}{i}$\lambda$1134.98& 1134.9803 & 0.04023000 & 1 & \ion{Cr}{ii}$\lambda$2066 & 2066.1610 & 0.05150000 & 11 \\
\ion{P}{ii}$\lambda$1152  & 1152.8180 & 0.23600000 & 1 & \ion{Fe}{ii}$\lambda$2249 & 2249.8768 & 0.00182100 & 13 \\
\ion{C}{i}$\lambda$1157   & 1157.1857 & 0.02440000 & 1 & \ion{Fe}{ii}$\lambda$2260 & 2260.7805 & 0.00244000 & 13 \\
\ion{S}{iii}$\lambda$1190 & 1190.2080 & 0.02217000 & 1 & \ion{Fe}{ii}$\lambda$2344 & 2344.2140 & 0.11400000 & 14 \\
\ion{N}{i}$\lambda$1199.55& 1199.5496 & 0.13280000 & 1 & \ion{Fe}{ii}$\lambda$2374 & 2374.4612 & 0.03130000 & 14 \\
\ion{N}{i}$\lambda$1200.22& 1200.2233 & 0.08849000 & 1 & \ion{Fe}{ii}$\lambda$2382 & 2382.7650 & 0.32000000 & 14 \\
\ion{N}{i}$\lambda$1200.71& 1200.7098 & 0.04423000 & 1 & \ion{Mn}{ii}$\lambda$2576 & 2576.8770 & 0.35080000 & 1 \\
\ion{Si}{iii}$\lambda$1206& 1206.5000 & 1.66000000 & 1 & \ion{Fe}{ii}$\lambda$2586 & 2586.6500 & 0.06910000 & 14 \\
Ly$\alpha$                & 1215.6701 & 0.41640000 & 1 & \ion{Mn}{ii}$\lambda$2594 & 2594.4990 & 0.27100000 & 1 \\
\ion{Mg}{ii}$\lambda$1239 & 1239.9250 & 0.00063000 & 3 & \ion{Fe}{ii}$\lambda$2600 & 2600.1729 & 0.23900000 & 14 \\
\ion{Mg}{ii}$\lambda$1240 & 1240.3950 & 0.00035000 & 3 & \ion{Mn}{ii}$\lambda$2606 & 2606.4620 & 0.19270000 & 1 \\
\ion{S}{ii}$\lambda$1250  & 1250.5840 & 0.00545300 & 1 & \ion{Mg}{ii}$\lambda$2796 & 2796.3520 & 0.61230000 & 15 \\
\ion{S}{ii}$\lambda$1253  & 1253.8110 & 0.01088000 & 1 & \ion{Mg}{ii}$\lambda$2803 & 2803.5310 & 0.30540000 & 15 \\
\ion{S}{ii}$\lambda$1259  & 1259.5190 & 0.01624000 & 1 & \ion{Mg}{i}$\lambda$2852  & 2852.9642 & 1.81000000 & 1 \\
\ion{C}{i}$\lambda$1277   & 1277.2450 & 0.09665000 & 1 & \ion{Ti}{ii}$\lambda$3067 & 3067.2451 & 0.04655000 & 1 \\
\ion{C}{i}$\lambda$1280   & 1280.1353 & 0.02432000 & 1 & \ion{Ti}{ii}$\lambda$3073 & 3073.8770 & 0.10910000 & 1 \\
\ion{Ti}{iii}$\lambda$1298& 1298.6970 & 0.09507000 & 1 & \ion{Ti}{ii}$\lambda$3230 & 3230.1310 & 0.05860000 & 1 \\
\ion{O}{i}$\lambda$1302   & 1302.1685 & 0.04887000 & 1 & \ion{Ti}{ii}$\lambda$3242 & 3242.9290 & 0.18320000 & 1 \\
\ion{Si}{ii}$\lambda$1304 & 1304.3702 & 0.09400000 & 4 & \ion{Ti}{ii}$\lambda$3384 & 3384.7400 & 0.34000000 & 1 \\
\ion{Ni}{ii}$\lambda$1317 & 1317.2170 & 0.07786000 & 5 & & & & \\
\hline
\end{tabular}
\begin{minipage}{160mm}
\smallskip
REFERENCES: (1)~\citet{morton91}; (2)~\citet{raassen98}; (3)~\citet{theodosiou99}; 
(4)~\citet{tripp96}; (5)~\citet{fedchak99}; (6)~\citet{schectman98}; (7)~\citet{bergeson96}; 
(8)~\citet{fedchak00}; (9)~\citet{bergeson93a}; (10)~\citet{wiese01}; (11)~\citet{bergeson93b}; 
(12)~\citet{verner94}; (13)~\citet{bergeson94}; (14)~\citet{morton00,morton01}; 
(15)~\citet{verner96}.
\end{minipage}
\end{center}
\end{table*}
%

Throughout the analysis we adopted the list of atomic data $-$ laboratory wavelengths and 
oscillator strengths~$-$ presented in Table~\ref{atomic-data}. The most recent measurements of 
$\lambda$ and $f$-values of the metal-ions that impact the abundances of DLA systems and their 
references are reported there. Information on the atomic data of additional useful metal-ions can 
be found in \citet{prochaska01} and on the web site of ``The HIRES Damped Ly$\alpha$ Abundance 
Database''\footnote{http://kingpin.ucsd.edu/$\sim$hiresdla/} maintained by J.~X. Prochaska and 
collaborators. Finally, we adopt the solar meteoritic abundances from \citet{grevesse98}.
%

\begin{table}[t]
\begin{center}
\caption{Component structure of the $z_{\rm abs} = 2.309$ DLA system toward Q0100+13} 
\label{Q0100-t}
\vspace{-0.4cm}
\begin{tabular}{l c c c l c}
\hline \hline
No & $z_{\rm abs}$ & $v_{\rm rel}^*$ & $b (\sigma_b)$ & Ion & $\log N (\sigma_{\log N})$ \\
   &               & km s$^{-1}$     & km s$^{-1}$    &     &                          
\smallskip
\\ 
\hline 
\multicolumn{5}{l}{\hspace{0.3cm} Low-ion transitions} & \\
\hline
1 & 2.309027 & 0     & \phantom{0}5.6{\scriptsize (0.1)} & \ion{Fe}{ii} & 14.86{\scriptsize (0.01)} \\
  &	     &       &  			         & \ion{Zn}{ii} & 12.28{\scriptsize (0.01)} \\
  &	     &       &  			         & \ion{Cr}{ii} & 13.15{\scriptsize (0.01)} \\
  &	     &       &  			         & \ion{S}{ii}  & 14.85{\scriptsize (0.05)} \\
  &	     &       &  			         & \ion{Ar}{i}  & 13.99{\scriptsize (0.15)} \\
  &	     &       &  			         & \ion{P}{ii}  & 12.88{\scriptsize (0.08)} \\
  &	     &       &  			         & \ion{N}{i}	& 14.79{\scriptsize (0.08)} \\
  &	     &       &  			         & \ion{Mg}{ii} & 15.41{\scriptsize (0.07)} \\
2 & 2.309161 & $+$12 & \phantom{0}5.3{\scriptsize (0.1)} & \ion{Fe}{ii} & 14.61{\scriptsize (0.01)} \\
  &	     &       &  			         & \ion{Zn}{ii} & 12.03{\scriptsize (0.02)} \\
  &	     &       &  			         & \ion{Cr}{ii} & 12.96{\scriptsize (0.01)} \\
  &	     &       &  			         & \ion{S}{ii}  & 14.73{\scriptsize (0.08)} \\
  &	     &       &  			         & \ion{Ar}{i}  & 13.81{\scriptsize (0.08)} \\
  &	     &       &  			         & \ion{P}{ii}  & 12.55{\scriptsize (0.10)} \\
  &	     &       &  			         & \ion{N}{i}	& 14.65{\scriptsize (0.12)} \\
  &	     &       &  			         & \ion{Mg}{ii} & 15.06{\scriptsize (0.13)} \\
3 & 2.309443 & $+$37 &           14.1{\scriptsize (0.2)} & \ion{Fe}{ii} & 14.02{\scriptsize (0.03)} \\
4 & 2.309486 & $+$52 & \phantom{0}2.8{\scriptsize (0.3)} & \ion{Fe}{ii} & 12.96{\scriptsize (0.03)} \\
\hline 
\multicolumn{6}{l}{\hspace{0.3cm} Intermediate-ion transitions} \\
\hline
1 & 2.308980 & \phantom{0}$-$4 & \phantom{0}6.0{\scriptsize (1.9)} & \ion{Al}{iii} & 12.18{\scriptsize (0.12)} \\
  &          &                 &                                   & \ion{Fe}{iii} & $< 13.28$ \\
  &          &                 &                                   & \ion{N}{ii}   & $< 13.24$ \\
2 & 2.309093 & \phantom{0}$+$6 &           11.2{\scriptsize (1.2)} & \ion{Al}{iii} & 12.22{\scriptsize (0.08)} \\
  &          &                 &                                   & \ion{Fe}{iii} & $< 13.22$ \\
  &          &                 &                                   & \ion{N}{ii}   & $< 13.64$ \\
3 & 2.309498 &           $+$43 &           13.0{\scriptsize (1.9)} & \ion{Al}{iii} & 12.12{\scriptsize (0.05)} \\
\hline 
\end{tabular}
\begin{minipage}{160mm}
\smallskip
$^*$ Velocity relative to $z=2.309027$
\end{minipage}
\end{center}
\end{table}
%

\begin{figure*}[t]
\centering
   \includegraphics[width=15cm]{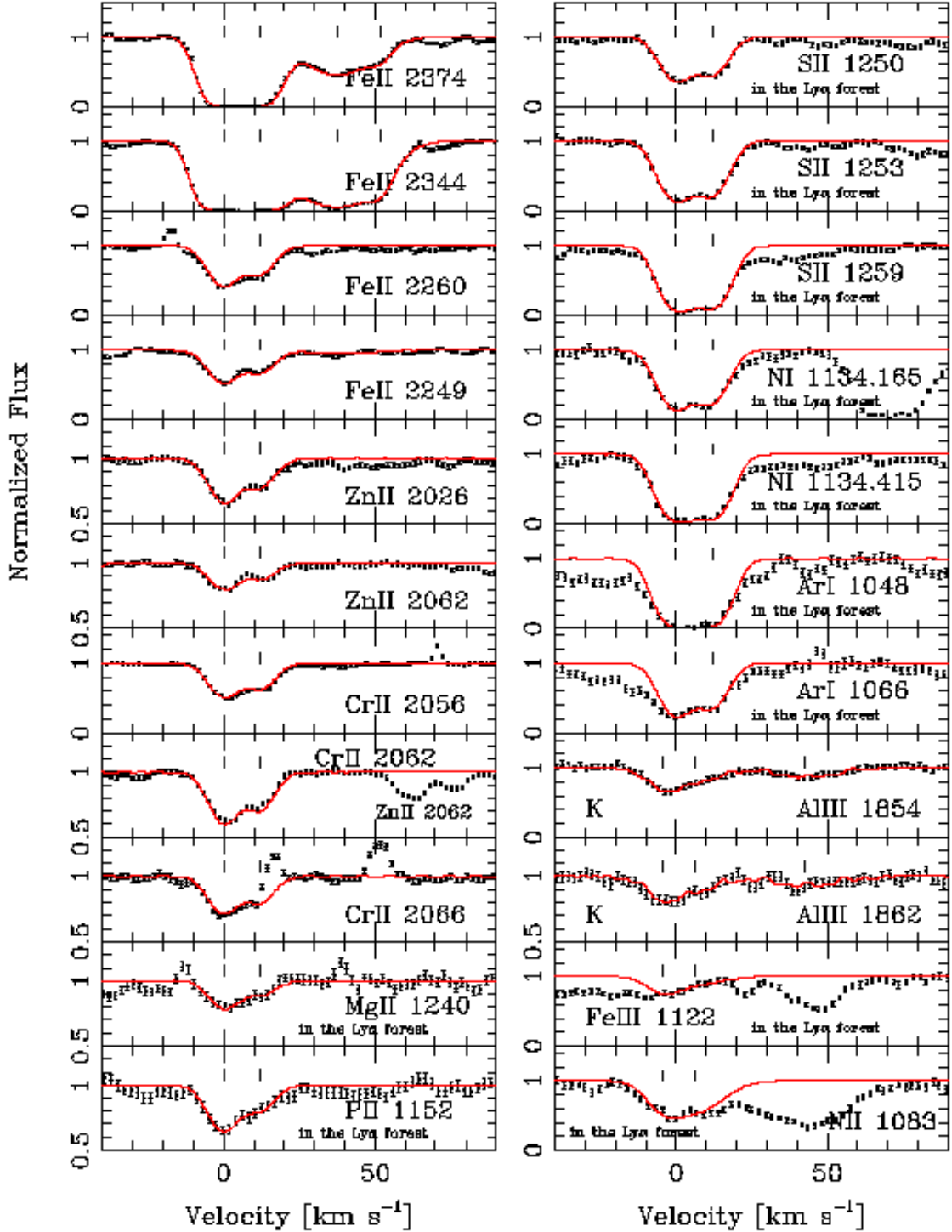}
\caption{Velocity plots of the metal line transitions (normalized intensities shown by dots
with $1 \sigma$ error bars) for the DLA toward Q0100+13. The zero velocity is fixed at $z = 
2.309027$. For this and all the following figures with velocity plots, the vertical bars mark the 
positions of the fitted velocity components and the symbols $\oplus$ correspond to the telluric 
lines. The letter K refers to the Keck spectra.}
\label{Q0100-ions-f}
\end{figure*}

\begin{figure*}[!]
\centering
   \includegraphics[width=17cm]{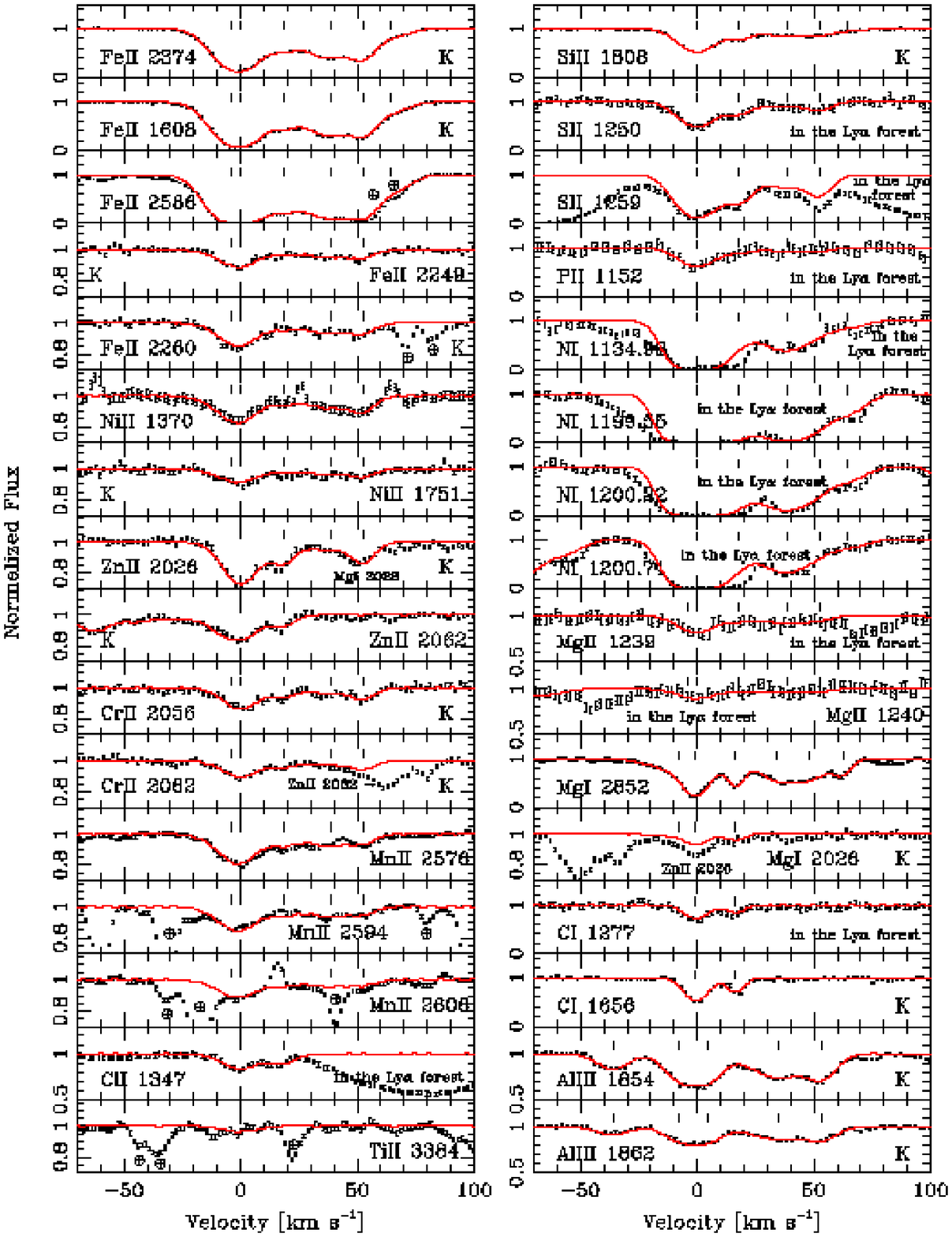}
\caption{Same as Fig.~\ref{Q0100-ions-f} for the DLA toward Q1331+17. The zero velocity is fixed 
at $z = 1.776370$.}
\label{Q1331-ions-f}
\end{figure*}

The measured component per component ionic column densities obtained from the fitting model 
solutions of the low- and intermediate-ion transitions are summarized in 
Tables~\ref{Q0100-t}$-$\ref{Q2343-t-cont}. The reported errors are the 1$\sigma$ errors on 
the fits computed by {\tt fitlyman}. They possibly underestimate the real error on the measure, 
since they do not take into account the uncertainty on the continuum level determination. For the 
components where the line profile is saturated, the column densities are listed as lower limits. 
The values reported as upper limits are cases with significant line blending with \ion{H}{i} clouds
of the Ly$\alpha$ forest or with telluric lines. The fitting solutions of the low- and 
intermediate-ion transitions are shown in Figs.~\ref{Q0100-ions-f}, \ref{Q1331-ions-f}, 
\ref{Q2231-ions-f} and \ref{Q2343-ions-f} for the four DLA. In these velocity plots, $v=0$ 
corresponds to an arbitrary component, and all the identified components are marked by small 
vertical bars. The thin solid line represent the best-fitting solution. The telluric lines have 
been identified thanks to the spectra of a hot, fast rotating star taken in the same nights as the 
scientific exposures.

\begin{figure}[!]
\centering
   \includegraphics[width=9cm]{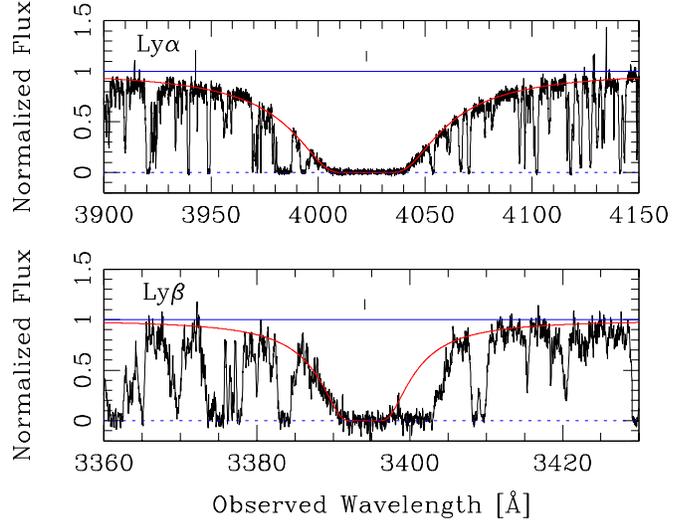}
\caption{Normalized UVES spectra of Q0100+13 showing the DLA Ly$\alpha$ and Ly$\beta$ profiles with
the Voigt profile fits. The vertical bar corresponds to the wavelength centroid of the component 
used for the best fit, $z = 2.309027$. The measured \ion{H}{i} column density is 
$\log N$(\ion{H}{i}) $= 21.37\pm 0.08$.}
\label{Q0100-Ly-f}
\end{figure}

\begin{figure}[!]
\centering
   \includegraphics[width=9cm]{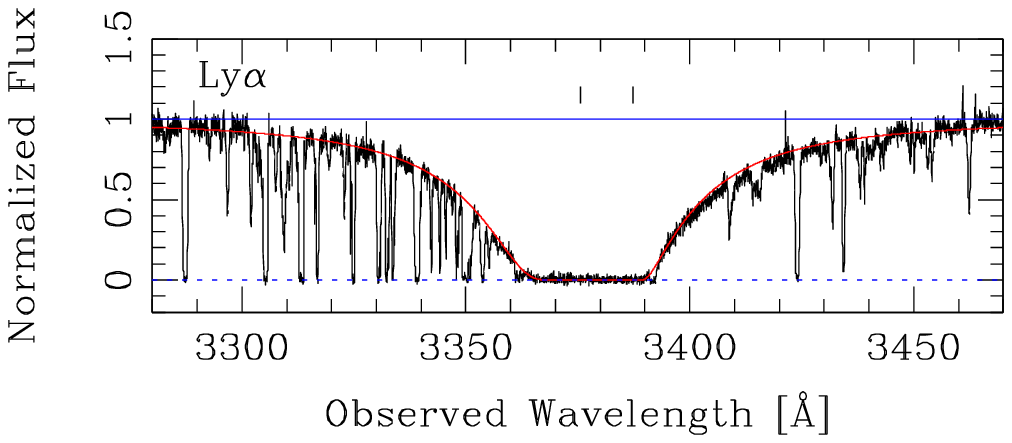}
\caption{Same as Fig.~\ref{Q0100-Ly-f} but for Q1331+17. The vertical bars correspond to the 
wavelength centroids of the components used for the best fit, $z = 1.776724$ and $z = 1.786345$, 
referring to the DLA system and an additional absorber, respectively. The measured \ion{H}{i} 
column densities are $\log N$(\ion{H}{i}) $= 21.14\pm 0.08$ and $19.80\pm 0.10$, respectively.}
\label{Q1331-Lya-f}
\end{figure}

The neutral hydrogen column densities were estimated from the fits of the Ly$\alpha$ damping line 
profiles. The $b$-values were fixed at 20 km~s$^{-1}$, and the redshift $z$ were left as a free 
parameter or fixed to the redshift of the strongest component of the metal-line profiles according
to the system (see comments in the following sub-Sections). When other lines of the Lyman series 
were accessible in our spectra, we used them to check the \ion{H}{i} column densities derived from 
the Ly$\alpha$ lines. Figures~\ref{Q0100-Ly-f}, \ref{Q1331-Lya-f}, \ref{Q2231-Lya-f} and 
\ref{Q2343-Ly-f} show the results of the \ion{H}{i} fitting solutions for the four DLA systems. The 
small vertical bar corresponds to the redshift obtained in the best-fitting solution and the thin 
solid line represents the best fit.

We now briefly describe the fitting results for the four individual DLA systems. Their \ion{Mn}{ii}
$\lambda$2576,2594,2606, \ion{Ti}{ii} $\lambda_{\rm rest} > 3000$ and \ion{Mg}{ii} 
$\lambda$1239,1240 fits and column density measurements have already been discussed in details in 
\citet[][hereafter Paper~I]{dessauges02a}. Let's just remind that the \ion{Mg}{ii} 
$\lambda$1239,1240 lines are detected in the DLA Ly$\alpha$ damping line red wing and that they 
have been fitted after a local renormalization of the spectrum around the \ion{Mg}{ii} lines with 
the fit of the Ly$\alpha$ damping wing profile. The errors on the measured Mg$^+$ column densities 
have been estimated by varying the continuum level by 5\%.
%

%

%

\subsection{Q0100+13, $z_{\rm abs} = 2.309$}

This system was carefully studied by \citet{wolfe94}, and subsequently by \citet{prochaska99} and 
\citet{prochaska01}. Thanks to the UVES spectra we confirm some of their column density 
measurements, namely $N$(Zn$^+$), $N$(Cr$^+$) and $N$(Fe$^+$) obtained from the \ion{Fe}{ii} 
$\lambda$2249,2260,2344,2374 lines not present in the HIRES spectra, and we obtain new column 
density measurements of \ion{S}{ii}, \ion{Ar}{i}, \ion{P}{ii}, \ion{N}{i} and \ion{Mg}{ii} (see 
Fig.~\ref{Q0100-ions-f}). We also get upper limits on the column densities of two intermediate-ions, 
\ion{N}{ii} and \ion{Fe}{iii} located in the Ly$\alpha$ forest, by using the profile shapes of the 
intermediate-ion lines \ion{Al}{iii} $\lambda$1854,1862 observed in the HIRES spectra to constrain 
the fitting parameters (see Fig.~\ref{Q0100-ions-f}).

The low-ion absorption line profiles of this system are characterized by 4 components. Their 
redshifts, $b$-values and column densities are presented in Table~\ref{Q0100-t}. Two of the 4 
components, the components 3 and 4, are too weak to be accurately identified in the weak metal-lines 
and in the metal-lines located in the Ly$\alpha$ forest. Thus, we prefer to avoid measuring them 
in most of the observed transitions. Being weak and since we are interested in the relative 
elemental abundances, this has no impact on the final conclusions on this DLA system. The 
intermediate-ion lines show very similar profiles to the low-ion profiles suggesting they are 
coming from the same absorption regions. Their fitting solutions are given in Table~\ref{Q0100-t}. 
%

\begin{table*}[t]
\begin{center}
\caption{Component structure of the $z_{\rm abs} = 1.776$ DLA system toward Q1331+17} 
\label{Q1331-t}
\begin{tabular}{l c c c l c | l c c c l c}
\hline\hline
No & $z_{\rm abs}$ & $v_{\rm rel}^*$ & $b (\sigma_b)$ & Ion & $\log N (\sigma_{\log N})$ & No & $z_{\rm abs}$ & $v_{\rm rel}^*$ & $b (\sigma_b)$ & Ion & $\log N (\sigma_{\log N})$ \\
   &               & km s$^{-1}$     & km s$^{-1}$    &     &                            &    &               & km s$^{-1}$     & km s$^{-1}$    &     &                            
\smallskip
\\ 
\hline
\multicolumn{5}{l}{\hspace{0.3cm} Low-ion transitions} & & & & & & & \\
\hline
1 & 1.776336 & \phantom{1}$-$4 & 12.0{\scriptsize (0.3)} & \ion{Fe}{ii} &           13.90{\scriptsize (0.04)} &  4 & 1.776724 & $+$38 & 	  13.4{\scriptsize (0.8)} & \ion{Fe}{ii} & 14.07{\scriptsize (0.02)} \\
  &	     &  	       &                         & \ion{Cr}{ii} & \phantom{1}9.00{\scriptsize (0.25)} &    &	      &       & 				  & \ion{Zn}{ii} & 11.50{\scriptsize (0.06)} \\
  &	     &  	       &                         & \ion{Ni}{ii} &           12.82{\scriptsize (0.13)} &    &	      &       & 				  & \ion{Cr}{ii} & 12.22{\scriptsize (0.06)} \\
  &	     &  	       &                         & \ion{Mn}{ii} &           11.84{\scriptsize (0.06)} &    &	      &       & 				  & \ion{Ni}{ii} & 12.87{\scriptsize (0.04)} \\
  &	     &  	       &                         & \ion{Mg}{ii} &           11.00{\scriptsize (0.25)} &    &	      &       & 				  & \ion{Mn}{ii} & 11.91{\scriptsize (0.01)} \\
2 & 1.776370 &   \phantom{+1}0 & \phantom{1}9.8{\scriptsize (0.1)} & \ion{Fe}{ii} & 14.16{\scriptsize (0.02)} &    &	      &       & 				  & \ion{Si}{ii} & 14.60{\scriptsize (0.01)} \\
  &	     &  	       &                                   & \ion{Zn}{ii} & 12.35{\scriptsize (0.01)} &    &	      &       & 				  & \ion{S}{ii}  & 14.15{\scriptsize (0.11)} \\
  &	     &  	       &                                   & \ion{Cr}{ii} & 12.67{\scriptsize (0.01)} &    &	      &       & 				  & \ion{P}{ii}  & 12.41{\scriptsize (0.21)} \\
  &	     &  	       &                                   & \ion{Ni}{ii} & 12.89{\scriptsize (0.10)} &    &	      &       & 				  & \ion{N}{i}   & $< 14.30$ \\ 	       
  &	     &  	       &                                   & \ion{Mn}{ii} & 12.04{\scriptsize (0.03)} &    &	      &       &           			  & \ion{Mg}{ii} & 14.99{\scriptsize (0.14)} \\
  &	     &  	       &                                   & \ion{Si}{ii} & 15.06{\scriptsize (0.01)} &  5 & 1.776859 & $+$53 & \phantom{1}5.5{\scriptsize (0.3)} & \ion{Fe}{ii} & 13.63{\scriptsize (0.05)} \\
  &	     &  	       &                                   & \ion{Ti}{ii} & 11.34{\scriptsize (0.10)} &    &	      &       & 				  & \ion{Zn}{ii} & 11.43{\scriptsize (0.04)} \\
  &	     &  	       &                                   & \ion{S}{ii}  & 14.88{\scriptsize (0.12)} &    &	      &       & 				  & \ion{Cr}{ii} & 12.12{\scriptsize (0.05)} \\
  &	     &  	       &                                   & \ion{P}{ii}  & 13.11{\scriptsize (0.04)} &    &	      &       & 				  & \ion{Ni}{ii} & 12.58{\scriptsize (0.05)} \\
  &	     &  	       &                                   & \ion{N}{i}   & $< 15.16$                 &    &	      &       & 				  & \ion{Mn}{ii} & 11.33{\scriptsize (0.04)} \\
  &          &                 &                                   & \ion{Cl}{i}  & 12.90{\scriptsize (0.03)} &    &	      &       & 				  & \ion{Si}{ii} & 14.05{\scriptsize (0.03)} \\
  &	     &  	       &                                   & \ion{Mg}{ii} & 15.27{\scriptsize (0.12)} &    &	      &       & 				  & \ion{S}{ii}  & 14.05{\scriptsize (0.10)} \\
3 & 1.776538 &           $+$18 & \phantom{1}6.0{\scriptsize (0.3)} & \ion{Fe}{ii} & 13.46{\scriptsize (0.04)} &    &	      &       & 				  & \ion{P}{ii}  & 12.00{\scriptsize (0.30)} \\
  &	     &  	       &                                   & \ion{Zn}{ii} & 11.81{\scriptsize (0.02)} &    &	      &       & 				  & \ion{N}{i}   & $< 12.82$ \\ 	       
  &	     &  	       &                                   & \ion{Cr}{ii} & 12.07{\scriptsize (0.04)} &    &	      &       &         			  & \ion{Mg}{ii} & 14.31{\scriptsize (0.26)} \\
  &	     &  	       &                                   & \ion{Ni}{ii} & 12.27{\scriptsize (0.10)} &  6 & 1.776964 & $+$64 & \phantom{1}9.4{\scriptsize (0.9)} & \ion{Fe}{ii} & 13.23{\scriptsize (0.07)} \\
  &	     &  	       &                                   & \ion{Mn}{ii} & 11.53{\scriptsize (0.02)} &    &	      &       & 				  & \ion{Si}{ii} & 13.92{\scriptsize (0.05)} \\
  &	     &  	       &                                   & \ion{Si}{ii} & 14.40{\scriptsize (0.01)} &    &	      &       & 				  & \ion{N}{i}   & $< 13.33$ \\ 	       
  &	     &  	       &                                   & \ion{S}{ii}  & 14.27{\scriptsize (0.07)} &  & & & & & \\											       
  &	     &  	       &                                   & \ion{P}{ii}  & 12.19{\scriptsize (0.25)} &  & & & & & \\											       
  &	     &  	       &                                   & \ion{N}{i}   & $< 13.60$                 &  & & & & & \\											       
  &          &                 &                                   & \ion{Cl}{i}  & 12.50{\scriptsize (0.06)} &  & & & & & \\
  &	     &  	       &                                   & \ion{Mg}{ii} & 14.52{\scriptsize (0.19)} &  & & & & & \\
\hline
1 & 1.776299 & \phantom{0}$-$8\phantom{.0} &           15.0{\scriptsize (1.0)} & \ion{Mg}{i} & 11.72{\scriptsize (0.07)} & 4 & 1.776684 & $+$34 & 8.6{\scriptsize (0.4)} & \ion{Mg}{i} & 11.83{\scriptsize (0.07)} \\
2 & 1.776365 & \phantom{0}$-0.5$           & \phantom{0}6.0{\scriptsize (0.2)} & \ion{Mg}{i} & 12.08{\scriptsize (0.03)} & 5 & 1.776814 & $+$48 & 9.1{\scriptsize (1.1)} & \ion{Mg}{i} & 11.85{\scriptsize (0.07)} \\
  &	     &                             &                                   & \ion{C}{i}  & 13.12{\scriptsize (0.02)} & 6 & 1.776947 & $+$62 & 3.0{\scriptsize (1.3)} & \ion{Mg}{i} & 11.39{\scriptsize (0.05)} \\
3 & 1.776523 &           $+$17\phantom{.0} & \phantom{0}3.4{\scriptsize (0.3)} & \ion{Mg}{i} & 11.77{\scriptsize (0.01)} & & & & & & \\
  &	     &  	                   &                                   & \ion{C}{i}  & 12.72{\scriptsize (0.02)} & & & & & & \\
\hline
\multicolumn{6}{l}{\hspace{0.3cm} Intermediate-ion transitions} & & & & & & \\
\hline
1 & 1.776036 &           $-$36 & 8.4{\scriptsize (0.4)} & \ion{Al}{iii} & 12.03{\scriptsize (0.02)} & 4 & 1.776694 & $+$35 &	       15.1{\scriptsize (1.4)} & \ion{Al}{iii} & 12.49{\scriptsize (0.04)} \\
2 & 1.776300 & \phantom{0}$-$7 & 8.1{\scriptsize (1.1)} & \ion{Al}{iii} & 12.29{\scriptsize (0.13)} & 5 & 1.776866 & $+$54 & \phantom{0}7.2{\scriptsize (0.5)} & \ion{Al}{iii} & 12.16{\scriptsize (0.06)} \\
3 & 1.776414 & \phantom{0}$+$5 & 7.8{\scriptsize (1.4)} & \ion{Al}{iii} & 12.29{\scriptsize (0.14)} & & & & & & \\
\hline
\end{tabular}
\begin{minipage}{160mm}
\smallskip
$^*$ Velocity relative to $z=1.776370$ 
\end{minipage}
\end{center}										     
\end{table*}
%

Figure~\ref{Q0100-Ly-f} shows the fitting solutions of two Lyman lines of this DLA, Ly$\alpha$ and 
Ly$\beta$, observed in the UVES spectra. The fits were obtained by fixing the $b$-value at 20 
km~s$^{-1}$ and the redshift at $z = 2.309027$, i.e. at the redshift of the strongest metal-line 
component (the component 1). The measured \ion{H}{i} column density, $\log N$(\ion{H}{i}) $= 
21.37\pm 0.08$, is very similar to the one derived by \citet{pettini90} from low-resolution spectra. 
Giving the high H$^0$ column density, the red wing of the Ly$\alpha$ damping profile extends over 
the \ion{S}{ii} $\lambda$1250,1253,1259 lines. Similarly to the \ion{Mg}{ii} $\lambda$1240 line 
(the \ion{Mg}{ii} $\lambda$1239 line being heavily blended), we had first to locally renormalize 
the spectrum with the fit of the Ly$\alpha$ damping wing profile before measuring the S$^+$ column 
density.
%
					     
%

%

\subsection{Q1331+17, $z_{\rm abs} = 1.776$}\label{q1331}

This famous DLA system has been studied by a large number of authors. We mention here only the work
of \citet{prochaska99} and \citet{prochaska01} who obtained the highest quality data and made the 
most accurate analysis. By extending their wavelength coverage with our UVES data, we obtained 
additional column density measurements of \ion{S}{ii}, \ion{P}{ii}, \ion{N}{i}, \ion{Mg}{ii}, 
\ion{Mn}{ii} and \ion{Cl}{i}. We confirm the \citet{prochaska99} and \citet{prochaska01} column 
density measurements of \ion{Si}{ii}, \ion{Fe}{ii}, \ion{C}{i}, \ion{Mg}{i} and \ion{Al}{iii} (see 
Fig.~\ref{Q1331-ions-f}). But, their $N$(Ni$^+$) and $N$(Cr$^+$) measurements differ from ours by 
almost 0.1~dex. We also confirm the revised value of $N$(Zn$^+$) of \citet{prochaska01} obtained by 
correcting the contamination of \ion{Zn}{ii} $\lambda$2026 by the \ion{Mg}{i} $\lambda$2026 profile. 
The detection of the relatively strong \ion{Mg}{i} $\lambda$2852 line in the UVES spectra has 
allowed us to measure even more reliably this contamination. Finally, we obtain a more constraining 
upper limit for $N$(Ti$^+$) from the \ion{Ti}{ii} $\lambda$3384 line than the higher value deduced 
by \citet{prochaska01} from the \ion{Ti}{ii} $\lambda$1910 lines having a 3 times lower oscillator 
strength.

The fitting solutions of the detected components in the low-ion metal-lines are presented in 
Table~\ref{Q1331-t}. A surprising particularity of this system known as a system exhibiting the 
largest dust depletion level of any DLA is that the fits of the refractory element lines require 6 
components (2 components at $v\sim 0$ km~s$^{-1}$, see Fig.~\ref{Q1331-ions-f}), while the 
non-refractory and mildly refractory element lines are well fitted with 5 components only (1 
component at $v = 0$ km~s$^{-1}$, see Fig.~\ref{Q1331-ions-f}). The refractory elements showing 
weak absorption lines, like the \ion{Mg}{ii} $\lambda$1239,1240 and \ion{Cr}{ii} $\lambda$2056,2062
lines, are borderline cases, in the sense that the component 1 is detected but only marginally. 
Concerning the component 6, it is only observed in the strongest metal-line transitions.
%

\begin{figure}[!]
\centering
   \includegraphics[width=9cm]{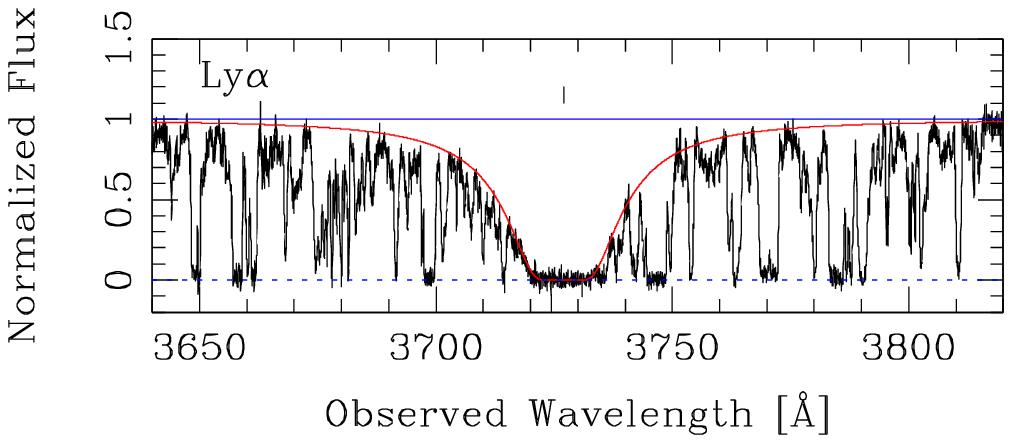}
\caption{Same as Fig.~\ref{Q0100-Ly-f} but for Q2231$-$00. The vertical bar corresponds to the 
wavelength centroid of the component used for the best fit left as a free parameter, $z = 2.065774$.
The measured \ion{H}{i} column density is $\log N$(\ion{H}{i}) $= 20.53\pm 0.08$.}
\label{Q2231-Lya-f}
\end{figure}
%

The detection of the \ion{Cl}{i} $\lambda$1347 line is exceptional. Cl$^0$ has previously been 
observed only by \citet{ledoux02} and \citet{prochaska03b} in the DLA toward Q0551$-$366 and in the
DLA toward FJ081240.6+320808, respectively. In the DLA system studied in this paper, we detect only 
the components 2 and 3 of the \ion{Cl}{i} $\lambda$1347 line, the other components 4, 5 and 6 are 
blended with Ly$\alpha$ forest absorptions. However, since the line profiles of \ion{S}{ii} 
$\lambda$1250, \ion{Si}{ii} $\lambda$1808 and \ion{C}{i} $\lambda$1656 show very little absorption 
at $v > 25$ km~s$^{-1}$ (only $\sim 20$\% of the total column density for \ion{S}{ii} and 
\ion{Si}{ii}), we consider the \ion{Cl}{i} column density derived from the components 2 and 3 as a 
value. We adopt an error on $N$(\ion{Cl}{i}) of 0.10~dex to account for the possible contribution 
from the blended components 4, 5 and 6. Nevertheless, the Cl$^0$ column density measurement provides 
only a strict lower limit on the Cl absolute abundance of [Cl/H] $> -1.37$, since the dominant state 
of Cl should be \ion{Cl}{ii} in DLAs (the ionization potential of Cl$^0$ is 13.01 eV which is lower 
than that of hydrogen). The \ion{Cl}{ii} column density cannot unfortunately be determined in this 
DLA system due to the low rest-wavelengths of the \ion{Cl}{ii} lines.

The \ion{N}{i} column density is not easy to derive in this DLA. First, given the high \ion{H}{i} 
column density of this DLA system, the \ion{N}{i} triplet at $\lambda_{\rm rest} \sim 1200$ \AA\ is 
blended with the blue damping wing of the DLA Ly$\alpha$ absorption line. We thus have to 
renormalize the spectral region covering the \ion{N}{i} $\lambda$1200 triplet with the fit of the 
Ly$\alpha$ damping wing profile, before fitting the \ion{N}{i} lines. Second, it is difficult to 
determine whether the derived \ion{N}{i} column density is a lower or an upper limit. Indeed, the 
component 2 of all the detected \ion{N}{i} lines is saturated, even in the weakest \ion{N}{i} 
transition at $\lambda_{\rm rest} = 1134.98$ \AA. Thus, we are inclined to assume that the derived 
column density is a lower limit. But, when looking carefully at the \ion{N}{i} line profiles, they 
all show a stronger component 4 than the component 5, on the opposite to what is observed in other 
low-ionization transition profiles. This suggests that the \ion{N}{i} lines are likely blended with 
Ly$\alpha$ forest absorptions. Therefore, we finally assume that the derived \ion{N}{i} column 
density is an upper limit.

While fitting the Ly$\alpha$ damping line, we found it necessary to include, besides the 
contribution of the DLA at $z = 1.776724$ corresponding to the redshift of the metal-line component 
4, a second absorber shifted by about 1079 km~s$^{-1}$ redwards of the DLA system (see 
Fig.~\ref{Q1331-Lya-f}). The redshift of this second absorber, $z_{\rm abs} = 1.786$, is accurately 
determined thanks to the detections of several associated metal-lines \citep[see also][]{lopez03}. 
The derived \ion{H}{i} column densities of the DLA and the second absorber are $\log 
N$(\ion{H}{i}) $= 21.14\pm 0.08$ and $19.80\pm 0.10$, respectively. They confirm the previous 
measurements made by \citet{pettini94}.
%

\begin{table*}[t]
\begin{center}
\caption{Component structure of the $z_{\rm abs} = 2.066$ DLA system toward Q2231$-$00} 
\label{Q2231-t}
\begin{tabular}{l c c c l c | l c c c l c}
\hline\hline
No & $z_{\rm abs}$ & $v_{\rm rel}^*$ & $b (\sigma_b)$ & Ion & $\log N (\sigma_{\log N})$ & No & $z_{\rm abs}$ & $v_{\rm rel}^*$ & $b (\sigma_b)$ & Ion & $\log N (\sigma_{\log N})$ \\
   &               & km s$^{-1}$     & km s$^{-1}$    &     &                            &    &               & km s$^{-1}$     & km s$^{-1}$    &     &                            
\smallskip
\\ 
\hline
\multicolumn{5}{l}{\hspace{0.3cm} Low-ion transitions} & & & & & & & \\
\hline
1 & 2.064746 &            $-$138 & \phantom{0}6.6{\scriptsize (0.5)} & \ion{Fe}{ii} & 13.06{\scriptsize (0.04)}	& 8  & 2.065887 &	     $-$27 & 4.9{\scriptsize (0.1)} & \ion{Fe}{ii} & 13.41{\scriptsize (0.02)} \\
  &	     &  	         &                                   & \ion{Si}{ii} & 13.53{\scriptsize (0.12)}	&    &  	&		   &			    & \ion{Si}{ii} & 13.87{\scriptsize (0.01)} \\
2 & 2.064897 &            $-$123 & \phantom{0}9.0{\scriptsize (0.7)} & \ion{Fe}{ii} & 13.20{\scriptsize (0.07)} &    &  	&		   &			    & \ion{S}{ii}  & 13.80{\scriptsize (0.26)} \\
  &	     &  	         &                                   & \ion{Si}{ii} & 13.79{\scriptsize (0.06)}	&    &  	&		   &			    & \ion{N}{i}   & 13.59{\scriptsize (0.22)} \\
3 & 2.065087 &            $-$105 &           13.8{\scriptsize (1.4)} & \ion{Fe}{ii} & 13.58{\scriptsize (0.04)} & 9  & 2.066072 & \phantom{0}$-$9  & 5.5{\scriptsize (0.4)} & \ion{Fe}{ii} & 14.00{\scriptsize (0.05)} \\
  &	     &  	         &                                   & \ion{Si}{ii} & 13.97{\scriptsize (0.10)}	&    &  	&		   &			    & \ion{Zn}{ii} & 11.71{\scriptsize (0.06)} \\
4 & 2.065266 & \phantom{0}$-$88  & \phantom{0}8.6{\scriptsize (0.2)} & \ion{Fe}{ii} & 14.19{\scriptsize (0.01)} &    &  	&		   &			    & \ion{Cr}{ii} & 12.12{\scriptsize (0.08)} \\
  &          &  	         &                                   & \ion{Zn}{ii} & 11.63{\scriptsize (0.04)} &    &  	&		   &			    & \ion{Ni}{ii} & 12.73{\scriptsize (0.03)} \\
  &	     &  	         &                                   & \ion{Cr}{ii} & 12.46{\scriptsize (0.03)} &    &  	&		   &			    & \ion{Mn}{ii} & 11.92{\scriptsize (0.05)} \\
  &	     &  	         &                                   & \ion{Ni}{ii} & 13.04{\scriptsize (0.10)} &    &  	&		   &			    & \ion{Si}{ii} & 14.59{\scriptsize (0.04)} \\
  &	     &  	         &                                   & \ion{Mn}{ii} & 11.97{\scriptsize (0.03)} &    &  	&		   &			    & \ion{Ti}{ii} & 11.88{\scriptsize (0.18)} \\
  &	     &  	         &                                   & \ion{Si}{ii} & 14.57{\scriptsize (0.04)} &    &  	&		   &			    & \ion{S}{ii}  & 14.33{\scriptsize (0.13)} \\
  &	     &  	         &                                   & \ion{S}{ii}  & 14.19{\scriptsize (0.13)} &    &  	&		   &			    & \ion{P}{ii}  & $< 12.81$ \\
  &	     &  	         &                                   & \ion{P}{ii}  & $< 12.90$                 &    &  	&		   &			    & \ion{N}{i}   & 14.43{\scriptsize (0.06)} \\
  &	     &  	         &                                   & \ion{N}{i}   & 13.95{\scriptsize (0.11)} & 10 & 2.066161 & \phantom{$-$00}0 & 3.8{\scriptsize (0.2)} & \ion{Fe}{ii} & 14.26{\scriptsize (0.03)} \\
5 & 2.065422 & \phantom{0}$-$72  & \phantom{0}4.2{\scriptsize (0.5)} & \ion{Fe}{ii} & 13.59{\scriptsize (0.03)} &    &  	&		   &			    & \ion{Zn}{ii} & 11.98{\scriptsize (0.03)} \\
  &          &  	         &                                   & \ion{Zn}{ii} & 10.92{\scriptsize (0.17)} &    &  	&		   &			    & \ion{Cr}{ii} & 12.60{\scriptsize (0.02)} \\
  &	     &  	         &                                   & \ion{Cr}{ii} & 11.77{\scriptsize (0.13)} &    &  	&		   &			    & \ion{Ni}{ii} & 13.01{\scriptsize (0.02)} \\
  &	     &  	         &                                   & \ion{Ni}{ii} & 12.44{\scriptsize (0.17)} &    &  	&		   &			    & \ion{Mn}{ii} & 12.16{\scriptsize (0.03)} \\
  &	     &  	         &                                   & \ion{Mn}{ii} & 11.46{\scriptsize (0.07)} &    &  	&		   &			    & \ion{Si}{ii} & 14.83{\scriptsize (0.03)} \\
  &	     &  	         &                                   & \ion{Si}{ii} & 13.90{\scriptsize (0.06)} &    &  	&		   &			    & \ion{Ti}{ii} & 12.58{\scriptsize (0.06)} \\
  &	     &  	         &                                   & \ion{S}{ii}  & 13.87{\scriptsize (0.16)} &    &  	&		   &			    & \ion{S}{ii}  & 14.66{\scriptsize (0.13)} \\
  &	     &  	         &                                   & \ion{P}{ii}  & $< 12.32$                 &    &  	&		   &			    & \ion{P}{ii}  & $< 13.03$ \\
  &	     &  	         &                                   & \ion{N}{i}   & 13.22{\scriptsize (0.03)} &    &  	&		   &			    & \ion{N}{i}   & 14.76{\scriptsize (0.22)} \\
6 & 2.065546 & \phantom{0}$-$60  & \phantom{0}6.4{\scriptsize (0.2)} & \ion{Fe}{ii} & 13.58{\scriptsize (0.02)} & 11 & 2.066275 &	     $+$11 & 5.4{\scriptsize (0.1)} & \ion{Fe}{ii} & 13.72{\scriptsize (0.01)} \\
  &	     &  	         &                                   & \ion{Si}{ii} & 13.83{\scriptsize (0.08)} &    &  	&		   &			    & \ion{Cr}{ii} & 12.07{\scriptsize (0.08)} \\
  &	     &  	         &                                   & \ion{S}{ii}  & 13.99{\scriptsize (0.14)} &    &  	&		   &			    & \ion{Ni}{ii} & 12.70{\scriptsize (0.04)} \\
  &	     &  	         &                                   & \ion{N}{i}   & 13.25{\scriptsize (0.13)} &    &  	&		   &			    & \ion{Mn}{ii} & 11.60{\scriptsize (0.06)} \\
7 & 2.065698 & \phantom{0}$-$45  & \phantom{0}4.7{\scriptsize (0.3)} & \ion{Fe}{ii} & 13.18{\scriptsize (0.02)} &    &  	&		   &			    & \ion{Si}{ii} & 13.94{\scriptsize (0.03)} \\
  &	     &  	         &                                   & \ion{Si}{ii} & 13.57{\scriptsize (0.01)} &    &  	&		   &			    & \ion{S}{ii}  & 14.06{\scriptsize (0.21)} \\
  &	     &  	         &                                   & \ion{S}{ii}  & 13.89{\scriptsize (0.25)} &    &  	&		   &			    & \ion{P}{ii}  & $< 12.74$ \\
  &	     &  	         &                                   & \ion{N}{i}   & 13.30{\scriptsize (0.23)} &    &  	&		   &			    & \ion{N}{i}   & 13.50{\scriptsize (0.23)} \\
\hline 
1 & 2.064788 &            $-$134 & \phantom{0}7.6{\scriptsize (3.0)} & \ion{Mg}{i}  & 11.04{\scriptsize (0.12)} &  6 & 2.065903 &	     $-$25\phantom{.0} & 7.0{\scriptsize (0.7)} & \ion{Mg}{i} & 11.36{\scriptsize (0.03)} \\ 
2 & 2.064981 &            $-$115 & \phantom{0}7.9{\scriptsize (2.1)} & \ion{Mg}{i}  & 11.12{\scriptsize (0.09)} &  7 & 2.066064 & \phantom{0}$-$9\phantom{.0}  & 5.2{\scriptsize (0.4)} & \ion{Mg}{i} & 11.72{\scriptsize (0.05)} \\ 
3 & 2.065233 & \phantom{0}$-$91  & \phantom{0}8.8{\scriptsize (0.5)} & \ion{Mg}{i}  & 11.87{\scriptsize (0.03)} &  8 & 2.066155 & \phantom{0}$-$0.6            & 3.1{\scriptsize (1.3)} & \ion{Mg}{i} & 12.11{\scriptsize (0.06)} \\
4 & 2.065461 & \phantom{0}$-$68  &           14.7{\scriptsize (1.7)} & \ion{Mg}{i}  & 11.65{\scriptsize (0.05)} &  9 & 2.066274 &	     $+$11\phantom{.0} & 4.6{\scriptsize (1.0)} & \ion{Mg}{i} & 11.59{\scriptsize (0.06)} \\ 
5 & 2.065692 & \phantom{0}$-$46  & \phantom{0}3.0{\scriptsize (1.0)} & \ion{Mg}{i}  & 11.05{\scriptsize (0.06)} &    &	        & &		 &			     \\
\hline
\multicolumn{6}{l}{\hspace{0.3cm} Intermediate-ion transitions} & & & & & & \\
\hline
1 & 2.064758 &           $-$137 & 3.0{\scriptsize (1.2)} & \ion{Al}{iii} & 11.72{\scriptsize (0.07)} & 6  & 2.065648 &  	  $-$50 &	    10.7{\scriptsize (2.3)} & \ion{Al}{iii} & 11.88{\scriptsize (0.07)} \\
  &          &                  &                        & \ion{Fe}{iii} & $< 13.64$                 &    &	     &  		&				    & \ion{Fe}{iii} & $< 13.83$ \\
2 & 2.064909 &           $-$122 & 6.6{\scriptsize (1.5)} & \ion{Al}{iii} & 12.00{\scriptsize (0.05)} & 7  & 2.065925 &  	  $-$23 & \phantom{0}8.5{\scriptsize (2.7)} & \ion{Al}{iii} & 12.10{\scriptsize (0.10)} \\
  &          &                  &                        & \ion{Fe}{iii} & $< 13.92$                 &    &	     &  		&				    & \ion{Fe}{iii} & $< 13.83$ \\
3 & 2.065123 &           $-$101 & 8.5{\scriptsize (2.3)} & \ion{Al}{iii} & 12.01{\scriptsize (0.10)} & 8  & 2.066076 &  \phantom{0}$-$8 & \phantom{0}3.3{\scriptsize (1.0)} & \ion{Al}{iii} & 12.04{\scriptsize (0.19)} \\
  &          &                  &                        & \ion{Fe}{iii} & $< 13.78$                 &    &	     &  		&				    & \ion{Fe}{iii} & $< 13.59$ \\
4 & 2.065274 & \phantom{0}$-$87 & 6.5{\scriptsize (0.7)} & \ion{Al}{iii} & 12.50{\scriptsize (0.03)} & 9  & 2.066165 & \phantom{$-$00}0 & \phantom{0}5.1{\scriptsize (2.5)} & \ion{Al}{iii} & 12.51{\scriptsize (0.09)} \\
  &          &                  &                        & \ion{Fe}{iii} & $< 13.59$                 &    &	     &  		&				    & \ion{Fe}{iii} & $< 13.74$ \\
5 & 2.065435 & \phantom{0}$-$71 & 3.0{\scriptsize (1.6)} & \ion{Al}{iii} & 11.97{\scriptsize (0.02)} & 10 & 2.066290 &  	  $+$13 & \phantom{0}4.8{\scriptsize (1.3)} & \ion{Al}{iii} & 11.96{\scriptsize (0.13)} \\
  &          &                  &                        & \ion{Fe}{iii} & $< 13.77$                 &    &	     &  	        &				   & \ion{Fe}{iii} & $< 13.59$ \\
\hline
\end{tabular}
\begin{minipage}{160mm}
\smallskip
$^*$ Velocity relative to $z=2.066161$ 
\end{minipage}
\end{center}										     
\end{table*}	
%

\begin{figure*}[!]
\centering
   \includegraphics[width=17cm]{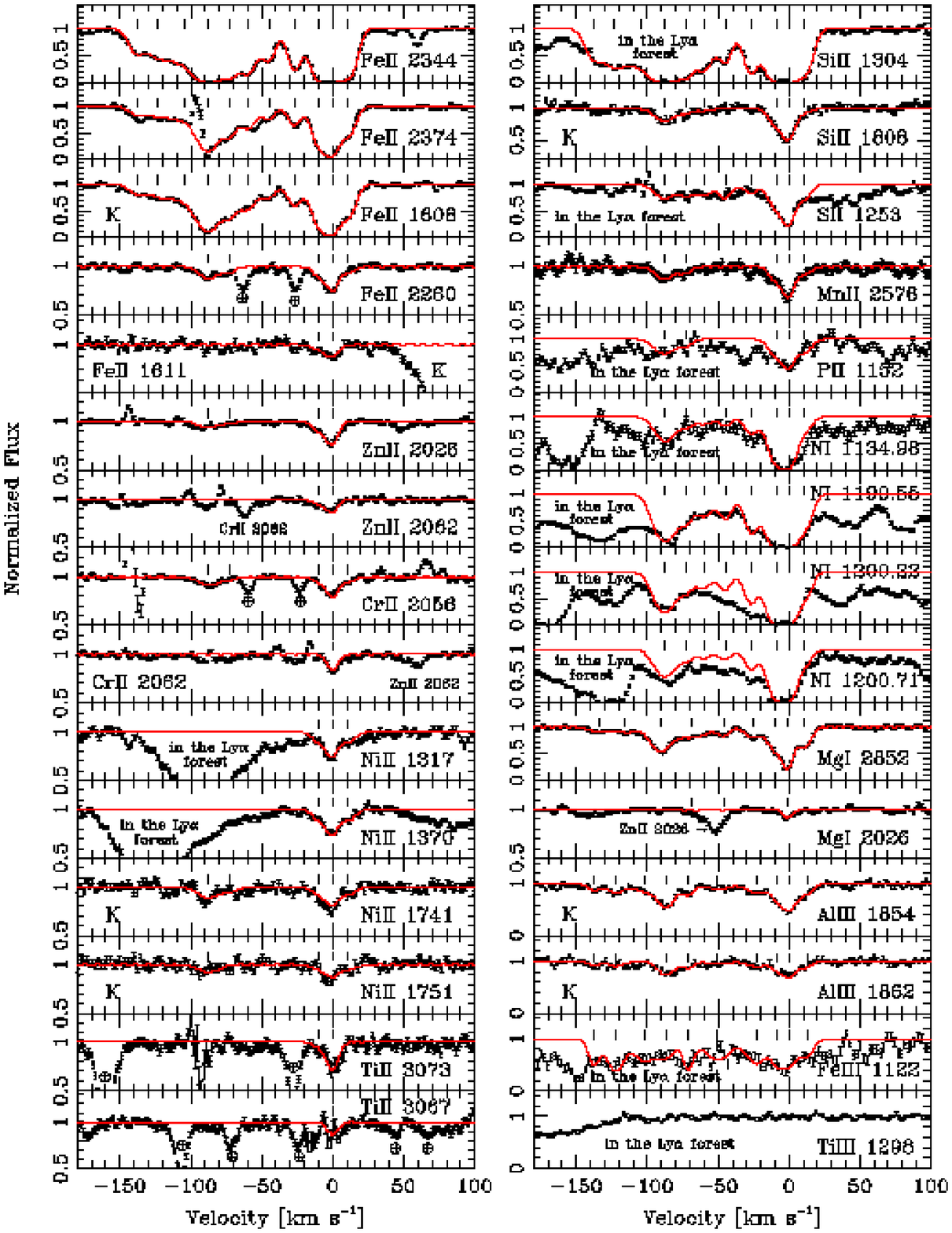}
\caption{Same as Fig.~\ref{Q0100-ions-f} for the DLA toward Q2231$-$00. The zero velocity is fixed 
at $z = 2.066161$.}
\label{Q2231-ions-f}
\end{figure*}
%

\subsection{Q2231$-$00, $z_{\rm abs} = 2.066$}\label{q2231}

This DLA system was previously observed and analyzed by \citet{lu96} and \citet{prochaska99}, and 
recently completed by \citet{prochaska01}. Throughout their analysis they adopted the \ion{H}{i} 
column density measured by \citet{lu94}. We obtain here a new value for the $\log N$(\ion{H}{i}) 
of $20.53\pm 0.08$ from the Ly$\alpha$ damping line observed in the UVES spectra (see 
Fig.~\ref{Q2231-Lya-f}). The fit was performed by fixing the $b$-value at 20 km~s$^{-1}$ and by 
leaving the redshift as a free parameter.

The UVES spectra provide column density measurements of the S$^+$, P$^+$, N$^0$, Mn$^+$ and Mg$^0$ 
ions (see Fig.~\ref{Q2231-ions-f}), and confirm the $N$(Si$^+$), $N$(Zn$^+$), $N$(Cr$^+$) and 
$N$(Ni$^+$) measurements of \citet{prochaska99}. The fit of the \ion{Mg}{i} $\lambda$2852 line 
clearly shows that the contamination of \ion{Zn}{ii} $\lambda$2026 by the \ion{Mg}{i} $\lambda$2026 
profile is negligible. We note that we had to fit the \ion{Mg}{i} profile with slightly different 
fitting parameters in comparison to the ones we have used for other low-ion metal lines (see 
Table~\ref{Q2231-t}). By combining the \ion{Fe}{ii} $\lambda$1608,1611 lines detected in the HIRES 
spectra and the \ion{Fe}{ii} $\lambda$2260,2344,2374 lines observed in the UVES spectra, we deduce 
a more accurate measurement of $N$(Fe$^+$). Finally, we obtain a revised value for $N$(\ion{Ti}{ii}) 
of $12.66\pm 0.08$ from the \ion{Ti}{ii} $\lambda$3067,3073 lines. We succeed in deriving an upper 
limit on the column density of the intermediate-ion, \ion{Fe}{iii}, located in the Ly$\alpha$ 
forest, by using the \ion{Al}{iii} lines observed in the HIRES spectra to constrain the fitting 
parameters (see Fig.~\ref{Q2231-ions-f} and Table~\ref{Q2231-t}). The intermediate-ion lines show 
very similar profiles to the low-ion line profiles.
%

\begin{figure}[!]
\centering
   \includegraphics[width=9cm]{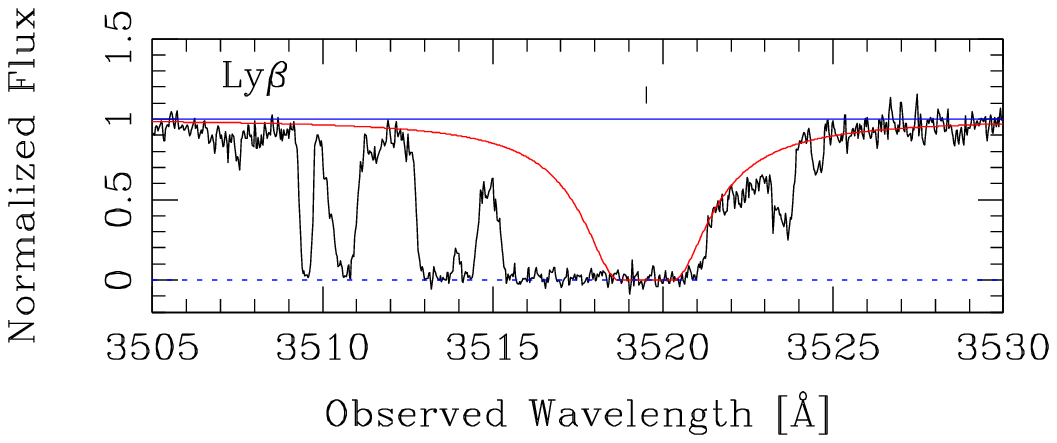}
\caption{Same as Fig.~\ref{Q0100-Ly-f} but for Q2343+12. The Voigt fit was performed with the
\ion{H}{i} column density measured by \citet[][Sect.~3.2]{dodorico02} from the Ly$\alpha$ line
($\log N$(\ion{H}{i}) $= 20.35\pm 0.05$). The vertical bar corresponds to the wavelength centroid 
of the component used for the best fit by these authors, $z = 2.43125$.}
\label{Q2343-Ly-f}
\end{figure}	
%

\begin{table*}[!]
\begin{center}
\caption{Component structure of the $z_{\rm abs} = 2.431$ DLA system toward Q2343+12} 
\label{Q2343-t}
\begin{tabular}{l c c c l c | l c c c l c}
\hline\hline
No & $z_{\rm abs}$ & $v_{\rm rel}^*$ & $b (\sigma_b)$ & Ion & $\log N (\sigma_{\log N})$ & No & $z_{\rm abs}$ & $v_{\rm rel}^*$ & $b (\sigma_b)$ & Ion & $\log N (\sigma_{\log N})$ \\
   &               & km s$^{-1}$     & km s$^{-1}$    &     &                            &    &               & km s$^{-1}$     & km s$^{-1}$    &     &                            
\smallskip
\\ 
\hline
\multicolumn{5}{l}{\hspace{0.3cm} Low-ion transitions} & & & & & & & \\
\hline
1  & 2.428181 &           $-$154 & 3.0{\scriptsize (1.0)} & \ion{Fe}{ii} & 12.44{\scriptsize (0.03)} & 15 & 2.430674 & \phantom{0}$+$64 & 2.8{\scriptsize (0.2)} & \ion{Fe}{ii} & 12.38{\scriptsize (0.03)} \\
   &          &                  &      		  & \ion{O}{i}   & 14.38{\scriptsize (0.03)} &    &	     &  		&			 & \ion{O}{i}	& 13.93{\scriptsize (0.05)} \\
   &          &                  &                        & \ion{Si}{ii} & 13.59{\scriptsize (0.02)} &    &	     &  		&			 & \ion{Si}{ii} & 13.13{\scriptsize (0.01)} \\
2  & 2.428277 &           $-$146 & 3.0{\scriptsize (0.9)} & \ion{Fe}{ii} & 12.94{\scriptsize (0.02)} & 16 & 2.430808 & \phantom{0}$+$76 & 3.2{\scriptsize (0.1)} & \ion{Fe}{ii} & 12.65{\scriptsize (0.02)} \\
   &          &                  &      		  & \ion{O}{i}   & 14.98{\scriptsize (0.04)} &    &	     &  		&			 & \ion{O}{i}	& 14.37{\scriptsize (0.03)} \\
   &          &                  &      		  & \ion{Si}{ii} & 13.81{\scriptsize (0.06)} &    &	     &  		&			 & \ion{Si}{ii} & 13.32{\scriptsize (0.01)} \\
3  & 2.428383 &           $-$136 & 4.1{\scriptsize (0.3)} & \ion{Fe}{ii} & 13.18{\scriptsize (0.01)} & 17 & 2.431018 & \phantom{0}$+$94 & 8.3{\scriptsize (0.4)} & \ion{Fe}{ii} & 13.45{\scriptsize (0.02)} \\
   &          &                  &      		  & \ion{O}{i}   & 14.52{\scriptsize (0.02)} &    &	     &  		&			 & \ion{O}{i}	& 15.27{\scriptsize (0.05)} \\
   &          &                  &      		  & \ion{Si}{ii} & 14.12{\scriptsize (0.06)} &    &	     &  		&			 & \ion{Si}{ii} & 14.01{\scriptsize (0.09)} \\
4  & 2.428499 &           $-$126 & 4.4{\scriptsize (0.5)} & \ion{Fe}{ii} & 12.49{\scriptsize (0.04)} & 18 & 2.431157 &  	 $+$106 & 5.1{\scriptsize (0.4)} & \ion{Fe}{ii} & 13.98{\scriptsize (0.04)} \\
   &          &                  &      		  & \ion{O}{i}   & 13.30{\scriptsize (0.19)} &    &	     &  		&			 & \ion{Zn}{ii} & 11.72{\scriptsize (0.03)} \\
   &          &                  &                        & \ion{Si}{ii} & 13.37{\scriptsize (0.01)} &    &	     &  		&			 & \ion{Cr}{ii} & 12.37{\scriptsize (0.03)} \\
5  & 2.428749 &           $-$104 & 8.5{\scriptsize (0.3)} & \ion{Fe}{ii} & 12.74{\scriptsize (0.01)} &    &	     &  		&			 & \ion{Mn}{ii} & 11.82{\scriptsize (0.04)} \\
   &          &                  &      		  & \ion{O}{i}   & 14.51{\scriptsize (0.02)} &    &	     &  		&			 & \ion{O}{i}	& 15.76{\scriptsize (0.06)} \\
   &          &                  &                        & \ion{Si}{ii} & 13.08{\scriptsize (0.05)} &    &	     &  		&			 & \ion{Si}{ii} & 14.55{\scriptsize (0.03)} \\
6  & 2.428884 & \phantom{0}$-$93 & 3.0{\scriptsize (1.0)} & \ion{Fe}{ii} & 12.68{\scriptsize (0.02)} &    &	     &  		&			 & \ion{Ar}{i}  & 12.85{\scriptsize (0.18)} \\
   &          &                  &      		  & \ion{O}{i}   & 14.29{\scriptsize (0.03)} &    &	     &  		&			 & \ion{N}{i}	& 14.31{\scriptsize (0.22)} \\
   &          &                  &                        & \ion{Si}{ii} & 12.95{\scriptsize (0.02)} &    &	     &  		&			 & \ion{P}{ii}  & $<12.67$ \\
7  & 2.429009 & \phantom{0}$-$82 & 5.9{\scriptsize (0.4)} & \ion{Fe}{ii} & 12.45{\scriptsize (0.02)} & 19 & 2.431288 &  	 $+$118 & 6.2{\scriptsize (0.6)} & \ion{Fe}{ii} & 14.09{\scriptsize (0.03)} \\
   &          &                  &      		  & \ion{O}{i}   & 14.16{\scriptsize (0.03)} &    &	     &  		&			 & \ion{Zn}{ii} & 11.87{\scriptsize (0.02)} \\  
   &          &                  &                        & \ion{Si}{ii} & 12.87{\scriptsize (0.02)} &    &	     &  		&			 & \ion{Cr}{ii} & 12.38{\scriptsize (0.03)} \\
8  & 2.429249 & \phantom{0}$-$61 & 5.1{\scriptsize (0.4)} & \ion{Fe}{ii} & 12.50{\scriptsize (0.02)} &    &	     &  		&			 & \ion{Mn}{ii} & 11.89{\scriptsize (0.03)} \\
   &          &                  &      		  & \ion{O}{i}   & 14.15{\scriptsize (0.03)} &    &	     &  		&			 & \ion{O}{i}	& 15.90{\scriptsize (0.05)} \\ 
   &          &                  &                        & \ion{Si}{ii} & 13.14{\scriptsize (0.05)} &    &	     &  		&			 & \ion{Si}{ii} & 14.71{\scriptsize (0.02)} \\
9  & 2.429355 & \phantom{0}$-$51 & 3.0{\scriptsize (0.8)} & \ion{Fe}{ii} & 12.85{\scriptsize (0.03)} &    &	     &  		&			 & \ion{Ar}{i}  & 12.86{\scriptsize (0.10)} \\
   &          &                  &      		  & \ion{O}{i}   & 14.54{\scriptsize (0.02)} &    &	     &  		&			 & \ion{N}{i}	& 14.32{\scriptsize (0.15)} \\
   &          &                  &                        & \ion{Si}{ii} & 13.71{\scriptsize (0.02)} &    &	     &  		&			 & \ion{P}{ii}  & $<12.55$ \\
10 & 2.429500 & \phantom{0}$-$39 & 5.2{\scriptsize (0.2)} & \ion{Fe}{ii} & 12.65{\scriptsize (0.01)} & 20 & 2.431439 &  	 $+$131 & 4.7{\scriptsize (0.4)} & \ion{Fe}{ii} & 13.82{\scriptsize (0.03)} \\
   &          &                  &      		  & \ion{O}{i}   & 14.22{\scriptsize (0.03)} &    &	     &  		&			 & \ion{Zn}{ii} & 11.67{\scriptsize (0.03)} \\
   &          &                  &                        & \ion{Si}{ii} & 13.28{\scriptsize (0.09)} &    &	     &  		&			 & \ion{Cr}{ii} & 12.20{\scriptsize (0.05)} \\
11 & 2.429702 & \phantom{0}$-$21 & 4.4{\scriptsize (0.5)} & \ion{Fe}{ii} & 12.20{\scriptsize (0.02)} &    &	     &  		&			 & \ion{Mn}{ii} & 11.64{\scriptsize (0.05)} \\
   &          &                  &      		  & \ion{O}{i}   & 13.97{\scriptsize (0.05)} &    &	     &  		&			 & \ion{O}{i}	& 15.80{\scriptsize (0.05)} \\   
   &          &                  &                        & \ion{Si}{ii} & 12.81{\scriptsize (0.06)} &    &	     &  		&			 & \ion{Si}{ii} & 14.54{\scriptsize (0.03)} \\
12 & 2.429942 & \phantom{$-$00}0 & 7.3{\scriptsize (0.2)} & \ion{Fe}{ii} & 13.28{\scriptsize (0.01)} &    &	     &  		&			 & \ion{Ar}{i}  & 12.44{\scriptsize (0.22)} \\
   &          &                  &      		  & \ion{O}{i}   & 14.89{\scriptsize (0.01)} &    &	     &  		&			 & \ion{N}{i}	& 13.58{\scriptsize (0.25)} \\
   &          &                  &      		  & \ion{Si}{ii} & 13.73{\scriptsize (0.11)} &    &	     &  		&			 & \ion{P}{ii}  & $<12.45$ \\
13 & 2.430086 & \phantom{0}$+$13 & 9.6{\scriptsize (0.8)} & \ion{Fe}{ii} & 12.86{\scriptsize (0.06)} & 21 & 2.431570 &  	 $+$142 & 5.7{\scriptsize (0.2)} & \ion{Fe}{ii} & 13.48{\scriptsize (0.04)} \\
   &          &                  &      		  & \ion{O}{i}   & 14.59{\scriptsize (0.02)} &    &	     &  		&			 & \ion{O}{i}	& 15.09{\scriptsize (0.07)} \\
   &          &                  &                        & \ion{Si}{ii} & 13.33{\scriptsize (0.01)} &    &	     &  		&			 & \ion{Si}{ii} & 13.76{\scriptsize (0.16)} \\
14 & 2.430394 & \phantom{0}$+$40 & 9.7{\scriptsize (0.2)} & \ion{Fe}{ii} & 13.19{\scriptsize (0.01)} & 22 & 2.431826 &  	 $+$165 & 6.9{\scriptsize (0.1)} & \ion{Fe}{ii} & 13.02{\scriptsize (0.01)} \\
   &          &                  &      		  & \ion{O}{i}   & 14.75{\scriptsize (0.01)} &    &	     &  		&			 & \ion{Si}{ii} & 13.77{\scriptsize (0.01)} \\
   &          &                  &                        & \ion{Si}{ii} & 13.68{\scriptsize (0.01)} & 23 & 2.432126 &  	 $+$191 & 6.9{\scriptsize (0.5)} & \ion{Fe}{ii} & 12.27{\scriptsize (0.03)} \\
   &          &                  &                        &              &                           &    &	     &  		&			 & \ion{Si}{ii} & 12.58{\scriptsize (0.03)} \\
\hline
\end{tabular}
\begin{minipage}{160mm}
\smallskip
$^*$ Velocity relative to $z=2.429942$ 
\end{minipage}
\end{center}										     
\end{table*}
%

The low-ion line profiles are characterized by a relatively complex velocity structure composed of 
11 components presented in Table~\ref{Q2231-t} and spread over 180 km~s$^{-1}$ in velocity space. 
But, only the 5 strongest components, the components 4, 5, 9, 10 and 11, are detected in the weak 
metal-lines. They contain only about 75\% of the total column density obtained by summing the 
contributions of the 11 components. This shows that when computing the abundance ratios [X/Y], it 
is necessary to consider only the column densities of the velocity components detected in both the 
X and Y profiles to avoid under/overestimations of the relative abundances. 

An additional difficulty which appears in absorption systems with complex metal-line profiles is 
that the probability of blending in the Ly$\alpha$ forest is higher over a larger velocity interval. 
For this reason it is more sensible to consider the measured P$^+$ and N$^0$ column densities as 
upper limits, the \ion{P}{ii} and \ion{N}{i} lines being located in a region of the Ly$\alpha$ 
forest with numerous absorption lines (see Fig.~\ref{Q2231-ions-f}). The S$^+$ column density is a 
borderline case, and unfortunately only the \ion{S}{ii} $\lambda$1253 line from the S$^+$ triplet 
can be used to determine $N$(S$^+$), the two other lines are heavily blended with Ly$\alpha$ clouds. 
We consider the measured $N$(S$^+$) as a value, assuming that the adopted error on $N$(S$^+$) 
covers the possible \ion{H}{i} blendings.
%

\subsection{Q2343+12, $z_{\rm abs} = 2.431$}

This DLA system has first been studied by \citet{sargent88}. The chemical abundance measurements 
obtained from HIRES spectra were mainly used in statistical samples 
\citep[e.g.][]{lu98,prochaska98,vladilo03}. Recently, \citet{dodorico02} and \citet{vladilo03}
reported some additional abundance measurements from UVES spectra. We present here the first 
complete set of elemental abundances of this DLA.
%

\begin{table*}[!]
\begin{center}
\caption{Component structure of the $z_{\rm abs} = 2.431$ DLA system toward Q2343+12 $-$ Continued} 
\label{Q2343-t-cont}
\begin{tabular}{l c c c l c | l c c c l c}
\hline\hline
No & $z_{\rm abs}$ & $v_{\rm rel}^*$ & $b (\sigma_b)$ & Ion & $\log N (\sigma_{\log N})$ & No & $z_{\rm abs}$ & $v_{\rm rel}^*$ & $b (\sigma_b)$ & Ion & $\log N (\sigma_{\log N})$ \\
   &               & km s$^{-1}$     & km s$^{-1}$    &     &                            &    &               & km s$^{-1}$     & km s$^{-1}$    &     &                            
\smallskip
\\ 
\hline
\multicolumn{6}{l}{\hspace{0.3cm} Intermediate-ion transitions} & & & & & & \\
\hline
1  & 2.428178 &           $-$154 & \phantom{0}6.5{\scriptsize (2.2)} & \ion{Al}{iii} & 11.93{\scriptsize (0.25)} & 7  & 2.430392 & \phantom{0}$+$39 &		10.8{\scriptsize (0.5)} & \ion{Al}{iii} & 11.92{\scriptsize (0.03)} \\
   &          &                  &                                   & \ion{Fe}{iii} & 13.06{\scriptsize (0.06)} &    & 	 &		    &					& \ion{Fe}{iii} & $< 13.40$ \\
   &          &                  &                                   & \ion{N}{ii}   & 13.99{\scriptsize (0.01)} &    & 	 &		    &					& \ion{N}{ii}	& 13.66{\scriptsize (0.02)} \\
   &          &                  &                                   & \ion{S}{iii}  & $< 14.17$                 &    & 	 &		    &					& \ion{S}{iii}  & $< 13.84$ \\   
2  & 2.428288 &           $-$144 & \phantom{0}3.6{\scriptsize (1.6)} & \ion{Al}{iii} & 11.90{\scriptsize (0.27)} & 8  & 2.430696 & \phantom{0}$+$66 &		10.8{\scriptsize (0.4)} & \ion{Al}{iii} & 11.72{\scriptsize (0.05)} \\
   &          &                  &                                   & \ion{Fe}{iii} & 13.19{\scriptsize (0.04)} &    & 	 &		    &					& \ion{N}{ii}	& 13.71{\scriptsize (0.01)} \\
   &          &                  &                                   & \ion{N}{ii}   & 14.48{\scriptsize (0.21)} &    & 	 &		    &					& \ion{S}{iii}  & $< 13.90$ \\  	  
   &          &                  &                                   & \ion{S}{iii}  & $< 13.56$                 & 9  & 2.431065 & \phantom{0}$+$98 &		15.0{\scriptsize (0.2)} & \ion{Al}{iii} & 12.08{\scriptsize (0.26)} \\
3  & 2.428426 &           $-$133 & \phantom{0}9.3{\scriptsize (0.6)} & \ion{Al}{iii} & 12.59{\scriptsize (0.03)} &    & 	 &		    &					& \ion{N}{ii}	& 14.04{\scriptsize (0.01)} \\
   &          &                  &                                   & \ion{Fe}{iii} & 13.54{\scriptsize (0.02)} &    & 	 &		    &					& \ion{S}{iii}  & $< 14.35$ \\
   &          &                  &                                   & \ion{N}{ii}   & 14.36{\scriptsize (0.01)} & 10 & 2.431185 &	     $+$109 & \phantom{0}8.9{\scriptsize (1.0)} & \ion{Al}{iii} & 12.45{\scriptsize (0.08)} \\
   &          &                  &                                   & \ion{S}{iii}  & $< 14.22$                 &    & 	 &		    &					& \ion{N}{ii}	& $> 14.25$ \\ 
4  & 2.429358 & \phantom{0}$-$51 & \phantom{0}8.3{\scriptsize (0.5)} & \ion{Al}{iii} & 11.51{\scriptsize (0.07)} &    & 	 &		    &					& \ion{S}{iii}  & $< 13.71$ \\
   &          &                  &                                   & \ion{Fe}{iii} & $< 10.00$                 & 11 & 2.431323 &	     $+$121 & \phantom{0}6.7{\scriptsize (0.9)} & \ion{Al}{iii} & 12.39{\scriptsize (0.08)} \\
   &          &                  &                                   & \ion{N}{ii}   & 13.89{\scriptsize (0.04)} &    & 	 &		    &					& \ion{N}{ii}	& $> 14.11$ \\ 
   &          &                  &                                   & \ion{S}{iii}  & $< 13.97$                 &    & 	 &		    &					& \ion{S}{iii}  & $< 14.36$ \\
5  & 2.429487 & \phantom{0}$-$40 &           19.5{\scriptsize (1.3)} & \ion{Al}{iii} & 11.83{\scriptsize (0.05)} & 12 & 2.431443 &	     $+$131 & \phantom{0}3.0{\scriptsize (1.5)} & \ion{Al}{iii} & 11.92{\scriptsize (0.08)} \\
   &          &                  &                                   & \ion{Fe}{iii} & $< 14.09$                 &    & 	 &		    &					& \ion{N}{ii}	& $> 13.66$ \\ 
   &          &                  &                                   & \ion{N}{ii}   & 13.80{\scriptsize (0.04)} &    & 	 &		    &					& \ion{S}{iii}  & $< 13.45$ \\
   &          &                  &                                   & \ion{S}{iii}  & $< 13.62$                 & 13 & 2.431545 &	     $+$140 &		10.5{\scriptsize (1.5)} & \ion{Al}{iii} & 12.23{\scriptsize (0.06)} \\
6  & 2.429995 & \phantom{00}$+$5 &           14.4{\scriptsize (0.8)} & \ion{Al}{iii} & 12.09{\scriptsize (0.02)} &    & 	 &		    &					& \ion{N}{ii}	& $> 14.27$ \\ 
   &          &                  &                                   & \ion{Fe}{iii} & $< 13.36$                 &    & 	 &		    &					& \ion{S}{iii}  & $< 14.59$ \\
   &          &                  &                                   & \ion{N}{ii}   & 13.71{\scriptsize (0.02)} & 14 & 2.431849 &	     $+$167 & \phantom{0}7.7{\scriptsize (0.2)} & \ion{Al}{iii} & 11.72{\scriptsize (0.04)} \\
   &          &                  &                                   & \ion{S}{iii}  & $< 13.79$                 &    & 	 &		    &					& \ion{N}{ii}	& 14.01{\scriptsize (0.01)} \\
   &          &                  &                                   &               &                           &    & 	 &		    &					& \ion{S}{iii}  & $< 14.34$ \\
   &          &                  &                                   &               &                           & 15 & 2.432135 &	     $+$192 &		15.0{\scriptsize (2.1)} & \ion{N}{ii}	& 13.72{\scriptsize (0.01)} \\
\hline
\end{tabular}
\begin{minipage}{160mm}
\smallskip
$^*$ Velocity relative to $z=2.429942$ 
\end{minipage}
\end{center}										     
\end{table*}
%

\begin{figure*}[!]
\centering
   \includegraphics[width=17cm]{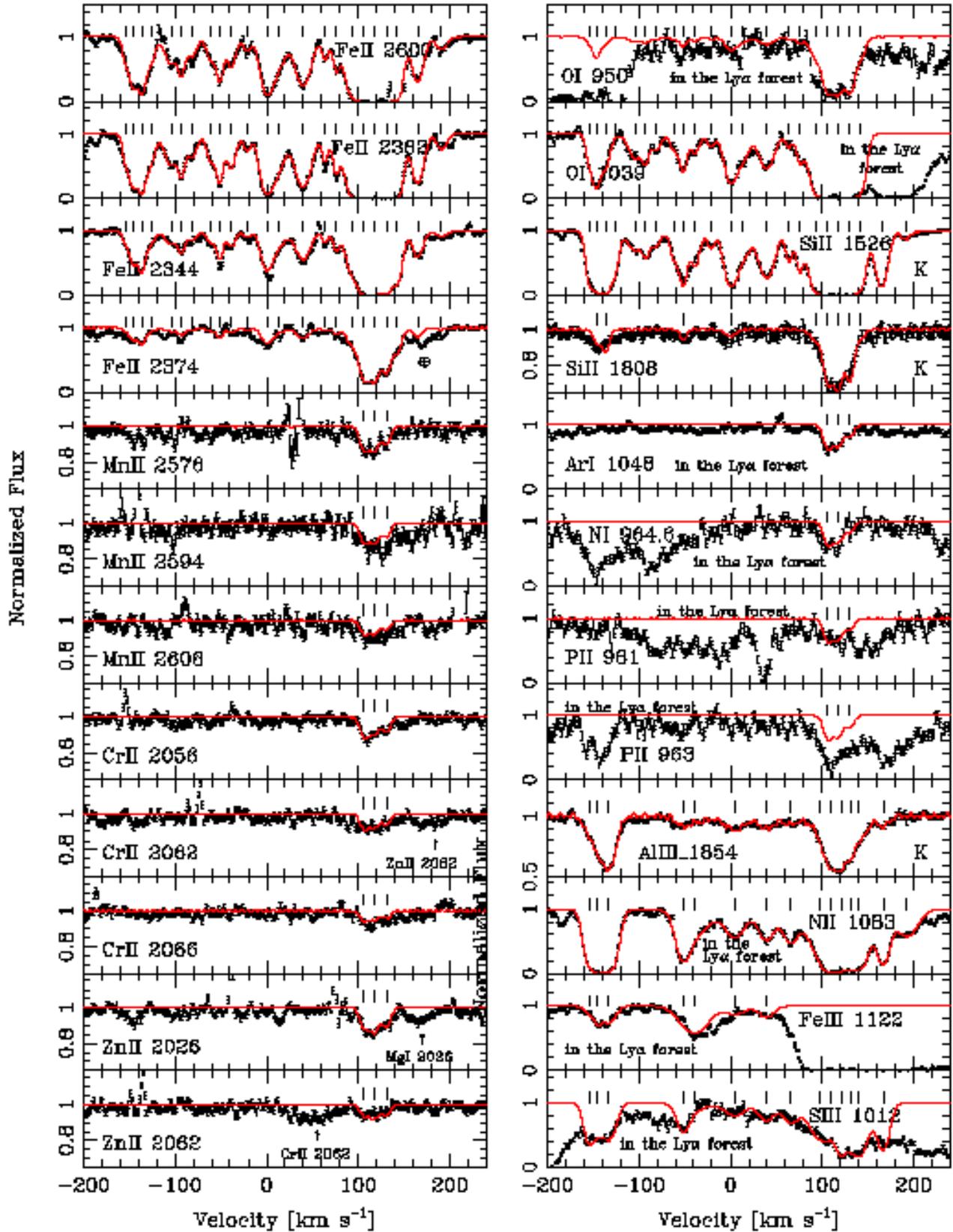}
\caption{Same as Fig.~\ref{Q0100-ions-f} for the DLA toward Q2343+12. The zero velocity is fixed 
at $z = 2.429942$.}
\label{Q2343-ions-f}
\end{figure*}

From our UVES spectra, we measured the column densities of \ion{O}{i}, \ion{Ar}{i}, \ion{N}{i}, 
\ion{Fe}{ii}, \ion{Zn}{ii}, \ion{Cr}{ii} and \ion{Mn}{ii}, and obtained an upper limit on the 
column density of \ion{P}{ii} contaminated by the Ly$\alpha$ clouds. The metal absorption profiles 
are characterized by an extremely complex velocity structure extended over 350 km~s$^{-1}$ in 
velocity space and composed of 23 components (see Fig.~\ref{Q2343-ions-f} and Table~\ref{Q2343-t}). 
The dominant feature is the group of 5 components around $v\sim +120$ km~s$^{-1}$, with the 
components 18, 19 and 20 being the strongest ones. These components are heavily saturated in strong 
metal-lines, but are the only components detected in weak metal-lines. They contain about 62\% of 
the total column density. 

The Ly$\alpha$ line is outside our UVES wavelength coverage, therefore we adopt the \ion{H}{i} 
column density measurement ($\log N$(\ion{H}{i}) $= 20.35\pm 0.05$) obtained by \citet{dodorico02}. 
Fig.~\ref{Q2343-Ly-f} shows the fit of the Ly$\beta$ line which confirms this $N$(H$^0$) 
measurement. This relatively low H$^0$ column density and the presence of a strong \ion{N}{ii} 
$\lambda$1083 line as well as of other intermediate-ion lines $-$ \ion{Al}{iii} $\lambda$1854, 
\ion{Fe}{iii} $\lambda$1122 and \ion{S}{iii} $\lambda$1012 $-$ with very similar profiles to the 
low-ion metal line profiles (see Fig.~\ref{Q2343-ions-f} and Table~\ref{Q2343-t-cont}) suggest that 
some photoionization corrections may be required in this DLA system. A discussion on the 
photoionization corrections is presented in Sect.~\ref{ionization}.
%

\section{Photoionization corrections}\label{ionization}

The photoionization effects have to be carefully examined in absorption metal-line systems, when 
the main goal is to provide a complete and unbiased interpretation of the chemical abundance 
patterns. Indeed, as we are studying gas-phase abundances, a fraction of the gas may be ionized and 
we need to determine this fraction to be sure that we are measuring the {\it intrinsic} abundances. 
The dust depletion effects also affect the gas-phase abundance measurements. The manner we tackle 
this issue is described in the next Section.
%

%

It is generally assumed that in the DLA systems the ionization fraction is low, so that the 
dominant ionization state in \ion{H}{i} regions is the neutral one for elements with the first 
ionization potential $> 13.6$~eV (e.g. O$^0$, N$^0$) and the singly ionized one for elements with 
the first ionization potential $< 13.6$~eV and the second $> 13.6$~eV (e.g. Fe$^+$, Si$^+$). The 
reason for this is that the bulk of the H$^0$ gas is self-shielded from $h\,\nu > 13.6$~eV photons, 
but transparent to $h\,\nu < 13.6$~eV photons. Under this assumption, the correction factors for 
ionization are small, and thus the column densities of low-ions are equal to the column densities 
of elements, e.g. $N$(Si$^+$) $\simeq$ $N$(Si). However, the detection of intermediate-ionization 
transitions, like Al$^{++}$, in the DLA systems provided doubts on the ionization levels in DLAs. 
The ionization potential of Al$^+$ is 18.8~eV, i.e. greater than that of hydrogen, therefore 
Al$^{++}$ is likely present in ionized and not in neutral gas, since photons with $h\,\nu > 
18.8$~eV cannot so easily penetrate gas clouds with large \ion{H}{i} column densities. In addition, 
the observations show a good correlation between the velocity structures of \ion{Al}{iii} and 
singly ionized and neutral species \citep[e.g.][]{lu96,prochaska99,wolfe00}. To explain the 
similarity between the \ion{Al}{iii} and low-ionization species line profiles, \citet{howk99a} and 
\citet{izotov99} proposed that these lines originate in the same ionized region or in a mix of 
neutral and ionized clouds, and stressed the importance of abundance corrections for ionization 
effects. Furthermore, indirect arguments to consider ionized gas in DLAs came from studies of the 
warm ionized medium in the Milky Way and other nearby galaxies \citep{howk99b,sembach00,jenkins00}.

Several authors have investigated the ionization effects in the DLA systems 
\citep{viegas95,howk99a,izotov01,vladilo01}. Although the different approaches used to deal with 
the problem of photoionization in DLAs have led to slightly different conclusions, it was 
generally found that ionization corrections in DLAs are negligible, being below the measurement 
errors, except for some particular systems like the DLA toward GB\,1759+7539 studied by 
\citet{prochaska02b}. While the authors stress that this particular DLA system has several
characteristics which separate it from the majority of DLAs, its properties highlight the 
importance of assessing the ionization state of each DLA system. The photoionization model 
computations are generally performed using the CLOUDY photoionization equilibrium software package 
\citep[e.g.][]{ferland98}. To avoid computing heavy photoionization models, \citet{prochaska02b} 
defined a number of {\it photoionization diagnostics} which provides a qualitative ``first-look'' 
analysis of the ionization state in a DLA system and of the level of required ionization 
corrections. In the following sub-Sections, we apply these photoionization diagnostics to each 
of the four DLAs studied and qualitatively evaluate the importance of ionization corrections.
%

\subsection{Q0100+13, $z_{\rm abs} = 2.309$}

According to the anti-correlation between $\log N$(H$^0$) and $\log N$(Al$^{++}$)/$N$(Al$^+$) found 
by \citet{vladilo01}, where the Al$^{++}$/Al$^+$ column density ratio is assumed to be at a first
approximation an indicator of the ionization level in the gas, the high \ion{H}{i} column density, 
$\log N$(\ion{H}{i}) $= 21.37\pm 0.08$, of this DLA system suggests that the ionization corrections 
are low. The following three additional and more reliable ionization indicators are at our disposal
in this DLA system.

Thanks to the high efficiency of UVES in the blue, we have the spectral coverage for this DLA
system of the rarely detected intermediate-ion transitions, \ion{Fe}{iii} and \ion{N}{ii} at 
$\lambda_{\rm rest} = 1122$ and $1083$ \AA, respectively. The column density ratios 
Fe$^{++}$/Fe$^+$ and N$^+$/N$^0$ are the most reliable ionization indicators, since the 
recombination coefficient of Al$^+$ is rather uncertain, being likely overestimated 
\citep{nussbaumer86}. The observed \ion{Fe}{iii} and \ion{N}{ii} lines show the same profiles as 
the low-ion profiles and are well fitted with the fitting parameters deduced for the low-ion lines. 
Located in the Ly$\alpha$ forest, one of the 2 components is clearly blended in both 
intermediate-ion lines (see Fig.~\ref{Q0100-ions-f}), therefore we derived only upper limits on 
their column densities. These limits lead to $\log N$(Fe$^{++}$)/$N$(Fe$^+$) $< -1.54$ and 
$\log N$(N$^+$)/$N$(N$^0$) $< -1.24$. According to the photoionization diagnostics established by
\citet{prochaska02b}, in a system where Fe$^{++}$/Fe$^+$ $< -1.60$~dex and N$^+$/N$^0$ 
$< -1.00$~dex, the ionization fraction, $x$, defined as the ratio of H$^+$ over (H$^0$+H$^+$), is 
expected to be lower than 10\%. Thus, the measured Fe$^{++}$ and N$^+$ column densities are in 
agreement with the expectations for a DLA with a \ion{H}{i} column density higher than $10^{21}$ 
cm$^{-2}$.

Another ionization indicator that we can consider in this system is argon. Indeed, Ar has a very 
high photoionization cross-section for photons with energy higher than 13.6~eV which is about ten 
times higher than the one of \ion{H}{i}, and thus Ar is very sensitive to ionization effects 
\citep{sofia98}. The comparison of the Ar abundance with the abundance of another $\alpha$-element 
$-$ O, S, or Si~$-$ provides indication on the ionization level in the gas. In this system, we 
measure [Ar/S] $= -0.20\pm 0.14$, and according to \citet{prochaska02b}, an observed value of 
[Ar$^0$/S$^+$] $>-0.20$~dex indicates $x < 10$\%. 

In summary, the high \ion{H}{i} column density, the Al$^{++}$/Al$^{+}$, Fe$^{++}$/Fe$^+$ and 
N$^+$/N$^0$ column density ratios, and the Ar$^0$/S$^+$ abundance ratio, all point to a low 
ionization fraction in this DLA system, and thus to low ionization corrections. The ionization 
corrections should indeed be lower than 0.1~dex for all the elements, except probably for Ar.
%

\subsection{Q1331+17, $z_{\rm abs} = 1.776$}

This is another very high \ion{H}{i} column density system with $\log N$(\ion{H}{i}) $= 21.14\pm
0.08$. The high \ion{H}{i} column density suggests that the ionization level is low in this DLA 
\citep{vladilo01}. However, we do not detect any very reliable ionization indicator in this system 
to confirm this statement. Only the Al$^{++}$/Al$^+$ column density ratio is available for which we 
derive an upper limit of $< -0.76$~dex from the HIRES spectra. This limit indicates that the 
ionization corrections should be low in this system.
%

\subsection{Q2231$-$00, $z_{\rm abs} = 2.066$}

Two ionization indicators are accessible in this DLA with a \ion{H}{i} column density of $\log 
N$(\ion{H}{i}) $= 20.53\pm 0.08$. The first one is the Al$^{++}$/Al$^+$ column density ratio 
obtained from the HIRES spectra, and the second one is the upper limit on the Fe$^{++}$/Fe$^+$ 
column density ratio obtained from the UVES spectra. However, the \ion{Al}{ii} line is so heavily 
saturated that no reliable $N$(Al$^+$) lower limit can be deduced. We thus derive an estimation of 
the Al$^{++}$/Al$^+$ column density ratio by using Si$^+$ as a proxy of Al$^+$ assuming the $\log 
N$(Si$^+$) versus $\log N$(Al$^+$) correlation found by \citet{vladilo01}. We obtain 
$\log N$(Al$^{++}$)/$N$(Al$^+$) $= -0.76\pm 0.19$. With regard to the second ionization indicator, 
the \ion{Fe}{iii} $\lambda$1122 line is located in the Ly$\alpha$ forest and is strongly blended 
with \ion{H}{i} clouds. Consequently, the deduced Fe$^{++}$ column density upper limit is not 
stringent. We get $\log N$(Fe$^{++}$)/$N$(Fe$^+$) $\ll -0.09$. 

Unfortunately the \ion{Ar}{i} lines as well as the \ion{N}{ii} line are beyond the quasar flux 
cut-off, no other ionization indicator is hence accessible in this DLA system. If we trust the 
derived Al$^{++}$/Al$^+$ column density ratio measurement, the ionization corrections should not 
be significant.
%

\subsection{Q2343+12, $z_{\rm abs} = 2.431$}

The low \ion{H}{i} column density of $\log N$(\ion{H}{i}) $= 20.35\pm 0.05$ suggests that the
ionization corrections might be relevant in this system. Several strong intermediate-ion lines with 
very similar profiles to the low-ion line profiles are observed in this system, namely 
\ion{Al}{iii}, \ion{Fe}{iii}, \ion{N}{ii} and \ion{S}{iii} (see Fig.~\ref{Q2343-ions-f}). This is
indeed a first indication that photoionization is playing an important role here. The second 
straightforward indication is provided by the Ar/O and Ar/Si abundance ratio measurements. They are 
very low in this system. We find [Ar/O] $= -0.72\pm 0.17$ and [Ar/Si] $= -0.81\pm 0.16$, in 
agreement with the measurements by \citet{vladilo03}. The Ar/Si abundance ratio in this DLA system 
is even lower than the [Ar/Si] abundance ratio of $-0.68\pm 0.04$ measured in the DLA system toward 
GB\,1759+7539 by \citet{prochaska02b} whose analysis showed important ionization corrections.

The measured $\log N$(Al$^{++}$)/$N$(Al$^+$) $= -0.62\pm 0.19$ obtained by taking Si$^+$ as a 
proxy of Al$^+$ according to the correlation identified by \citet{vladilo01}, 
$\log N$(Fe$^{++}$)/$N$(Fe$^+$) $< 0.37$ derived from the \ion{Fe}{iii} $\lambda$1122 line which 
is only partially blended with Ly$\alpha$ clouds (in the components 4 and 5), and 
$\log N$(N$^+$)/$N$(N$^0$) $> -0.10$ confirm that the ionization corrections are important in this 
DLA system. Indeed, a system with Fe$^{++}$/Fe$^+$ $> -1$~dex and N$^+$/N$^0$ $> -0.2$~dex is
expected to have an ionization fraction, $x$, higher than 50\% \citep{prochaska02b}. We suspect 
that the ionization corrections in this DLA system are of similar magnitude as the ones of the DLA 
toward GB\,1759+7539, namely between $0.1$ to $>0.5$~dex depending on the ion. Such high ionization 
corrections prevent us from deriving reliable chemical abundances in this DLA system. Therefore, we 
do not include this DLA in the following Sections where we discuss the chemical abundances of 
individual systems and make a detailed comparison with chemical evolution models. 
%

\section{Chemical abundances}\label{abundances}

The small sample of DLAs studied in this paper is unique, because by combining the UVES-VLT data 
with the HIRES-Keck data we could measure the abundances of up to 15 elements $-$ N, O, Mg, Al, 
Si, P, S, Cl, Ar, Ti, Cr, Mn, Fe, Ni, Zn. This contrasts with the majority of DLAs for which only a 
handful of elements (Si, Fe, occasionally Cr, Zn, Ni) is usually detected 
\citep[e.g.][]{lu96,prochaska99,prochaska01}. This low amount of information on individual systems 
has until now severely limited the interpretation of the DLA abundance patterns.
%

\begin{table*}[!]
\begin{center}
\caption{Summary of the absolute abundances in the three DLA systems studied} 
\label{abs-abundance}
{\scriptsize
\begin{tabular}{l l | l l | l l | l l}
\hline\hline
\vspace{-0.2cm}\\
\multicolumn{2}{c}{{\footnotesize Quasar}}    & 
\multicolumn{2}{c}{{\footnotesize Q0100+13}}  & 
\multicolumn{2}{c}{{\footnotesize Q1331+17}}  & 
\multicolumn{2}{c}{{\footnotesize Q2231$-$00}} 
\smallskip
\\ 
\hline
\vspace{-0.2cm}\\
\multicolumn{2}{c}{{\footnotesize $z_{\rm abs}$}} & 
\multicolumn{2}{c}{{\footnotesize 2.309}}         & 
\multicolumn{2}{c}{{\footnotesize 1.776}}         & 
\multicolumn{2}{c}{{\footnotesize 2.066}} \\
\multicolumn{2}{c}{{\footnotesize $\log N$(\ion{H}{i})}}      & 
\multicolumn{2}{c}{{\footnotesize 21.37}{\scriptsize (0.08)}} & 
\multicolumn{2}{c}{{\footnotesize 21.14}{\scriptsize (0.08)}} & 
\multicolumn{2}{c}{{\footnotesize 20.53}{\scriptsize (0.08)}} 
\smallskip
\\
\hline
\\
{\footnotesize $\log N$(\ion{Zn}{ii})} & {\footnotesize [Zn/H]$_{\rm obs}$} & 
{\footnotesize 12.47(0.01)} & {\footnotesize $-1.57$(0.09)} & 
{\footnotesize 12.54(0.02)} & {\footnotesize $-1.27$(0.09)} & 
{\footnotesize 12.30(0.05)} & {\footnotesize $-0.90$(0.10)} \\[0.1cm]
{\footnotesize $\log N$(\ion{Fe}{ii})} & {\footnotesize [Fe/H]$_{\rm obs}$} &
{\footnotesize 15.09(0.01)} & {\footnotesize $-1.78$(0.08)} & 
{\footnotesize 14.63(0.03)} & {\footnotesize $-2.01$(0.09)} &
{\footnotesize 14.83(0.03)} & {\footnotesize $-1.20$(0.09)} \\
                                       & {\footnotesize [Fe/H]$_{\rm cor}$ (E00)} &
                            & {\footnotesize $-1.63$(0.10)} &
			    & {\footnotesize $-1.23$(0.12)} &
                            & {\footnotesize $-0.83$(0.14)} \\
			               & {\footnotesize [Fe/H]$_{\rm cor}$ (E11)} &
		            & {\footnotesize $-1.63$(0.10)} &
			    & {\footnotesize $-1.15$(0.13)} & 
			    & {\footnotesize $-0.83$(0.14)} \\
			               & {\footnotesize [Fe/H]$_{\rm cor}$ (S00)} &
			    & {\footnotesize $-1.52$(0.10)} & 
			    & {\footnotesize $-0.95$(0.13)} & 
			    & {\footnotesize $-0.69$(0.14)} \\
			               & {\footnotesize [Fe/H]$_{\rm cor}$ (S11)} &
			    & {\footnotesize $-1.52$(0.10)} & 
			    & {\footnotesize $-0.95$(0.13)} & 
			    & {\footnotesize $-0.69$(0.14)} \\[0.1cm]
{\footnotesize $\log N$(\ion{Cr}{ii})} & {\footnotesize [Cr/H]$_{\rm obs}$} &
{\footnotesize 13.37(0.01)} & {\footnotesize $-1.69$(0.08)} & 
{\footnotesize 12.95(0.03)} & {\footnotesize $-1.88$(0.09)} & 
{\footnotesize 13.00(0.04)} & {\footnotesize $-1.22$(0.09)} \\
                                       & {\footnotesize [Cr/H]$_{\rm cor}$ (E00)} &
		            & {\footnotesize $-1.61$(0.09)} & 
			    & {\footnotesize $-1.23$(0.13)} & 
			    & {\footnotesize $-0.94$(0.13)} \\
			               & {\footnotesize [Cr/H]$_{\rm cor}$ (E11)} &
			    & {\footnotesize $-1.60$(0.10)} & 
			    & {\footnotesize $-1.14$(0.15)} & 
			    & {\footnotesize $-0.94$(0.13)} \\
			               & {\footnotesize [Cr/H]$_{\rm cor}$ (S00)} &
			    & {\footnotesize $-1.53$(0.10)} & 
			    & {\footnotesize $-1.07$(0.12)} & 
			    & {\footnotesize $-0.80$(0.13)} \\
			               & {\footnotesize [Cr/H]$_{\rm cor}$ (S11)} &
		            & {\footnotesize $-1.51$(0.10)} & 
			    & {\footnotesize $-0.94$(0.13)} & 
			    & {\footnotesize $-0.82$(0.13)} \\[0.1cm]
{\footnotesize $\log N$(\ion{Ni}{ii})} & {\footnotesize [Ni/H]$_{\rm obs}$} &
{\footnotesize 13.87(0.01)$^a$} & {\footnotesize $-1.75$(0.08)$^a$} & 
{\footnotesize 13.44(0.08)}     & {\footnotesize $-1.95$(0.11)}     & 
{\footnotesize 13.54(0.06)}     & {\footnotesize $-1.24$(0.10)} \\
                                       & {\footnotesize [Ni/H]$_{\rm cor}$ (E00)} &
				& {\footnotesize $-1.62$(0.10)} & 
				& {\footnotesize $-1.23$(0.15)} & 
				& {\footnotesize $-0.91$(0.14)} \\
				       & {\footnotesize [Ni/H]$_{\rm cor}$ (E11)} &
				& {\footnotesize $-1.60$(0.10)} & 
				& {\footnotesize $-1.09$(0.17)} & 
				& {\footnotesize $-0.87$(0.15)} \\
				       & {\footnotesize [Ni/H]$_{\rm cor}$ (S00)} &
				& {\footnotesize $-1.52$(0.10)} & 
				& {\footnotesize $-0.95$(0.14)} & 
				& {\footnotesize $-0.78$(0.14)} \\
				       & {\footnotesize [Ni/H]$_{\rm cor}$ (S11)} &
				& {\footnotesize $-1.50$(0.10)} & 
				& {\footnotesize $-0.88$(0.15)} & 
				& {\footnotesize $-0.73$(0.14)} \\[0.1cm]
{\footnotesize $\log N$(\ion{Mn}{ii})} & {\footnotesize [Mn/H]$_{\rm obs}$$^{\dag}$} &
{\footnotesize ...}         & {\footnotesize ...}           & 
{\footnotesize 12.50(0.03)} & {\footnotesize $-2.17$(0.09)} & 
{\footnotesize 12.59(0.04)} & {\footnotesize $-1.47$(0.09)} \\[0.1cm]
{\footnotesize $\log N$(\ion{Si}{ii})} & {\footnotesize [Si/H]$_{\rm obs}$} &
{\footnotesize $>14.72^a$}  & {\footnotesize $>-2.21^a$}    & 
{\footnotesize 15.30(0.01)} & {\footnotesize $-1.40$(0.08)} & 
{\footnotesize 15.29(0.04)} & {\footnotesize $-0.80$(0.09)} \\
                                       & {\footnotesize [Si/H]$_{\rm cor}$ (E00)} &
			    & {\footnotesize $>-2.21$}      & 
			    & {\footnotesize $-1.22$(0.11)} & 
			    & {\footnotesize $-0.78$(0.10)} \\
			               & {\footnotesize [Si/H]$_{\rm cor}$ (E11)} &
			    & {\footnotesize $>-2.21$}      & 
			    & {\footnotesize $-1.16$(0.12)} & 
			    & {\footnotesize $-0.78$(0.10)} \\
			               & {\footnotesize [Si/H]$_{\rm cor}$ (S00)} &
			    & {\footnotesize $>-2.21$}      & 
			    & {\footnotesize $-0.98$(0.11)} & 
			    & {\footnotesize $-0.74$(0.11)} \\
			               & {\footnotesize [Si/H]$_{\rm cor}$ (S11)} &
		            & {\footnotesize $>-2.21$}      & 
			    & {\footnotesize $-1.01$(0.11)} & 
			    & {\footnotesize $-0.74$(0.10)} \\[0.1cm]
{\footnotesize $\log N$(\ion{S}{ii})} & {\footnotesize [S/H]$_{\rm obs}$} &
{\footnotesize 15.09(0.06)} & {\footnotesize $-1.48$(0.11)} & 
{\footnotesize 15.08(0.11)} & {\footnotesize $-1.26$(0.14)} & 
{\footnotesize 15.10(0.15)} & {\footnotesize $-0.63$(0.17)} \\[0.1cm]
{\footnotesize $\log N$(\ion{Mg}{ii})} & {\footnotesize [Mg/H]$_{\rm obs}$} &
{\footnotesize 15.57(0.09)} & {\footnotesize $-1.38$(0.11)} & 
{\footnotesize 15.53(0.14)} & {\footnotesize $-1.19$(0.15)} &
{\footnotesize ...}         & {\footnotesize ...} \\
                                       & {\footnotesize [Mg/H]$_{\rm cor}$ (E00)} &
			    & {\footnotesize $-1.38$(0.11)} & 
			    & {\footnotesize $-1.05$(0.16)} &
			    & {\footnotesize ...} \\
			               & {\footnotesize [Mg/H]$_{\rm cor}$ (E11)} &
			    & {\footnotesize $-1.38$(0.11)} & 
			    & {\footnotesize $-0.90$(0.19)} &
			    & {\footnotesize ...} \\
			               & {\footnotesize [Mg/H]$_{\rm cor}$ (S00)} &
			    & {\footnotesize $-1.37$(0.11)} & 
			    & {\footnotesize $-0.87$(0.17)} &
			    & {\footnotesize ...} \\
			               & {\footnotesize [Mg/H]$_{\rm cor}$ (S11)} &
			    & {\footnotesize $-1.37$(0.11)} & 
			    & {\footnotesize $-0.76$(0.18)} &
			    & {\footnotesize ...} \\[0.1cm]
{\footnotesize $\log N$(\ion{Ti}{ii})} & {\footnotesize [Ti/H]$_{\rm obs}$$^*$} &
{\footnotesize $<12.21^b$}  & {\footnotesize $<-2.10^b$}    & 
{\footnotesize $<11.34$}    & {\footnotesize $<-2.74$}      & 
{\footnotesize 12.66(0.08)} & {\footnotesize $-0.81$(0.11)} \\[0.1cm]
{\footnotesize $\log N$(\ion{Ar}{i})} & {\footnotesize [Ar/H]$_{\rm obs}$} &
{\footnotesize 14.21(0.12)} & {\footnotesize $-1.68$(0.15)} & 
{\footnotesize ...}         & {\footnotesize ...}           & 
{\footnotesize ...}         & {\footnotesize ...} \\[0.1cm]
{\footnotesize $\log N$(\ion{N}{i})} & {\footnotesize [N/H]$_{\rm obs}$} &
{\footnotesize 15.03(0.10)} & {\footnotesize $-2.31$(0.14)} & 
{\footnotesize $<15.23$}    & {\footnotesize $<-1.88$}      & 
{\footnotesize $<15.02$}    & {\footnotesize $<-1.48$} \\[0.1cm]
{\footnotesize $\log N$(\ion{P}{ii})} & {\footnotesize [P/H]$_{\rm obs}$} &
{\footnotesize 13.05(0.09)} & {\footnotesize $-1.85$(0.13)} & 
{\footnotesize 13.25(0.10)} & {\footnotesize $-1.42$(0.13)} & 
{\footnotesize $<13.51$}    & {\footnotesize $<-0.55$} \\[0.1cm]
{\footnotesize $\log N$(\ion{Al}{ii})} & {\footnotesize [Al/H]$_{\rm obs}$} &
{\footnotesize ...}         & {\footnotesize ...}           & 
{\footnotesize $>13.74^a$}  & {\footnotesize $>-1.89^a$}    & 
{\footnotesize ...}         & {\footnotesize ...} \\[0.1cm]
{\footnotesize $\log N$(\ion{Cl}{i})} & {\footnotesize [Cl/H]$_{\rm obs}$} &
{\footnotesize ...}         & {\footnotesize ...}           & 
{\footnotesize 13.05(0.10)} & {\footnotesize $>-1.37^c$} & 
{\footnotesize ...}         & {\footnotesize ...} \\[0.1cm]
\hline
\end{tabular}}
\begin{minipage}{175mm}
\smallskip
$^a$ \citet{prochaska99} \\
$^b$ \citet{prochaska01} \\
$^c$ The [Cl/H] absolute abundance derived from the \ion{Cl}{i} column density has to be 
considered as a strict lower limit, since the dominant state of Cl should be \ion{Cl}{ii} in DLAs. \\
$^{\dag}$ We do not compute the dust corrections for [Mn/H], since this element when analyzed is 
compared with Fe, and these two elements have very similar dust depletions. \\
$^*$ We do not compute the dust corrections for [Ti/H], since this element when analyzed is 
compared with Fe, and in this case dust depletion and nucleosynthesis tend to work in the opposite 
sense (see Paper~I).\\
In the dust correction models E00 and E11, \citet{vladilo02a} assumes that the intrinsic 
[Zn/Fe] ratio is equal to $= +0.10$~dex. \\
In the dust correction models S00 and S11, \citet{vladilo02a} assumes that the intrinsic 
[Zn/Fe] ratio is equal to $= +0.00$~dex.
\end{minipage}
\end{center}										     
\end{table*}
%

\subsection{Dust content}\label{dust}

The interpretation of the elemental abundance patterns in DLAs is by far not straightforward, the 
principal difficulty is to disentangle the nucleosynthetic contributions from the dust depletion 
effects. Several pieces of evidence show that dust is indeed present in DLAs 
\citep[e.g.][]{pei91,pettini94,hou01,prochaska02c}. Therefore, as we are measuring gas-phase 
elemental abundances in DLAs, in presence of dust the observed abundances may not represent the 
intrinsic chemical composition of the system if part of the elements is removed from the gas to the 
solid phase, as it happens in the interstellar medium of our Galaxy \citep{savage96}. Consequently, 
the refractory elements (e.g. Si, Fe, Cr, Ni) preferentially incorporated into dust grains are not 
the best diagnostic elements and their relative ratios have to be cautiously interpreted, since 
their differential depletion can mimic the expected nucleosynthetic abundance patterns.
%

\begin{table*}[!]
\begin{center}
\caption{Summary of the relative abundances in the three DLA systems studied} 
\label{rel-abundance}
{\scriptsize
\begin{tabular}{l l | l l | l l | l l}
\hline\hline
\vspace{-0.2cm}\\
\multicolumn{2}{c}{{\footnotesize Quasar}}    & 
\multicolumn{2}{c}{{\footnotesize Q0100+13}}  & 
\multicolumn{2}{c}{{\footnotesize Q1331+17}}  & 
\multicolumn{2}{c}{{\footnotesize Q2231$-$00}} 
\smallskip
\\ 
\hline
\vspace{-0.2cm}\\
\multicolumn{2}{c}{{\footnotesize $z_{\rm abs}$}} & 
\multicolumn{2}{c}{{\footnotesize 2.309}}         & 
\multicolumn{2}{c}{{\footnotesize 1.776}}         & 
\multicolumn{2}{c}{{\footnotesize 2.066}} \\
\multicolumn{2}{c}{{\footnotesize $\log N$(\ion{H}{i})}}      & 
\multicolumn{2}{c}{{\footnotesize 21.37}{\scriptsize (0.08)}} & 
\multicolumn{2}{c}{{\footnotesize 21.14}{\scriptsize (0.08)}} & 
\multicolumn{2}{c}{{\footnotesize 20.53}{\scriptsize (0.08)}} 
\smallskip
\\
\hline
\\
 & {\footnotesize [Zn/Fe]$_{\rm obs}$} & 
 & {\footnotesize $+0.25$(0.04)} & 
 & {\footnotesize $+0.75$(0.05)} & 
 & {\footnotesize $+0.45$(0.07)} \\
 & {\footnotesize [Zn/Fe]$_{\rm cor}$ (E)} &
 & {\footnotesize $+0.10$} & & {\footnotesize $+0.10$} & & {\footnotesize $+0.10$} \\
 & {\footnotesize [Zn/Fe]$_{\rm cor}$ (S)} &
 & {\footnotesize $+0.00$} & & {\footnotesize $+0.00$} & & {\footnotesize $+0.00$} \\[0.1cm]
 & {\footnotesize [Ni/Fe]$_{\rm obs}$} & 
 & {\footnotesize $+0.07$(0.02)} & 
 & {\footnotesize $+0.07$(0.09)} & 
 & {\footnotesize $+0.07$(0.07)} \\
 & {\footnotesize [Ni/Fe]$_{\rm cor}$ (E00)} &
 & {\footnotesize $+0.05$(0.02)} & 
 & {\footnotesize $+0.01$(0.02)} & 
 & {\footnotesize $+0.03$(0.04)} \\
 & {\footnotesize [Ni/Fe]$_{\rm cor}$ (E11)} &
 & {\footnotesize $+0.07$(0.02)} & 
 & {\footnotesize $+0.07$(0.09)} & 
 & {\footnotesize $+0.07$(0.07)} \\
 & {\footnotesize [Ni/Fe]$_{\rm cor}$ (S00)} &
 & {\footnotesize $+0.04$(0.01)} & 
 & {\footnotesize $+0.00$(0.01)} & 
 & {\footnotesize $+0.02$(0.03)} \\
 & {\footnotesize [Ni/Fe]$_{\rm cor}$ (S11)} &
 & {\footnotesize $+0.07$(0.02)} & 
 & {\footnotesize $+0.08$(0.09)} & 
 & {\footnotesize $+0.07$(0.07)} \\[0.1cm]
 & {\footnotesize [Mn/Fe]$_{\rm obs}$} &
 & {\footnotesize ...}           & 
 & {\footnotesize $-0.15$(0.04)} & 
 & {\footnotesize $-0.16$(0.05)} \\[0.1cm]
 & {\footnotesize [Si/Fe]$_{\rm obs}$} &
 & {\footnotesize $>-0.43$}      & 
 & {\footnotesize $+0.61$(0.03)} & 
 & {\footnotesize $+0.40$(0.05)} \\
 & {\footnotesize [Si/Fe]$_{\rm cor}$ (E00)} &
 & {\footnotesize $>-0.58$}      & 
 & {\footnotesize $+0.01$(0.03)} & 
 & {\footnotesize $+0.05$(0.09)} \\
 & {\footnotesize [Si/Fe]$_{\rm cor}$ (E11)} &
 & {\footnotesize $>-0.58$}      & 
 & {\footnotesize $-0.01$(0.05)} & 
 & {\footnotesize $+0.05$(0.09)} \\
 & {\footnotesize [Si/Fe]$_{\rm cor}$ (S00)} &
 & {\footnotesize $>-0.67$}      & 
 & {\footnotesize $-0.02$(0.01)} & 
 & {\footnotesize $-0.05$(0.07)} \\
 & {\footnotesize [Si/Fe]$_{\rm cor}$ (S11)} &
 & {\footnotesize $>-0.68$}      & 
 & {\footnotesize $-0.06$(0.04)} & 
 & {\footnotesize $-0.05$(0.08)} \\[0.1cm]
 & {\footnotesize [Mg/Fe]$_{\rm obs}$} &
 & {\footnotesize $+0.44$(0.09)} & 
 & {\footnotesize $+0.83$(0.14)} & 
 & {\footnotesize ...} \\
 & {\footnotesize [Mg/Fe]$_{\rm cor}$ (E00)} &
 & {\footnotesize $+0.29$(0.11)} & 
 & {\footnotesize $+0.19$(0.14)} &
 & {\footnotesize ...} \\
 & {\footnotesize [Mg/Fe]$_{\rm cor}$ (E11)} &
 & {\footnotesize $+0.29$(0.11)} & 
 & {\footnotesize $+0.26$(0.16)} &
 & {\footnotesize ...} \\
 & {\footnotesize [Mg/Fe]$_{\rm cor}$ (S00)} &
 & {\footnotesize $+0.19$(0.11)} & 
 & {\footnotesize $+0.09$(0.12)} &
 & {\footnotesize ...} \\
 & {\footnotesize [Mg/Fe]$_{\rm cor}$ (S11)} &
 & {\footnotesize $+0.19$(0.11)} & 
 & {\footnotesize $+0.20$(0.17)} &
 & {\footnotesize ...} \\[0.1cm]
 & {\footnotesize [Ti/Fe]$_{\rm obs}$} &
 & {\footnotesize $<-0.09$}      & 
 & {\footnotesize $<-0.45$}      & 
 & {\footnotesize $+0.70$(0.09)} \\[0.1cm]
 & {\footnotesize [S/Zn]$_{\rm obs}$} &
 & {\footnotesize $+0.09$(0.07)} & 
 & {\footnotesize $+0.01$(0.12)} & 
 & {\footnotesize $+0.12$(0.14)} \\[0.1cm]
 & {\footnotesize [S/Fe]$_{\rm obs}$} &
 & {\footnotesize $+0.34$(0.07)} & 
 & {\footnotesize $+0.76$(0.12)} & 
 & {\footnotesize $+0.62$(0.15)} \\
 & {\footnotesize [S/Fe]$_{\rm cor}$ (E00)} &
 & {\footnotesize $+0.19$(0.09)} & 
 & {\footnotesize $-0.02$(0.16)} & 
 & {\footnotesize $+0.25$(0.19)} \\
 & {\footnotesize [S/Fe]$_{\rm cor}$ (E11)} &
 & {\footnotesize $+0.19$(0.09)} & 
 & {\footnotesize $-0.10$(0.17)} & 
 & {\footnotesize $+0.25$ 0.19)} \\
 & {\footnotesize [S/Fe]$_{\rm cor}$ (S00)} &
 & {\footnotesize $+0.08$(0.09)} & 
 & {\footnotesize $-0.30$(0.17)} & 
 & {\footnotesize $+0.11$(0.19)} \\
 & {\footnotesize [S/Fe]$_{\rm cor}$ (S11)} &
 & {\footnotesize $+0.08$(0.09)} & 
 & {\footnotesize $-0.30$(0.17)} & 
 & {\footnotesize $+0.11$(0.19)} \\[0.1cm]
 & {\footnotesize [N/S]$_{\rm obs}$} &
 & {\footnotesize $-0.83$(0.14)} & 
 & {\footnotesize $<-0.63$}      & 
 & {\footnotesize $<-0.85$} \\[0.1cm]
 & {\footnotesize [N/Si]$_{\rm obs}$} &
 & {\footnotesize ...}      & 
 & {\footnotesize $<-0.48$} &
 & {\footnotesize $<-0.64$} \\
 & {\footnotesize [N/Si]$_{\rm cor}$ (E00)} &
 & {\footnotesize ...}      &
 & {\footnotesize $<-0.66$} & 
 & {\footnotesize $<-0.66$} \\
 & {\footnotesize [N/Si]$_{\rm cor}$ (E11)} &
 & {\footnotesize ...}      &
 & {\footnotesize $<-0.72$} & 
 & {\footnotesize $<-0.66$} \\
 & {\footnotesize [N/Si]$_{\rm cor}$ (S00)} &
 & {\footnotesize ...}      &
 & {\footnotesize $<-0.90$} & 
 & {\footnotesize $<-0.70$} \\
 & {\footnotesize [N/Si]$_{\rm cor}$ (S11)} &
 & {\footnotesize ...}      &
 & {\footnotesize $<-0.87$} & 
 & {\footnotesize $<-0.70$} \\
 & {\footnotesize [Cl/S]$_{\rm obs}$} &
 & {\footnotesize ...}       &
 & {\footnotesize $> -0.11$} & 
 & {\footnotesize ...}           \\[0.1cm]
\hline
\end{tabular}}
\end{center}										     
\end{table*}
%

In Sect.~\ref{ionization} we have seen that at least the ionization effects are playing a 
negligible role in three out of the four DLAs studied in this paper. To constrain now the dust 
depletion effects, the access to a large number of elements is crucial, since elements with the 
same nucleosynthetic origin provide information on the dust depletion level in DLAs.

\noindent {\it Fe-peak elements and Zn.} When comparing the absolute abundances of different 
iron-peak elements, Cr, Fe, Ni, and of Zn measured in the DLAs toward Q0100+13, Q1331+17 and
Q2231$-$00, we notice variations, while in the Galactic stars, Cr, Fe, Ni and Zn track each other 
and have solar values relative to Fe 
\citep[apart from Zn which seems to be enhanced by at maximum $0.1-0.2$~dex relative to Fe, see][]
{prochaska00,mishenina02}. These absolute abundance variations are in line with the variations 
observed in the Galactic ISM due to the differential depletion \citep[e.g.][]{savage96}, and are 
thus suggestive as being the result of a differential depletion present in the DLA systems. Hence, 
the measured [Zn/Fe,Cr,Ni] ratios, when assumed as being the result of the different degree of 
incorporation into dust grains of Zn, a volatile element (almost undepleted), and of Fe, Cr and Ni, 
refractory elements, clearly show that dust depletion has to be seriously taken into account in 
the DLA systems toward Q1331+17 and Q2231$-$00, for which we find [Zn/Fe] $= +0.75\pm 0.05$ and 
$+0.45\pm 0.07$, respectively. The DLA toward Q1331+17, in particular, exhibits one of the largest 
dust depletion level of any DLA. On the contrary, the DLA toward Q0100+13 shows a low [Zn/Fe] ratio 
of $+0.25\pm 0.04$, and thus requires negligible dust corrections. 

\noindent {\it Alpha-elements.} Another similar way to highlight the presence of dust in the DLA 
systems is the comparison of the absolute abundances of different $\alpha$-elements. In the Galactic 
stars the $\alpha$-elements track more or less each other (within $\pm 0.10$~dex). Thus, by 
comparing $\alpha$-elements of different dust depletion levels, their relative ratios when 
differing from the solar value are indicative of the presence of dust. In the DLA toward Q0100+13 
we measure [S/Mg] $= -0.10\pm 0.12$, where Mg is relatively strongly depleted in the Galactic ISM 
and S is a non-refractory element \citep{savage96}. This ratio is in agreement with the measured 
[Zn/Fe] ratio, both of them suggest a small amount of dust in this DLA system. In the DLA toward 
Q1331+17 we measure [S/Si] $= +0.16\pm 0.12$. Since Si is only a mildly refractory element 
\citep{savage96}, the observed S overabundance relative to Si indicates important amount of dust. 
We also have the measurement of [S/Mg] $= -0.07\pm 0.18$ in this DLA system. This abundance 
ratio rather indicates no presence of dust, but we ascribe this result to the weak accuracy of the 
\ion{Mg}{ii} column density measurement, which may be slightly overestimated due to some blends in 
the Ly$\alpha$ forest (see Paper~I and Sect.~\ref{q1331}). Finally, in the DLA toward Q2231$-$00 
we measure [S/Si] $= +0.21\pm 0.16$ and [Si/Ti] $= -0.25\pm 0.09$. The [S/Si] ratio is in agreement 
with the conclusions derived from the [Zn/Fe] ratio, which highlights the presence of dust in this
DLA system. On the other hand, the measured undersolar [Si/Ti] ratio, while Ti is a strongly 
refractory element, does not confirm these findings. At a first glance, the undersolar [Si/Ti] 
ratio can only be the result of an overestimation of the \ion{Ti}{ii} column density.
%

\subsection{Dust corrections}

Three main approaches can be used to circumvent the problem of dust depletion in the studies of 
DLA abundance patterns. The first one consists in considering only the DLA systems with {\it no} 
or {\it low} dust depletion, [Zn/Cr] $\lesssim 0.3$ \citep{pettini00,molaro00,lopez02}. Indeed, if
the amount of dust is small, the dust depletion levels are negligible. This is valid for the DLA 
toward Q0100+13, which has a low [Zn/Cr] $= +0.12\pm 0.02$ ratio. However, for the two other DLAs, 
other solutions have to be found. The second approach consists in trying to quantify the dust 
depletion effects and provide dust corrections to the measured abundances. Several authors have 
considered different methods to compute the dust depletion corrections 
\citep[e.g.][]{vladilo98,savaglio00}. Recently, \citet{vladilo02a,vladilo02b} elaborated a new and 
more complete method for correcting for dust depletion. Finally, the third approach consists in 
focusing on non-refractory and mildly refractory elements, such as N, O, S, Ar, and Zn. In this way 
one has directly access to the intrinsic abundances of DLAs 
\citep{centurion00,molaro00,dessauges02b}. This approach to tackle the problem of dust depletion is 
by far the most accurate one. It is applicable to {\it all} DLAs whatever their amount of dust, and 
is independent of any assumption on the properties of dust in DLAs and of errors that may be 
introduced by a complex dust correction procedure.
%

\begin{figure*}[!]
\centering
   \includegraphics[width=14cm]{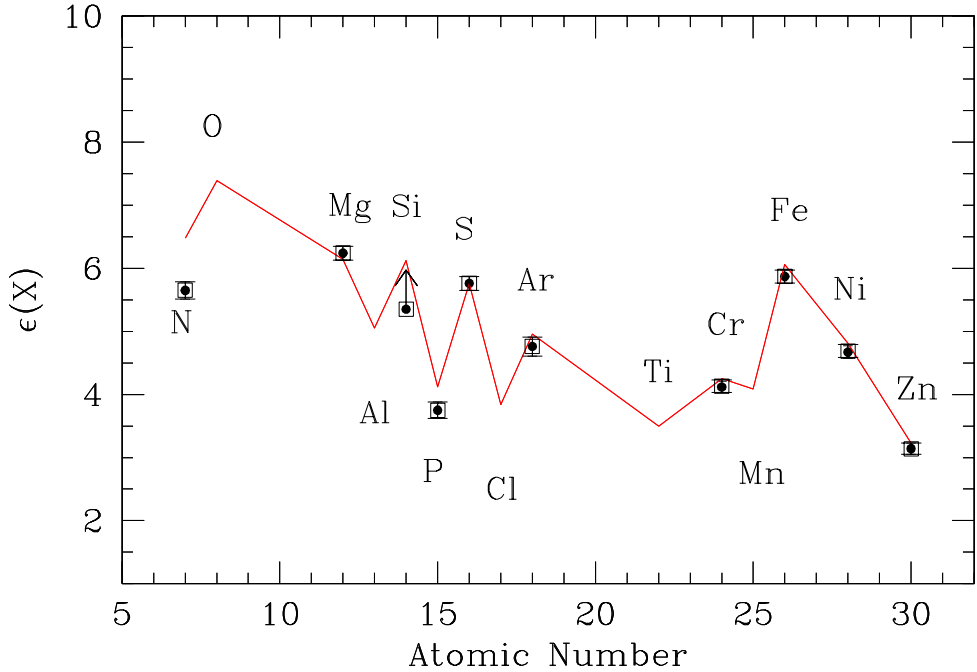}
\caption{The nucleosynthetic abundance pattern of the DLA at $z_{\rm abs} = 2.309$ toward Q0100+13. 
The dust-corrected elemental abundances are represented on a logarithmic scale, where hydrogen is 
defined to have $\epsilon({\rm H}) = 12$ and in a general way, for any element X, 
$\epsilon({\rm X}) = \log ({\rm X/H}) + 12$. They are compared with the solar abundance pattern 
from \citet{grevesse98} shown by the solid line and scaled to match the observed sulfur 
metallicity, [S/H] $= -1.44$, of the DLA system. The dust corrections were obtained via the 
\citet{vladilo02a,vladilo02b} method. The filled circles correspond to the dust-corrected abundances 
obtained with the E00 dust correction model and the open squares to the dust-corrected abundances 
obtained with the E11 dust correction model.}
\label{Q0100-pattern}
\end{figure*}
%

In our study we considered the second and the third approach to circumvent the problem of dust
depletion. In Table~\ref{abs-abundance} we summarize all the absolute abundances, and in 
Table~\ref{rel-abundance} we summarize the relative abundances which are further used in the paper. 
Both the observed and dust-corrected values are reported. The dust corrections were performed using 
the method developed by \citet{vladilo02a,vladilo02b}. This method groups together several dust
correction models based on different assumptions. They are labeled E00, E11 when one assumes that
the intrinsic [Zn/Fe] ratio is equal to $+0.10$~dex and S00, S11 when the intrinsic [Zn/Fe] ratio 
is equal to $+0.00$~dex \citep[see more details in][]{vladilo02a}. The absolute abundances, [X/H], 
were obtained by summing the contributions of all the components reported in 
Tables~\ref{Q0100-t}$-$\ref{Q2343-t}. The relative abundances, [X/Y], were computed by considering 
only the column densities of the components detected both in the X and Y profiles. In this way one 
avoids an overestimation of abundances derived from strong metal-line profiles relative to the 
abundances derived from weaker metal-line profiles in which generally only the strongest components 
are detected. In the case of very weak lines, like the \ion{Ti}{ii} lines for instance, one can 
indeed underestimate the [X/Fe] ratios by up to $0.3-0.4$~dex by considering the total Fe abundance 
generally derived from strong \ion{Fe}{ii} lines. In the three DLAs studied this effect is 
particularly important in the DLA toward Q2231$-$00 which show complex metal-line profiles, with a 
large number of components. The same approach has already been used in Paper~I.
%

\subsection{Intrinsic abundance patterns}\label{intrinsic-abundance}

In this Section we describe the derived {\it intrinsic} abundance patterns of the DLAs toward 
Q0100+13, Q1331+17 and Q2231$-$00. Their interpretation will be presented in Sect.~\ref{SFH}. To
present the abundance patterns we use the same type of diagrams as the ones used by 
\citet[][]{prochaska03b}. The dust-corrected elemental abundances are plotted on a logarithmic 
scale, where hydrogen is defined to have $\epsilon({\rm H}) = 12$ and, in a general way, for any 
element X, $\epsilon({\rm X}) = \log ({\rm X/H}) + 12$, and are compared with the solar abundance 
pattern from \citet{grevesse98} (shown by the solid line) scaled to match the observed sulfur 
metallicity of each DLA system. This way of showing the elemental abundances as a function of the 
atomic number, Z, has the advantage that we see in the same plot all the measured abundances in a 
system, and this allows us to directly identify possible deviations from solar values.

\begin{figure*}[!]
\centering
   \includegraphics[width=14cm]{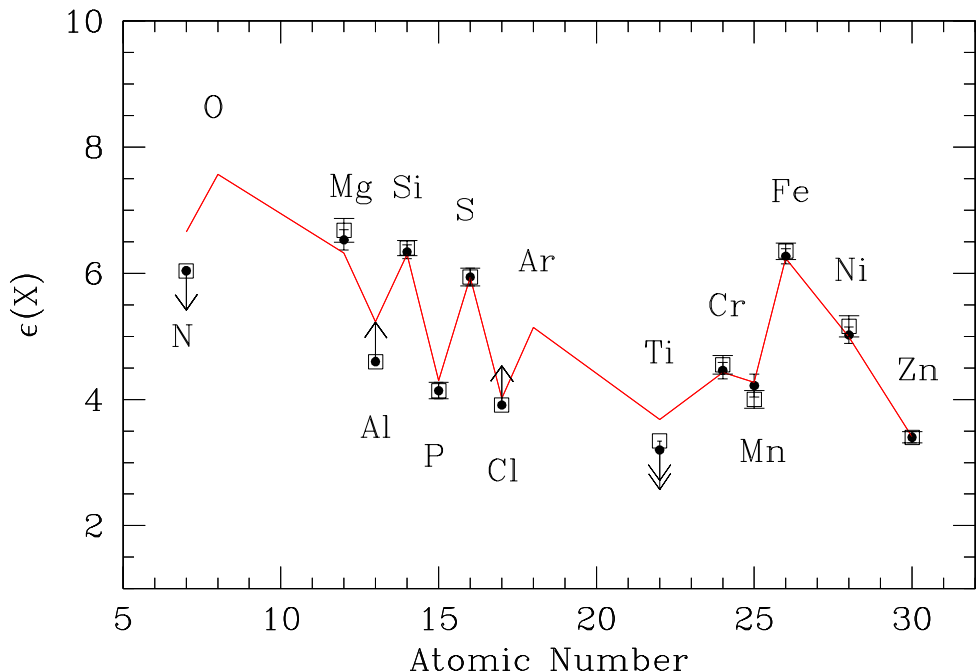}
\caption{The nucleosynthetic abundance pattern of the DLA at $z_{\rm abs} = 1.776$ toward Q1331+17.
The solar values were scaled to match the observed sulfur metallicity, [S/H] $= -1.26$. Same as 
Fig.~\ref{Q0100-pattern}.}
\label{Q1331-pattern}
\end{figure*}

Figures~\ref{Q0100-pattern}, \ref{Q1331-pattern} and \ref{Q2231-pattern} show the nucleosynthetic 
patterns of the DLAs toward Q0100+13, Q1331+17 and Q2231$-$00, respectively. At a first glance, the
abundance patterns of these high redshift galaxies resemble that of the solar neighborhood 
indicating that their nucleosynthetic enrichment histories are not too dissimilar from our Galaxy. 
However, at closer inspection, one notes several important differences.
%

\subsubsection{Q0100+13, $z_{\rm abs} = 2.309$}

The DLA toward Q0100+13 (Fig.~\ref{Q0100-pattern}) shows a slight enhancement of the 
$\alpha$-elements S and Mg abundances relative to the Zn and Fe-peak element abundances, namely 
[S/Zn] $= +0.19\pm 0.09$ and [Mg/Fe]$_{\rm cor} = +0.29\pm 0.11$. This $\alpha$/Fe-peak enhancement 
is suggestive of an enrichment by massive stars. Indeed, the $\alpha$-elements are produced in less 
than $2\times 10^7$ yrs by Type II supernovae (SNe) resulting from massive stars and the Fe-peak 
elements are mainly produced by Type Ia SNe on longer timescales, between $10^8-10^9$ yrs. 

The dust-corrected ratio of the two $\alpha$-elements Mg and S, [Mg/S]$_{\rm cor} = +0.10\pm 0.12$, 
shows that these two elements closely track each other in the DLA. On the other hand, the 
$\alpha$-element Ar is slightly underabundant relative to S, as noted in Sect.~\ref{ionization}. 
The Fe-peak elements Fe, Ni and Cr show solar values one relative to the other, as observed in the 
Galactic stars with similar metallicities (see Table~\ref{abs-abundance}). 

In this system, one also observes the odd-even effect, which corresponds to an underabundance of 
odd-Z elements relative to the even-Z elements of the same nucleosynthetic origin. Indeed, we 
obtained [P/S] $= -0.37\pm 0.12$ at [P/H] $= -1.85\pm 0.13$, which even shows evidence for an 
enhanced odd-even effect. This value is similar to the [P/Si] $=-0.21$ at [P/H] $=-1.16$, [P/Si] 
$=-0.40$ at [P/H] $=-2.30$ and [P/Si] $=-0.30$ at [P/H] $=-1.20$ values measured toward other DLAs 
by \citet{levshakov02}, \citet{molaro01} and \citet{outram99}, respectively, and is indicative of 
massive supernovae. Phosphorus abundance measurements in DLAs are very important to understand
the nucleosynthesis of this element, because no P abundance measurement exists in the Galactic 
stars. The observed odd-even effect in the P/S ratio is in agreement with the findings of
\citet{goswami00}.

Finally, the DLA shows undersolar [N/S] ratio of $-0.83\pm 0.14$ at [S/H] $= -1.48 \pm 0.11$. N is 
an element of particular interest, since it has a complex nucleosynthetic origin. Synthesized in 
the CNO cycle of stars, N is either a secondary element in the sense that it is produced in 
proportion to C and O originally present in the star, or a primary element in the sense that it can 
be produced starting from C and O manufactured by the star ``in situ''. In the N/$\alpha$ versus 
$\alpha$/H diagram, the secondary N over $\alpha$-element ratio is expected to increase steeply 
with the increasing metallicity, while the primary N over $\alpha$-element ratio remains constant 
when the metallicity increases \citep{talbot74}. The progenitors of primary N have not been fully 
constrained yet. The most recent stellar models of \citet{meynet02} including the stellar rotation 
indicate that the intermediate-mass stars are the main producers of N. The [N/S] ratio reached in 
the DLA toward Q0100+13 is a ``high'' value, in the sense that it is close to the primary N 
``plateau'' at [N/S] $\simeq -0.75$. It is at the upper end of the range of N/S values measured in 
DLAs and in good agreement with the typical values observed in \ion{H}{ii} regions at similar 
metallicities \citep[e.g.][]{izotov99,pilyugin02}.
%

%

\subsubsection{Q1331+17, $z_{\rm abs} = 1.776$}

The DLA toward Q1331+17 (Fig.~\ref{Q1331-pattern}) has almost solar abundance pattern. Very few 
differences are observed. Indeed, the Fe-peak elements Fe, Ni and Cr closely track each other, the
two $\alpha$-elements S and Si also (see Table~\ref{abs-abundance}), and the $\alpha$-element over 
Fe-peak element ratios $-$ Si/Fe, S/Zn and S/Fe~$-$ are all solar. Only the Mg/Fe ratio is slightly 
oversolar (see Table~\ref{rel-abundance}). We would like to note that the S00 and S11 dust 
correction models of \citet{vladilo02a} which assume a solar ratio of Zn/Fe, lead to 
[S/Fe]$_{\rm cor} = -0.30\pm 0.17$ (see Table~\ref{rel-abundance}). Such an undersolar 
$\alpha$-element over Fe-peak element ratio is neither observed in Galactic stars nor predicted by 
chemical evolution models. Thus, this provides evidence that [Zn/Fe] $> 0$ (intrinsically) in this 
DLA system. Ti exhibits a large underabundance relative to the Fe-peak elements and this, although 
Ti is an $\alpha$-element, and even after having applied the dust corrections. At a first glance, 
the only way to explain this observation is that we have underestimated the Ti upper limit. 

We measure the abundances of P and Mn, two odd-Z elements. The odd-even effect is less marked in 
this system than in the DLA toward Q0100+13. We get [P/Si]$_{\rm cor} = -0.20\pm 0.11$ when the E00 
dust correction is applied and [P/Si]$_{\rm cor} = -0.26\pm 0.12$ when the E11 dust correction is 
applied at [P/H] $= -1.42\pm 0.13$, and [Mn/Fe] $= -0.15\pm 0.04$ (without dust corrections, and 
$-0.07\pm 0.05$, $-0.37\pm 0.05$ when the E00, E11 dust corrections are applied, respectively) at 
[Fe/H]$_{\rm cor} = -1.23\pm 0.12$. 

We have detected the \ion{Cl}{i} $\lambda$1347 line in this DLA system, which allows us to 
measure the Cl$^0$ column density and to provide a lower limit on the total Cl absolute abundance. 
Cl is also an odd-Z element. This is the third Cl abundance upper limit derived in a DLA system, 
the other two were obtained by \citet{ledoux02} and \citet{prochaska03b}. Cl abundance measurements 
in DLAs are very important to constrain the Cl nucleosynthesis in the low metallicity regime, in 
particular because no Cl abundance measurement exists in Galactic stars. Cl measurements are 
available in planetary nebulae of the Galactic disk \citep[PNe; e.g.][]{costa96,kwitter03} 
and in the Galactic disk ISM \citep[e.g.][]{welty99}. Oversolar [Cl/S] ratios between $+0.09$ and
$+0.5$~dex are observed in Type II PNe, and undersolar [Cl/S] ratios of $-0.6$ and $-0.2$~dex are 
measured in the cold and warm Galactic disk, respetively, suggesting that Cl is dust depleted. In 
the DLA studied here we derive a lower limit on the [Cl/S] ratio of $> -0.11$ at [Cl/H] $> -1.37$, 
and in the DLA studied by \citet{ledoux02} and \citet{prochaska03b} [Cl/S] $> -0.33$ and $> -0.68$, 
respectively. These three lower limits measured in DLAs suggest that the Cl/S ratios likely are 
solar to oversolar similarly to what is observed in PNe, thus leaving very little space for an 
odd-even effect for Cl. The chemical evolution models of \citet{goswami00} computed with the 
\citet{woosley95} yields predict a solar Cl/S ratio, and hence a negligible odd-even effect for Cl. 

\citet{prochaska99} also reported the detection of \ion{C}{ii}$^*$ in the DLA toward Q1331+17. The 
presence of both strong \ion{C}{ii}$^*$ and \ion{Cl}{i} absorptions suggests that the gas resides 
in a cold neutral medium, characteristic of highly depleted gas in the Milky Way. This DLA precisely 
shows an important dust depletion level, with [Zn/Fe] $=+0.75\pm 0.05$. Furthermore, the observation 
of significant \ion{Cl}{i} requires at least a modest molecular hydrogen fraction. Indeed, in 
regions where H$_2$ is optically thick, \ion{Cl}{ii} reacts rapidly with H$_2$ to form HCl$^+$, 
which in turn leads to \ion{Cl}{i} and \ion{H}{i}. Because the conversion of \ion{Cl}{ii} to 
\ion{Cl}{i} is faster than the photoionization of \ion{Cl}{i}, Cl is primarily neutral in regions 
where H$_2$ is abundant but otherwise is primarily ionized \citep{jura78}. 

Finally, the derived upper limits on [N/S] and [N/Si] are at the upper end of the range of 
N/$\alpha$ values measured in DLAs (see Table~\ref{rel-abundance}). They are close to the primary N 
``plateau''.
%

\begin{figure*}[!]
\centering
   \includegraphics[width=14cm]{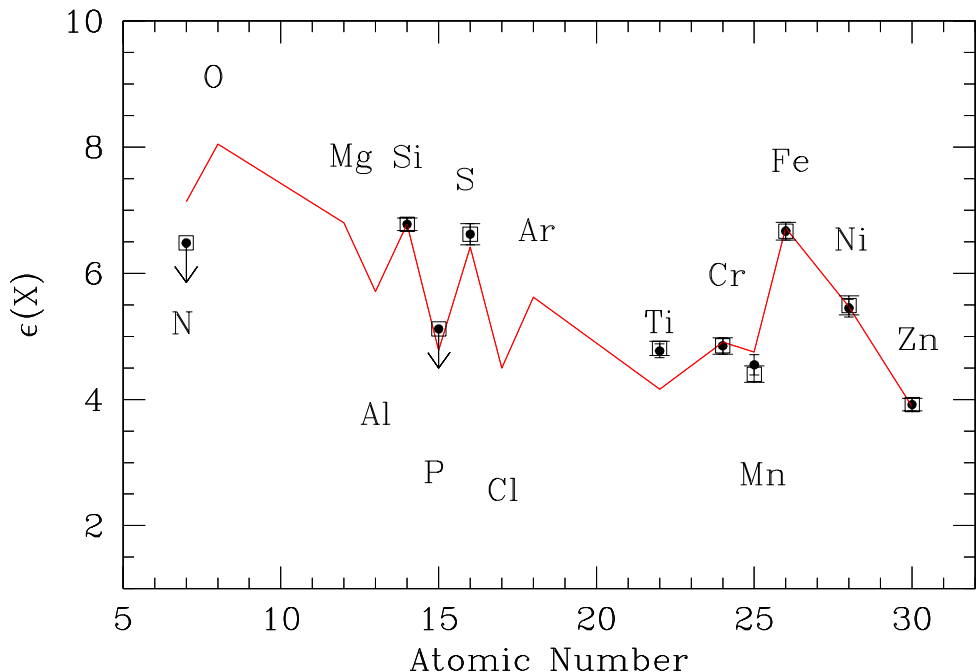}
\caption{The nucleosynthetic abundance pattern of the DLA at $z_{\rm abs} = 2.066$ toward 
Q2231$-$00. The solar values were scaled to match the dust-corrected silicon metallicity, 
[Si/H]$_{\rm cor} = -0.78$. Same as Fig.~\ref{Q0100-pattern}.}
\label{Q2231-pattern}
\end{figure*}
%

\subsubsection{Q2231$-$00, $z_{\rm abs} = 2.066$}

The DLA toward Q2231$-$00 (Fig.~\ref{Q2231-pattern}) shows a solar [Si/Fe] ratio, but enhanced
[S/Zn,Fe] and [Ti/Zn,Fe] ratios (see Table~\ref{rel-abundance}). The S abundance was derived from a 
single \ion{S}{ii} line located in the Ly$\alpha$ forest (see Sect.~\ref{q2231}), and we believe
that it may be partly blended with \ion{H}{i} clouds. The comparison of the dust-corrected absolute 
abundances of Si and S, two $\alpha$-elements, shows further evidence for a slight overestimation 
of the abundance of S, since [S/Si]$_{\rm cor} = +0.19\pm 0.16$ when the E00 dust correction is 
applied. Although the adopted large error on the S abundance measurement takes into account 
this slight overestimation, in Fig.~\ref{Q2231-pattern} we scale the solar values to match the dust 
corrected silicon metallicity, instead of the observed sulfur metallicity as done in 
Figs.~\ref{Q0100-pattern} and \ref{Q1331-pattern}. The Ti abundance is also very difficult to 
measure, thus it is very likely that it is overestimated, as already suspected in Sect.~\ref{dust}. 

The Fe-peak elements, Fe, Ni and Cr, track very well each other, as observed in the Galactic stars 
with similar metallicities (see Table~\ref{abs-abundance}). With the abundance measurement of the 
Fe-peak element Mn, we highlight a rather pronounced odd-even effect in this DLA system, [Mn/Fe] 
$ = -0.16\pm 0.05$ at [Fe/H]$_{\rm cor} = -0.83\pm 0.14$. Another odd-Z element detected in this 
system is P. Only an upper limit on the [P/Si] abundance ratio is derived, [P/Si]$_{\rm cor} < 
+0.34$, which is potentially in line with an odd-even effect for P. 

The obtained [N/Si,S] upper limits (see Table~\ref{rel-abundance}) are again ``high'', close to the 
primary N ``plateau''. We do not believe that the observed \ion{N}{i} lines are heavily blended 
with Ly$\alpha$ forest lines. 
%

\section{Determination of the star formation histories}\label{SFH}

Until now the DLA galaxy population has been analyzed as a whole. Several chemical evolution models 
were constructed in order to interpret the abundance patterns observed in DLAs as an ensemble, 
considering them as an evolutionary sequence, i.e. objects caught in a different phase of their
evolution \citep[e.g.][]{matteucci97,jimenez99,hou01,mathlin01}. However, several pieces of evidence 
$-$ like the low redshift deep imaging revealing a variety of morphological types belonging to the 
DLA population \citep[e.g.][]{lebrun97,nestor02}, the large scatter in the $\alpha$ over Fe-peak 
element abundance ratios at a given metallicity and the large scatter observed in the 
metallicities~$-$ indicate that the DLAs trace galaxies with different star formation histories. 
Some may have formed stars on timescales similar to that of the early Milky Way, while others 
apparently did so more slowly or intermittently, so that the Fe-peak elements could catch up with 
the $\alpha$-elements.

Thanks to the large number of elements detected in the DLA systems studied in this paper, we are 
for the first time in a very appropriate situation to analyze these high redshift galaxies 
individually. The different elemental abundances at disposal allow us, in particular, to constrain 
the star formation history (hereafter SFH, and SF for the star formation) and the age of the 
associated galaxies by means of a direct comparison with a grid of chemical evolution models for 
spiral and dwarf irregular galaxies \citep[see][]{calura03a} rather than just by means of a 
comparison with the solar abundance pattern as presented in Sect.~\ref{intrinsic-abundance}. We 
present below the approach which has been applied. The derived results for the DLAs toward Q0100+13, 
Q1331+17 and Q2231$-$00 are summarized in Table~\ref{summary-models}.
%

\begin{table*}[t]
\begin{center}
\caption{Nature of the galaxies associated with the DLA systems studied} 
\label{summary-models}
\begin{tabular}{l c c c c}
\hline \hline
\\[-0.3cm]
DLA system & Model & Star formation characteristics                      & $z_f$ & Age \\
           &       & $\nu$ [Gyr$^{-1}$] / $\Delta t$ [Gyr] / $t_b$ [Gyr] &       & [Gyr]
\smallskip 
\\     
\hline    
Q0100+13, $z_{\rm abs} = 2.309$     & Spiral          & solar neighborhood: $R=8$ kpc & $\sim 2.5$    & $0.25\pm 0.20$ \\
                                    & Dwarf irregular & bursting SF: 0.9 / 0.07 / 2.0 & ($2.35-2.47$) & $0.05-0.20$ \\
Q1331+17, $z_{\rm abs} = 1.776$     & Spiral          & outer regions: $R=16$ kpc     & $\sim 2.8$    & $1.5\pm 0.4$ \\
                                    & Dwarf irregular & continuous SF: 0.03 / 13 / ...& $\sim 10$     & $\gtrsim 3.5$ \\
Q2231$-$00, $z_{\rm abs} = 2.066$   & Dwarf irregular & bursting SF: 4.2 / 0.10 / 0.5 & ($2.10-2.16$) & $0.05-0.15$ \\
\hline
\end{tabular}
\begin{minipage}{153mm}
\smallskip
$z_f$ in brackets = Values obtained under the assumption that the galaxy has undergone a single 
burst of SF
\end{minipage}
\end{center}
\end{table*}
%
 
\subsection{Chemical evolution models}\label{models}

A chemical evolution model allows one to follow in detail the evolution of abundances of several 
chemical species, starting from the matter reprocessed by the stars and restored into the ISM 
through stellar winds and supernova explosions. Here we briefly summarize the major ingredients of 
the chemical evolutions models used in this work which we identify as ``spiral'' and ``dwarf 
irregular'' models according to the type of galaxies they do match best. These models are the same 
as the models used by \citet{calura03a} in their study of the DLA nature. A detailed description of 
the ``spiral'' model can be found in \citet{chiappini97,chiappini01} and of the ``dwarf irregular'' 
model in \citet{bradamante98} and \citet{recchi01,recchi02}. In both models no instantaneous 
recycling approximation is adopted, namely the stellar lifetimes are properly taken into account.
%

\subsubsection{``Spiral'' model}\label{spiral}

The spiral galaxies are assumed to form as a result of two main infall episodes \citep{chiappini97}. 
During the first episode the halo forms and the gas shed by the halo rapidly gathers in the center 
leading to the formation of the bulge. During the second episode, a slower infall of external gas 
gives rise to the disk with the gas accumulating faster in the inner than in the outer region 
\citep[``inside-out'' scenario,][]{matteucci89}. The process of disk formation is much longer than 
the halo and bulge formation, with timescales varying from $\sim 2$ Gyr in the inner disk to $\sim 
7$ Gyr in the solar region and up to $13$ Gyr in the outer disk. In particular, a timescale of 7 
Gyr in the solar neighborhood is required to fit the G-dwarf metallicity distribution 
\citep{chiappini97,boissier99}. The adopted SFR expression is:
\begin{equation}
\psi(R,t) = \nu [\frac{\sigma(R,t)}{\sigma(R_{\odot},t)}]^{2(k-1)} 
[\frac{\sigma(R,t_{Gal})}{\sigma(R,t)}]^{k-1} \sigma^{k}_{gas}(R,t)
\end{equation}
where $\nu$ is the star formation efficiency, namely the inverse of the typical timescale for star 
formation, $\sigma(R,t)$ is the total (gas + stars) mass surface density at a radius R and time t, 
$\sigma(R_{\odot},t)$ is the total mass surface density in the solar region, and $\sigma_{gas}(R,t)$ 
is the gas mass surface density. For the gas density exponent $k$ a value of 1.5 has been assumed 
by \citet{chiappini97} in order to ensure a good fit to the observational constraints in the solar 
vicinity. The star formation efficiency is set to $\nu = 1$ Gyr$^{-1}$, and becomes zero when the 
gas surface density drops below a certain critical threshold. A threshold density $\sigma_{\rm th} 
\sim 7$ M$_{\odot}$~pc$^{-2}$ is adopted in the disk as suggested by \citet{chiappini97}. The 
initial mass function (IMF) is taken from \citet{scalo86}. The galactic wind is considered not 
efficient \citep{matteucci01}. All the input parameters of this ``spiral'' model were constrained 
in order to successfully reproduce the observed properties first in the solar vicinity, such as the 
solar abundances, the G-dwarf metallicity distribution (see above) and the abundance ratios, and 
then at the global scale, such as the radial gas and stellar density profiles and the abundance 
gradients of various elements observed in the stars of the Milky Way.

Due to the ``inside-out'' scenario and the threshold density adopted in the ``spiral'' model, {\it 
the SFH is different at different galactocentric radii}. Indeed, in the inner regions of the disk 
where the rate of accretion of matter onto the disk is fast, a high gas surface density, well above 
the critical threshold, is reached and maintained during a long period, and hence {\it the SF is 
almost continuous} through the galaxy lifetime. On the other hand, in the outer regions of the disk 
(radius $R \gtrsim 12$ kpc) where the rate of accretion of matter onto the disk is slow, the SFH 
proceeds in a gasping way, due to the fact that in these regions the gas is always close to the 
critical threshold. As a consequence, the external regions of the disk look like the Magellanic
irregulars where this kind of SF seems to take place \citep{tosi91}. These different star formation 
histories at different galactocentric radii are associated with different abundance patterns which 
can be compared with observations. 

When comparing the ``spiral'' model with our observations of DLA systems, we run a set of models
each of them corresponding to a different galactocentric radius R. This radius represents the
position at which the QSO line of sight crosses the disk of the observed DLA galaxy. Finally, we 
need to constrain the formation redshift, $z_f$, of the ``spiral'' model, which is {\it the single 
free parameter}.
%

\subsubsection{``Dwarf irregular'' model}\label{dwarf}

We adopt a model based on the work of \citet{bradamante98}, in which the dwarf irregular galaxies 
form owing to a continuous infall of pristine gas with an infall timescale of 0.5 Gyr, until a mass 
of $\sim 10^9$ M$_{\odot}$ is accumulated. The evolution of dwarf irregular galaxies is assumed to 
be characterized by {\it a bursting star formation history}. This particular model was built to 
reproduce the chemical properties of the local blue compact galaxies (BCGs). The parameters which 
need to be defined in this model are the number of bursts of SF that the galaxy has undergone, and 
for each burst, the star formation efficiency, $\nu$, (same definition as in the ``spiral'' model), 
the burst duration, $\Delta t$, and the time of occurrence of the burst, $t_b$, i.e. the time 
necessary for the infall of pristine gas before the SF starts. Hence, the star formation rate in 
the ``dwarf irregular'' model can proceed either in short bursts of a duration from 10 to 200 Myr 
separated by long quiescent periods, or at a low regime but continuously, namely in one or two long 
episodes of inefficient SF lasting between 3 and 13 Gyr. In this model, the adopted SFR is:
\begin{equation}
\psi(t) = \nu \sigma_{gas}(t) 
\end{equation}
where $\sigma_{gas}(t)$ represents the gas density at time $t$ and $\nu$ is the star formation 
efficiency. 

The dwarf irregular galaxies are particularly sensitive to outflows resulting from the energy 
injection from the SN explosions. Indeed, galactic winds develop when the thermal energy of the gas 
equals its binding energy \citep{matteucci87}. In the ``dwarf irregular'' model used in this paper, 
the rate of gas loss via galactic winds is assumed to be proportional to the SF, and the recent 
chemodynamical results of \citet{recchi01}, suggesting that the wind is differential, are adopted.
This implies that the ejecta of Type Ia SNe and intermediate-mass stars are lost from the parent 
galaxy more easily than Type II SN ejecta, so that the Fe and N ejection efficiencies are larger 
than the ejection efficiencies of $\alpha$-elements (e.g. O, Mg). This is due to the fact that 
the Type Ia SNe can transfer into the ISM more energy than the Type II SNe, since they explode in 
an already rarefied and hot medium. The IMF is taken from \citet{salpeter55}.

For the comparison of the ``dwarf irregular'' model with our observations of DLA systems, we assume 
that these high redshift galaxies have undergone a {\it single} burst of SF. This can be justified
by the fact that the DLA galaxies are probably young galaxies. We thus need to essentially constrain 
{\it three free parameters}, the burst star formation efficiency, $\nu$, the burst duration, 
$\Delta t$, and the time of occurrence of the burst, $t_b$. 
%

\subsubsection{Stellar yields}

We describe here the nucleosynthesis prescriptions adopted in both the ``spiral'' and ``dwarf
irregular'' models. They include the yields of \citet{nomoto97a} for massive stars (M $> 10$ 
M$_{\odot}$), the yields of \citet{hoeck97} for low- and intermediate-mass stars ($0.8 \le$ 
M/M$_{\odot}$ $\le 8$) and the yields of \citet{nomoto97b} for Type Ia SNe (model W7). For Zn and 
Ni we consider the following specific prescriptions.
%

\begin{figure}[!]
\centering
   \includegraphics[width=8.5cm]{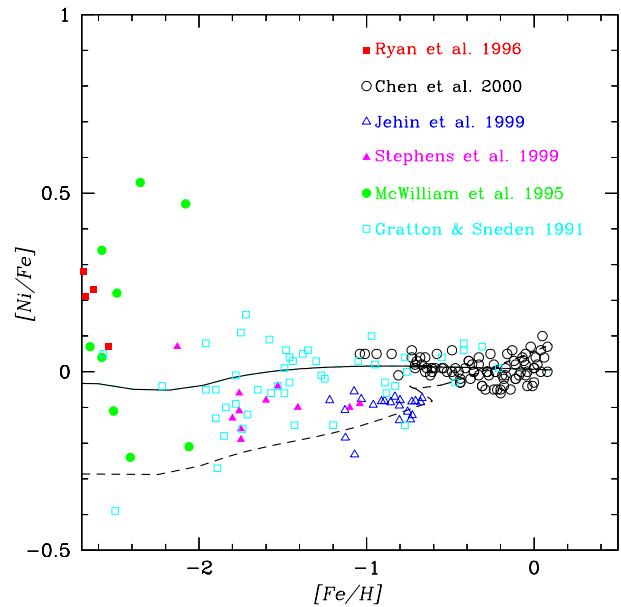}
\caption{Observed and predicted [Ni/Fe] versus [Fe/H] distributions in the solar neighborhood for 
two different nucleosynthetic prescriptions for the production of Ni. The {\it dashed line}
corresponds to the model with $\beta = 0.0065$ and the {\it solid line} to the model with $\beta = 
0.0049$, where $\beta$ is a multiplicative factor between the Fe and Ni yields (see text). The {\it 
solid squares} are the data from \citet{ryan96}, the {\it open circles} the ones from 
\citet{chen00}, the {\it open triangles} the ones from \citet{jehin99}, the {\it filled triangles} 
the ones from \citet{stephens99}, the {\it filled circles} the ones from \citet{mcwilliam95}, and 
the {\it open squares} the ones from \citet{gratton91}.}
\label{NiFe-Fe}
\end{figure}
%

The nucleosynthesis of Zn is a debated issue, since what process is making Zn is rather uncertain. 
According to stellar models, the production of Zn can ensue via s-processes in low- and high-mass
stars during He-core burning as well as during explosive nucleosynthesis events occurring in Type Ia
and II SNe \citep[][and references therein]{matteucci93}. This uncertainty in the Zn production has 
repercussions on the reliability of Zn yields. In this paper we use the Zn yields adopted by 
\citet{calura03a}, which were estimated by extrapolating the Galactic star abundance observations 
in the solar neighborhood to the high redshift Universe, assuming Fe and Zn track each other as 
observed in the metallicity range where the DLAs lie \citep{prochaska00,mishenina02}. 

The Ni yields have also some difficulties to reproduce the Galactic star abundance observations. 
Therefore, similarly to Zn, we assumed for the Ni production the results of \citet{matteucci93}, in 
which a good fit to the solar abundance of Ni is found if the bulk of its production is ascribed to 
Type Ia SNe. However, since a non-negligible fraction of Fe also comes from Type II SNe 
\citep{nomoto97a} and since the abundance of Fe seems to vary in lockstep with that of Ni in the 
solar vicinity, the Type II SNe should also produce some Ni. In our models the amount of Ni 
produced through explosive nucleosynthesis in massive stars scales with the Fe yields, according to 
$Y_{{\rm Ni}} = \beta \times Y_{{\rm Fe}}$, where $\beta$ represents a multiplicative factor. We 
have run a chemical evolution model for the solar vicinity varying $\beta$ in order to reproduce 
the [Ni/Fe] versus [Fe/H] distribution observed in Galactic stars of different metallicities. 
Figure~\ref{NiFe-Fe} shows this [Ni/Fe] versus [Fe/H] distribution compared with the predictions 
for the solar neighborhood model, when two values of $\beta$ are adopted (dashed line: $\beta = 
0.0065$; solid line: $\beta = 0.0049$). The choice of $\beta=0.0049$ gives a satisfactory agreement 
between the predictions and the observations, in particular in the metallicity range occupied by 
the DLAs ([Fe/H] $\le -2$). For Ni produced in Type Ia SNe, we assume a constant value M$_{\rm Ni} 
\sim 2.3 \times 10^{-2}$ M$_{\odot}$.
%

\subsection{Comparing the DLA abundances with the models}\label{model-comparison}

To be able to constrain the star formation history of DLAs from the comparison of their abundance
patterns with chemical evolution models, we first need to understand how the abundance patterns can 
provide information on the star formation history of galaxies. In few words, the absolute 
abundances depend on the model assumptions (i.e. the SFH), whereas the relative abundances depend 
only on the nucleosynthesis, the stellar lifetimes and the IMF. Relative abundances can therefore 
be used as cosmic clocks if they involve two elements formed on different timescales, like it is the 
case of the $\alpha$-element over Fe-peak element ratios and the N over $\alpha$-element ratios. As 
a consequence, these abundance ratios when examined together with the absolute abundances [Fe/H], 
or any other metallicity tracer such as [Zn/H], and [$\alpha$/H], respectively, completely determine 
{\it the star formation history} of a galaxy. On the other hand, when these abundance ratios are 
examined as a function of the redshift, they provide constraints on {\it the age} of a galaxy, 
defined as the epoch at which the galaxy has started to form stars \citep{matteucci01}. 

The relative abundances that we have at disposal in the DLA systems studied here are [Si/Fe], 
[S/Zn], [S/Fe], [Mg/Fe], [Ni/Fe], [N/Si] and [N/S]. For each DLA individually, we consider all of 
them at once to first constrain the SFH and then the age of the associated galaxy by means of a 
detailed comparison with a grid of chemical evolution models. This grid includes the ``spiral'' 
model computed at different galactocentric radii from $R = 2$ to 18 kpc and the ``dwarf irregular'' 
model computed for a single burst of SF with different characteristics defined by varying the three 
free parameters $\nu$, $\Delta t$ and $t_b$. The adopted cosmology is ${\rm H}_0 = 65$ 
km~s$^{-1}$~Mpc$^{-1}$ (and hence $h=0.65$), $\Omega_{\rm M} = 0.3$ and $\Omega_{\rm \Lambda} = 
0.7$.
%

\begin{figure*}[t]
\centering
   \includegraphics[width=16cm]{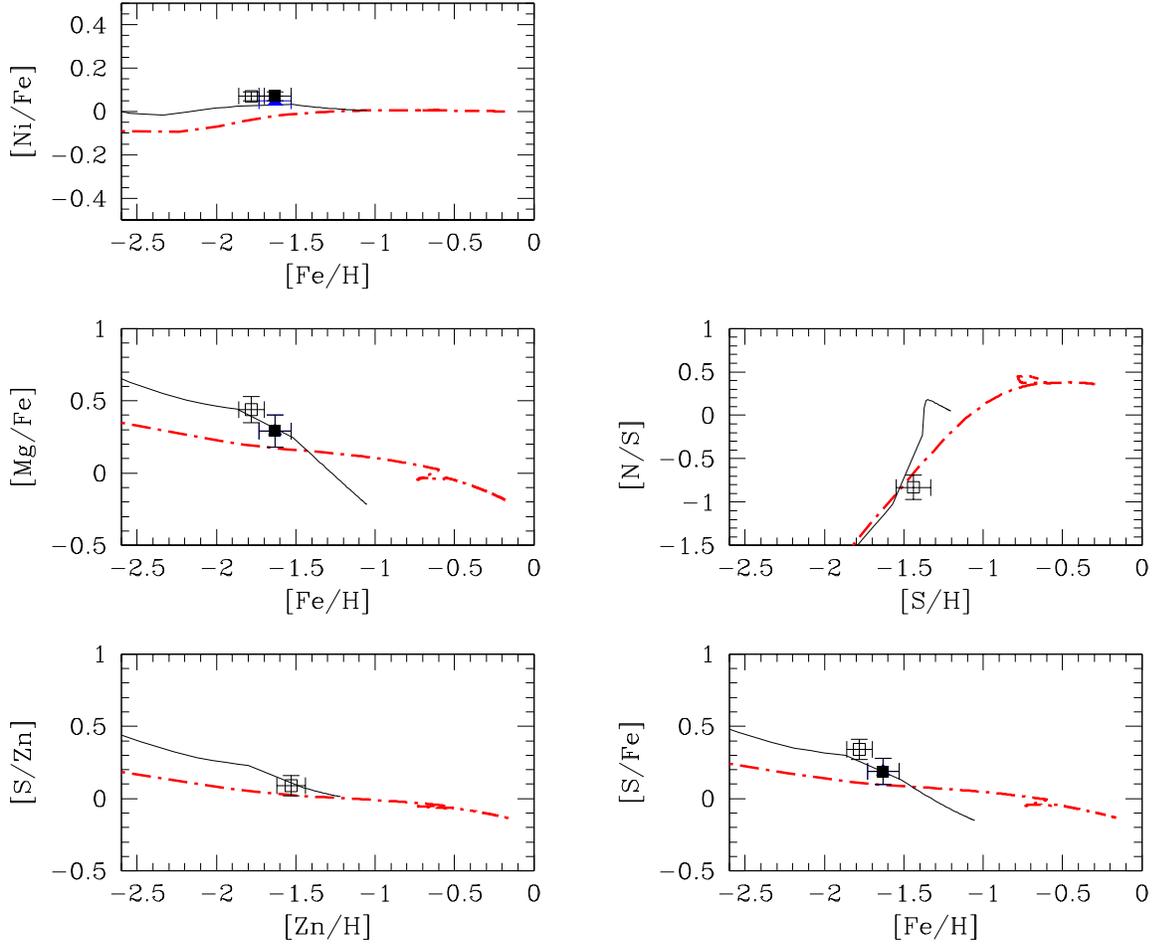}
\caption{Observed and predicted abundance ratios versus metallicity for the DLA at $z_{\rm abs} = 
2.309$ toward Q0100+13. The {\it thick dashed-dotted} curves correspond to the ``spiral'' model at 
$R = 8$ kpc. The {\it thin solid} curves correspond to the ``dwarf irregular'' model with a single 
burst of SF occurring at $t_b = 2$ Gyr and with a star formation efficiency $\nu = 0.9$ Gyr$^{-1}$ 
and a duration $\Delta t = 0.07$ Gyr. In this and all the following figures, the {\it open squares} 
represent the measured abundance ratio measurements, and the {\it filled triangles} and the {\it 
filled squares} represent the abundance ratios corrected for dust depletion according to the 
\citet{vladilo02a} method using the models E00 and E11, respectively.}
\label{alpha-Fe-Q0100}
\end{figure*}
%

To determine the best chemical evolution model reproducing the abundance patterns observed in each 
DLA system by taking into account all the different informations available on the system, we use a 
statistical test when the best solution cannot be clearly identified by eye. This test consists 
first in determining the minimal distance between the data point and the curve of a given chemical 
model in each abundance diagram at disposal. For this purpose, we consider the 1~$\sigma$ error 
on the data point (or more precisely the covariance matrix of the 1~$\sigma$ measurement error), 
and we derive this minimal distance by computing the distances $d_i$ from the data point to the 
points defining the considered model curve and by looking for the $d_i$ for which the 
$d_i$/$\sigma_i$ ratio is minimal. Second, once the minimal distances for all the abundance 
diagrams considered in each system and for a given model are derived, we compute their weighted 
mean. Finally, the comparison of the weighted means obtained for different chemical models 
determines the best chemical evolution model which represents the data points and thus the DLA 
galaxy. The upper and lower limits are not taken into account in this test.
%

\subsubsection{Q0100+13, $z_{\rm abs} = 2.309$}\label{SFH-q0100}

To determine the nature of this DLA galaxy, we have five different abundance ratios at disposal. 
We can thus try to constrain the best chemical evolution model which reproduces this DLA galaxy by 
taking into account the information provided by the following five diagrams\,: [S/Zn] versus [Zn/H], 
[S/Fe] versus [Fe/H], [Mg/Fe] versus [Fe/H], [Ni/Fe] versus [Fe/H] and [N/S] versus [S/H]. For the 
absolute and relative abundances of refractory elements we consider the dust-corrected values as 
derived with the dust correction method of 
\citet[][and see Tables~\ref{abs-abundance} and \ref{rel-abundance}]{vladilo02a,vladilo02b}. In 
this DLA system, however, the dust corrections are small, since the [Zn/Fe] ratio is close to solar
($= +0.25\pm 0.04$).
%

\begin{figure*}[t]
\centering
   \includegraphics[width=16cm]{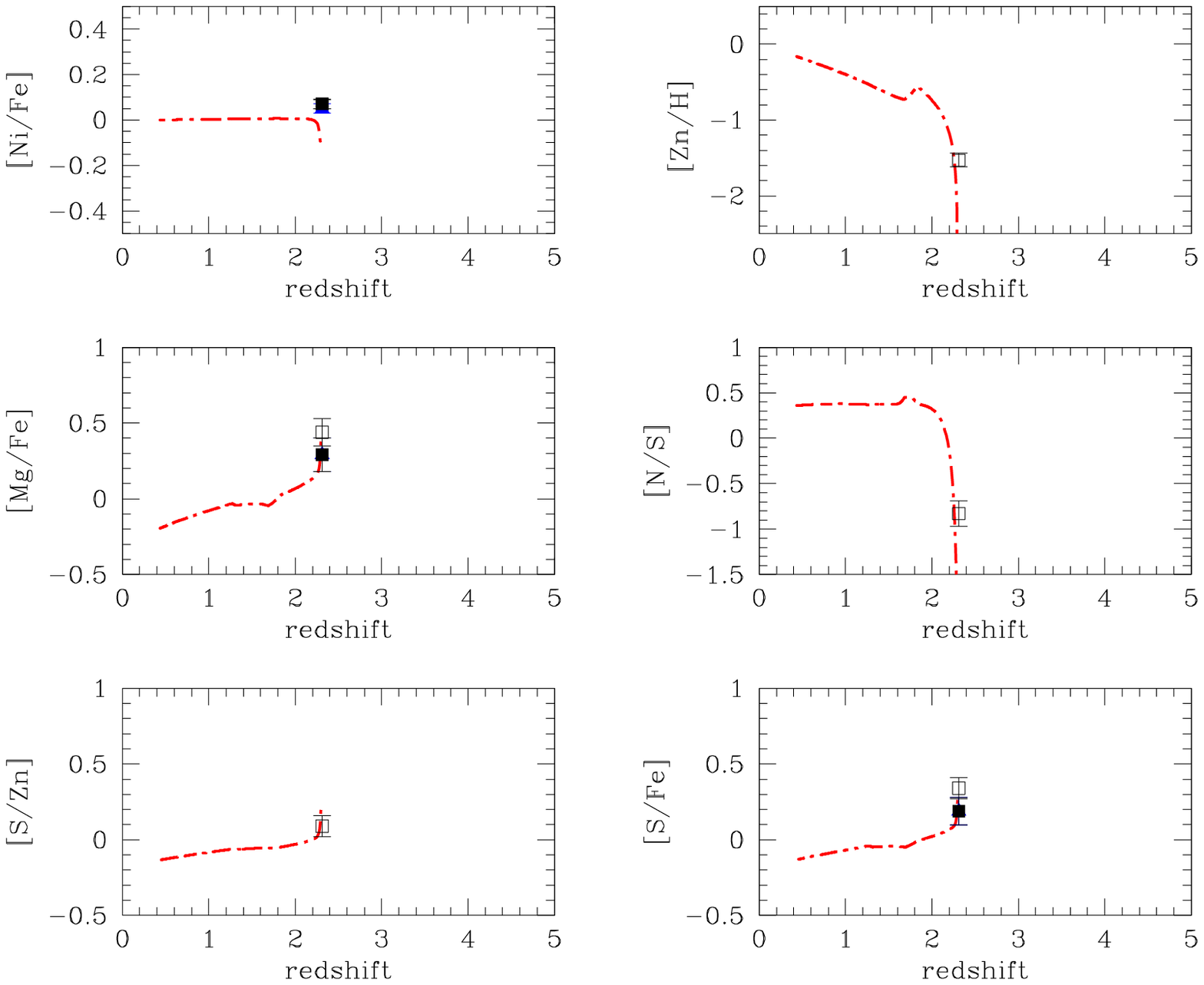}
\caption{Observed and predicted abundance ratios versus redshift for the DLA at $z_{\rm abs} = 
2.309$ toward Q0100+13. The {\it thick dashed-dotted} curves correspond to the ``spiral'' model at 
$R = 8$ kpc and a redshift of formation $z_f = 2.5$. For the definition of symbols, see 
Fig.~\ref{alpha-Fe-Q0100}.}
\label{alpha-Fe-z-Q0100}
\end{figure*}
%

We considered the ``spiral'' model at the galactocentric radii $R = 6,\,8,\,12,$ and $18$ kpc. The 
``spiral'' model at $R = 8$ kpc best reproduces the five abundance ratios measured in this high 
redshift galaxy (see the thick dashed-dotted curves in Fig.~\ref{alpha-Fe-Q0100}). We might be 
satisfied with this solution, since the model curves are in agreement with the data points within 
less than 1~$\sigma$ in four out of the five diagrams. In the case of the Ni/Fe ratio, the model is 
in agreement only within $2-3$~$\sigma$. But, to have a complete picture and to check the uniqueness 
of the derived solution, we investigated whether a ``dwarf irregular'' model can also correctly 
reproduce the data points. We considered a ``dwarf irregular'' model with a single burst of SF and 
explored the following values for the three parameters characterizing the burst: $\nu = 
0.5,\,0.7,\,0.9,\,1$ Gyr$^{-1}$, $\Delta t = 0.05,\,0.07,\,0.1,\,0.2$ Gyr and $t_b = 
1,\,1.1,\,1.5,\,2$ Gyr. The ``dwarf irregular'' model which best reproduces the data points has one 
burst with a star formation efficiency $\nu = 0.9$ Gyr$^{-1}$ and a short duration $\Delta t = 0.07$ 
Gyr (see the thin solid curves in Fig.~\ref{alpha-Fe-Q0100}). The burst occurs after $t_b = 2$ Gyr 
of continuous infall of pristine gas. These ``dwarf irregular'' model characteristics very 
well match the starburst parameters determined for the BCGs \citep[e.g.][]{lanfranchi03}. In 
summary, two chemical evolution models yet tracing different star formation histories correctly 
reproduce the abundance ratio versus metallicity distributions observed in the DLA toward Q0100+13. 
The ``spiral'' model at $R = 8$ kpc has a continuous SFH, whereas the ``dwarf irregular'' model 
constrained by the observations has a bursting SFH (see Sect.~\ref{models}). The identification of 
the best model among these two is very difficult, because at this intermediate metallicity both
models yield very similar chemical abundance patterns. The distinction is more pronounced at [Fe/H] 
$< -2$ or at [Fe/H] $> -1$.
%

\begin{figure*}[t]
\centering
   \includegraphics[width=16cm]{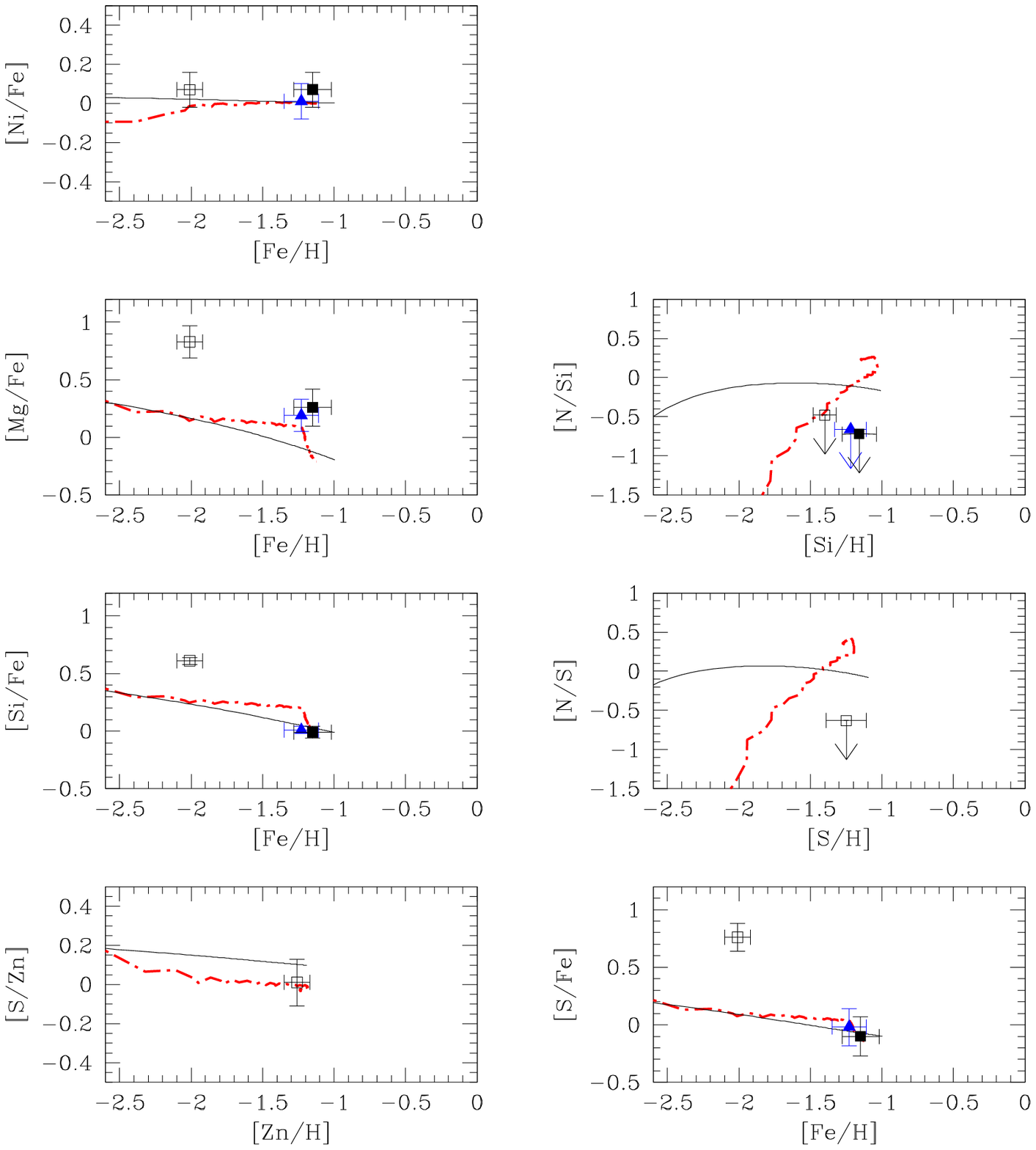}
\caption{Observed and predicted abundance ratios versus metallicity for the DLA at $z_{\rm abs} = 
1.776$ toward Q1331+17. The {\it thick dashed-dotted} curves correspond to the ``spiral'' model at 
$R = 16$ kpc and the {\it thin solid} curves correspond to the ``dwarf irregular'' model with a 
continuous SF at an efficiency $\nu = 0.03$ Gyr$^{-1}$. For the definition of symbols, see 
Fig.~\ref{alpha-Fe-Q0100}.}
\label{alpha-Fe-Q1331}
\end{figure*}
%

We carried out the investigation of the age of the system, namely the most likely redshift of 
formation, $z_f$, of the system, for the two models. For the ``spiral'' model at $R = 8$ kpc, we 
analyzed the [Zn/H], [Ni/Fe], [Mg/Fe], [S/Zn], [S/Fe] and [N/S] versus redshift diagrams. The 
estimated redshift of formation is between $z_f = 2.4-3$ with the best solution being at $z_f\sim 
2.5$ which is in excellent agreement with all the data points (see the thick dashed-dotted curves 
in Fig.~\ref{alpha-Fe-z-Q0100}). The DLA system is observed at $z_{\rm abs} = 2.309$, hence 
$z_f\sim 2.5$ corresponds to an age of the DLA galaxy of $0.25\pm 0.20$ Gyr. In the case of the
``dwarf irregular'' model, the determination of the age is more complex, because this model has 
several free parameters. The parameter having the lowest impact is the time of occurrence of the 
burst, $t_b$, according to the adopted definition of the age (see Sect.~\ref{model-comparison}). In 
addition, $t_b$ is difficult to determine, whereas the star formation efficiency, $\nu$, and the
burst duration, $\Delta t$, can be robustly constrained by the observations. It is the burst 
duration which has the largest weight in the determination of the age of the system. We 
satisfactorily reproduce the data points (within 1~$\sigma$) for $\Delta t$ values between 0.05 and 
0.20 Gyr. Hence, we feel confident in concluding that the possible age of the system is between 
$0.05-0.20$ Gyr. Given the assumption that the galaxy has undergone a single burst of SF, the 
corresponding redshift of formation is between $2.35-2.47$.
%
\begin{figure*}[t]
\centering
   \includegraphics[width=16cm]{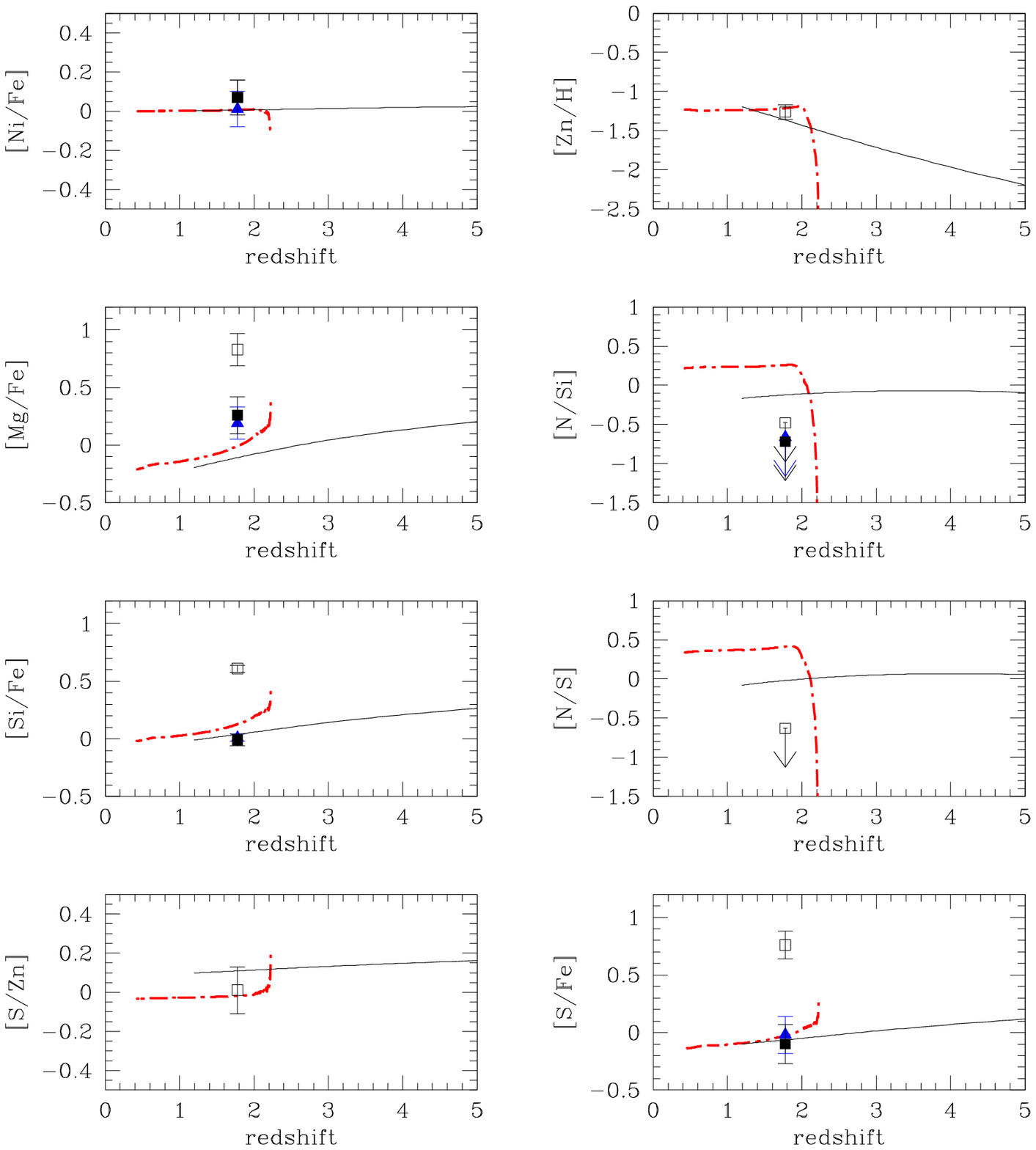}
\caption{Observed and predicted abundance ratios versus redshift for the DLA at $z_{\rm abs} = 
1.776$ toward Q1331+17. The {\it thick dashed-dotted} curves correspond to the ``spiral'' model at 
$R = 16$ kpc and a redshift of formation $z_f = 2.8$. The {\it thin solid} curves correspond to 
the ``dwarf irregular'' model with a continuous SF at an efficiency $\nu = 0.03$ Gyr$^{-1}$ and a 
redshift of formation $z_f = 10$. For the definition of symbols, see Fig.~\ref{alpha-Fe-Q0100}.}
\label{alpha-Fe-z-Q1331}
\end{figure*}
%

Both models point to a young age for the DLA galaxy, lower than 250 Myr. This age is consistent 
with the $\alpha$-element over Fe-peak element enhancement observed in this DLA system (see 
Sect.~\ref{intrinsic-abundance}), which indicates that the enrichment of this system is dominated 
by massive star products and requires a minimal contribution from Type Ia SNe. In addition, we saw 
in Sect.~\ref{intrinsic-abundance} that the N/S ratio in this DLA system is relatively high, very 
close to the primary N ``plateau'', [N/S] $= -0.83\pm 0.14$. This N/S ratio and the estimated 
age for the DLA galaxy provide important constraints on the N stellar progenitors. Indeed, such a 
high N/S value can be reached within less than 250 Myr, only if N is produced by intermediate-mass 
stars with masses between $5-8$ M$_{\odot}$, which have lifetimes between 30 and 70 Myr, and by 
massive stars. \citet{chiappini03} recently showed that the same chemical evolution models as those
used in this paper, but computed with the recent published stellar yields of \citet{meynet02} which 
take into account the effects of rotation in the stellar evolution, still reproduce such high N/S 
values within a timescale lower than 250 Myr.
%

\subsubsection{Q1331+17, $z_{\rm abs} = 1.776$}\label{SFH-q1331}

We have collected a lot of information on this DLA system. The observed abundance ratios at our 
disposal are [S/Zn], [S/Fe], [Si/Fe], [Mg/Fe], [Ni/Fe] and the upper limits on [N/S] and [N/Si]. 
This DLA system has a very high [Zn/Fe] $= 0.75\pm 0.05$ ratio and exhibits one of the largest dust 
depletion level of any DLA. In Fig.~\ref{alpha-Fe-Q1331} we can see the high differences between 
the observed and dust-corrected values. 

First, we examined the ``spiral'' model at the galactocentric radii $R = 8,\,12,\,16$ and $18$ kpc. 
The best model reproducing the data points in the [S/Zn] versus [Zn/H], [S/Fe] versus [Fe/H], 
[Si/Fe] versus [Fe/H], [Mg/Fe] versus [Fe/H] and [Ni/Fe] versus [Fe/H] diagrams is the ``spiral'' 
model of outer regions of the disk, at $R = 16$ kpc (see the thick dashed-dotted curves in 
Fig.~\ref{alpha-Fe-Q1331}). The model curves are in excellent agreement with the data points within 
less than 1~$\sigma$ in the five diagrams. Secondly, we investigated the ``dwarf irregular'' model 
to check the uniqueness of the solution. We considered the ``dwarf irregular'' model with a single 
burst of SF and explored a large range of values for the three parameters characterizing the burst. 
We found out that the ``dwarf irregular'' model with a bursting SFH cannot reproduce the 
observations, which show solar $\alpha$-element over Fe-peak element ratios, because the bursting 
SFH predicts $\alpha$-enhanced patterns. On the other hand, the ``dwarf irregular'' model with a 
continuous SFH characterized by a single burst of SF with a star formation efficiency in the range 
$\nu = 0.02-0.09$ Gyr$^{-1}$, with the best value being $\nu = 0.03$ Gyr$^{-1}$, and a duration 
over the whole Hubble time satisfactorily reproduces the observations (see the thin solid curves in 
Fig.~\ref{alpha-Fe-Q1331}). 
%

\begin{figure*}[t]
\centering
   \includegraphics[width=16cm]{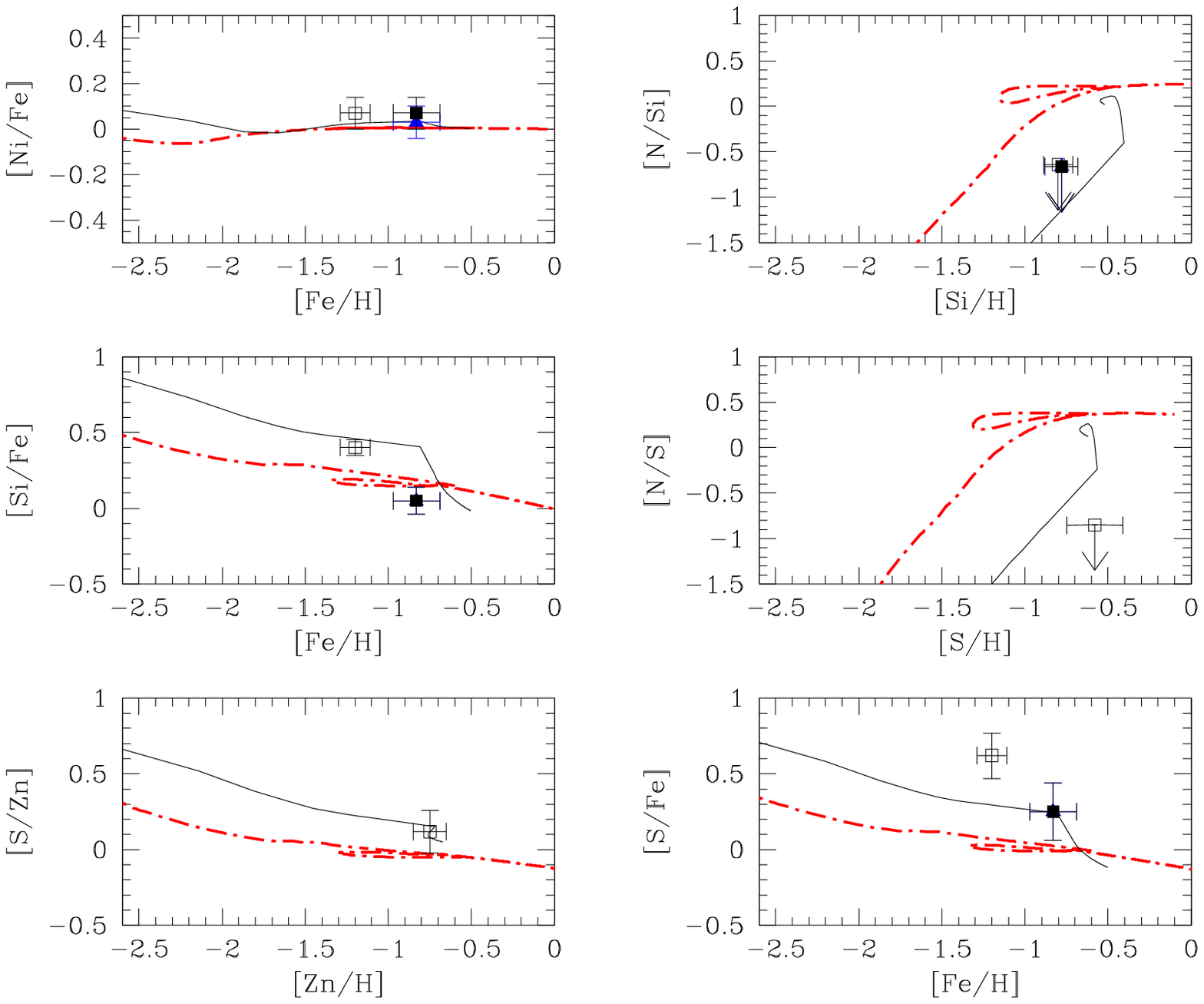}
\caption{Observed and predicted abundance ratios versus metallicity for the DLA at $z_{\rm abs} = 
2.066$ toward Q2231$-$00. The {\it thick dashed-dotted} curves correspond to the ``spiral'' model 
at $R = 2$ kpc. The {\it thin solid} curves correspond to the ``dwarf irregular'' model with a 
single burst of SF occurring at $t_b = 0.5$ Gyr and with a star formation efficiency $\nu = 4.2$ 
Gyr$^{-1}$ and a duration $\Delta t = 0.1$ Gyr. For the definition of symbols, see 
Fig.~\ref{alpha-Fe-Q0100}.}
\label{alpha-Fe-Q2231}
\end{figure*}
%

Again none of these two models yet having a different star formation history $-$ the ``spiral''
model of outer regions of the disk has a bursting SFH (see Sect.~\ref{spiral}), while the ``dwarf 
irregular'' model constrained by the observations has an inefficient continuous SFH~$-$ can be 
easily discarded, because they yield similar chemical abundance patterns due to the fact that they
both are characterized by a weak SF. The Mg/Fe ratio could be used as a discriminant between the
two models (see Fig.~\ref{alpha-Fe-Q1331}), but there are substantial uncertainties in the Mg yields 
\citep[e.g.][]{chiappini99}.

We investigated the age of the DLA galaxy for the two models. For this purpose, we considered the 
[Zn/H], [S/Zn], [S/Fe], [Si/Fe], [Mg/Fe] and [Ni/Fe] versus redshift diagrams. The range of 
redshifts of formation determined for the ``spiral'' model at $R = 16$ kpc is $z_f = 2.3-3$, with 
the best solution being at $z_f\sim 2.8$ (see the thick dashed-dotted curves in 
Fig.~\ref{alpha-Fe-z-Q1331}). For the ``dwarf irregular'' model with an inefficient continuous SF 
rate, the estimated redshift of formation is large, roughly around $z_f \sim 10$ (see the thin 
solid curves in Fig.~\ref{alpha-Fe-z-Q1331}). The DLA system is observed at $z_{\rm abs} = 1.776$, 
hence $z_f \sim 2.8$ and $z_f \sim 10$ correspond to an age of the DLA galaxy of $1.5\pm 0.4$ Gyr 
and $\gtrsim 3.5$ Gyr, respectively.

According to the recent WMAP results, the reionization seems to have occurred at a redshift 
considerably higher than what was thought previously, namely at $z_{\rm reion} = 20_{-9}^{+10}$ 
\citep{bennett03}. Therefore, if the intergalactic medium has been reionized by the first stars 
formed in very young galaxies, a redshift of the order of 10 could be reasonable for the appearance 
of the first galactic structures in the Universe. The derived solution that the associated galaxy 
with the DLA system at $z_{\rm abs} = 1.776$ toward Q1331+17 may be a dwarf irregular galaxy with a 
continuous SFH formed at $z_f \sim 10$ thus is possible. The solution that the associated galaxy is 
a spiral galaxy with the QSO line of sight crossing the outer regions of its disk at $R=16$ kpc and 
an age of 1.5 Gyr is also plausible. Both models point to a large age for this DLA galaxy. This 
reflects the long timescale necessary to accrete the gas as well as to reach the observed solar 
$\alpha$-element over Fe-peak element ratios (see Sect.~\ref{intrinsic-abundance}) which require a 
substantial contribution from Type Ia SNe, releasing the Fe-peak elements only after $10^8-10^9$ 
yrs. But, both models fail in reproducing the measured [N/S] and [N/Si] upper limits, since they 
predict almost solar [N/S,Si] ratios at [S,Si/H] $\simeq -1.25$. Such high N/$\alpha$ values at 
these low metallicities have never been observed in any  DLA and \ion{H}{ii} region. This suggests 
that the yields of \citet{hoeck97} may lead to an overestimation of the N production by 
intermediate-mass stars at the lower end of masses and by low-mass stars, which had time to release 
N in this DLA galaxy, given the inferred long age. In the case of the N production by massive 
intermediate-mass stars, the yields of \citet{hoeck97} are more reliable. 
%

\subsubsection{Q2231$-$00, $z_{\rm abs} = 2.066$}\label{SFH-q2231}

In this system, the measured abundance ratios at disposal are [S/Zn], [S/Fe], [Si/Fe], [Ni/Fe], and
the upper limits on [N/Si] and [N/S]. We consider the dust-corrected values, since non-negligible
dust corrections are required in this system with [Zn/Fe] $= + 0.45\pm 0.07$. We first investigated 
the ``spiral'' model to reproduce the [S/Zn] versus [Zn/H], [S/Fe] versus [Fe/H], [Si/Fe] versus 
[Fe/H] and [Ni/Fe] versus [Fe/H] diagrams. The favored ``spiral'' model turned out to be the one 
representing the inner regions of the galactic disk, i.e. at galactocentric radii $R<8$ kpc. In 
the inner regions of the disk compared to the outer regions, the infall of external gas is faster 
and leads to higher gas densities, and hence to a more efficient SF (see Sect.~\ref{spiral}). The 
$\alpha$-elements thus are more enhanced relative to the Fe-peak elements at metallicities [Fe/H] 
$< -1.5$ dex in the inner regions of the disk than in the outer regions. Despite this, the 
``spiral'' model of inner regions of the disk does not succeed in reproducing all the measured data 
points (see the thick dashed-dotted curves in Fig.~\ref{alpha-Fe-Q2231}). It mainly fails in 
correctly reproducing the [S/Zn] versus [Zn/H] and [S/Fe] versus [Fe/H] data points, and the [N/Si] 
versus [Si/H] and [N/S] versus [S/H] upper limits. 
%

\begin{table*}[t]
\begin{center}
\caption{Determination of the star formation rates} 
\label{SFR}
\begin{tabular}{l c c}
\hline \hline
\\[-0.3cm]
DLA system & SFR for the best ``spiral'' model  & SFR for the best ``dwarf irregular'' model \\
           & [M$_{\odot}$ yr$^{-1}$ kpc$^{-2}$] & [M$_{\odot}$ yr$^{-1}$ kpc$^{-2}$]
\smallskip 
\\     
\hline 
\\[-0.3cm]
Q0100+13, $z_{\rm abs} = 2.309$     & $8.5\times 10^{-3}$ & $1.0\times 10^{-2}$ \\
Q1331+17, $z_{\rm abs} = 1.776$     & $7.5\times 10^{-3}$ & $9.2\times 10^{-5}$ \\
Q2231$-$00, $z_{\rm abs} = 2.066$   & -- & $4.0\times 10^{-2}$ \\
\hline
\end{tabular}
\end{center}
\end{table*}
%

We then considered the ``dwarf irregular'' model. We assumed a single burst of SF and explored the 
following values for the three free parameters characterizing the burst: $\nu = 2.8-5.0$ Gyr$^{-1}$, 
$\Delta t = 0.05,\,0.1,\,0.15$ Gyr and $t_b = 0.2,\,0.5,\,0.8$ Gyr. The ``dwarf irregular'' model 
which best reproduces the data points has one burst with a very high star formation efficiency $\nu 
= 4.2$ Gyr$^{-1}$ and a duration $\Delta t = 0.1$ Gyr (see the thin solid curves in 
Fig.~\ref{alpha-Fe-Q2231}). The burst occurs after $t_b = 0.5$ Gyr of continuous infall of pristine 
gas. This model is the favored one in comparison with the ``spiral'' model of inner regions of the 
disk. It reproduces at 1~$\sigma$ all the abundance ratios. Only the [N/S] upper limit is at odds.

We investigated the age of the DLA galaxy assuming the constrained ``dwarf irregular'' model. As in 
the case of the DLA toward Q0100+13, the parameter having the highest impact in the age 
determination is the duration of the burst. We feel confident in assuming an age between 
$0.05-0.15$ Gyr, corresponding to the range of values explored for $\Delta t$. This yields a 
redshift of formation between $2.10-2.16$ given the assumption that the galaxy has undergone a 
single burst of SF.
%

\subsection{Star formation rates}

The star formation is a key parameter in the formation and evolution of galaxies. Therefore, the
knowledge of the star formation rate (SFR) of galaxies is very important. \citet{kennicutt83} 
provided the first precise diagnostics for the measure of the SFR, such as emission-line fluxes and
UV continuum luminosities. Later \citet{madau96} reconstructed the cosmic star formation history 
by measuring the comoving luminosity density of star-forming galaxies as a function of redshift. 
While the original work of \citet{madau96} showed a peak in the cosmic star formation at $z\approx 
1-2$, recent results, based on larger samples of galaxies and corrections for dust extinction of 
the emitted starlight, do not show such a peak. The SFR per unit comoving volume increases by a 
factor of $\sim 10$ in the redshift interval $z = [0,1]$, and then either remains constant at $z>1$ 
up to $z=6$ \citep{steidel99} or keeps increasing to even higher redshifts
\citep{lanzetta02,calura03b}. However, the galaxies from which these results are derived are 
unlikely to be representative of the bulk of the galaxy population in the Universe. Indeed, whereas 
the SFR per unit area for the Milky Way is $\psi \sim 4\times 10^{-3}$ M$_{\odot}$ yr$^{-1}$ 
kpc$^{-2}$ \citep{kennicutt98}, the comoving SFR at $\sim 3$ is inferred from Lyman-break galaxies, 
a highly luminous population of star-forming objects in which $\psi \geq 1$ M$_{\odot}$ yr$^{-1}$ 
kpc$^{-2}$ \citep{pettini01}. As a result, the existing measurements of the cosmic star 
formation history take into account the contribution of only highly luminous and star-forming 
galaxies. The access to the star formation rates in other types of galaxies thus is crucial.

The DLA galaxy population is ideal for this purpose. Indeed, these objects sample various types of 
galaxies over a large range of lookback times (see Sects.~\ref{SFH-q0100}$-$\ref{SFH-q2231}), since 
they are detected independently of their distance, luminosity and SFH. In addition, because the 
DLAs are not drawn from a flux limited sample of galaxies, we are able to derive the SFR values 
below those determined from radiation emitted by star-forming galaxies. Recently, 
\citet{wolfe03a,wolfe03b} provided the first estimations of SFRs per unit area in DLAs. Their 
technique to infer the SFRs consists in determining the rate at which the neutral gas in DLAs is 
heated. 

In this work, we also have access to the SFRs of DLAs. Indeed, in the case of the ``spiral'' model 
the SFR per unit area is a direct output of the model, and in the case of the ``dwarf irregular'' 
model the different physical quantities correspond to the absolute values, thus to derive the SFR 
per unit area we assume a spherical symmetry and a galactic radius of 5 kpc. The derived star 
formation rates per unit area for the DLAs studied are given in Table~\ref{SFR}, and are between 
$-2.1 < \log \psi < -1.5$ M$_{\odot}$ yr$^{-1}$ kpc$^{-2}$. They were obtained for the models 
constrained in the previous Sections (Sects.~\ref{SFH-q0100}$-$\ref{SFH-q2231}). They correspond in 
the case of the ``spiral'' model to the SFR that the DLA galaxy has at the time of its observations 
and in the case of the ``dwarf irregular'' model to the average SFR integrated over the time of the 
burst of SF, i.e. over the period when the SF is active. The errors on the derived SFRs are very 
difficult to estimate. Our SFR measurements are in agreement with the interval of SFR values 
obtained by \citet{wolfe03a,wolfe03b}\footnote{The SFRs obtained by \citet{wolfe03a,wolfe03b} 
correspond to the SFR values that the DLAs have at the time of their observations.} in DLAs using 
a completely different technique. We thus confirm that the SFRs per unit area in DLAs are moderate 
and similar to that measured in the Milky Way disk. 
%

\section{Conclusions}\label{conclusion}

The damped Ly$\alpha$ systems are our best laboratory to study the high redshift galaxies. Indeed, 
accurate chemical abundances of these systems can be obtained over a large interval of cosmic time, 
and they thus offer the best opportunity to track the chemical evolution of galaxies in the 
Universe. The DLA galaxy population has until now been analyzed as a whole and chemical evolution 
models were constructed in order to interpret the abundance patterns of these objects as an 
ensemble, considering them as an evolutionary sequence. However, several pieces of evidence show 
that the DLAs likely trace galaxies of different types and with different evolutionary histories. 
At this stage of knowledge, one would like to define more precisely the star formation histories 
and the chemical evolution stages sampled by these objects. Therefore, we aimed at constructing a 
sample of DLAs for which one would be able to constrain the star formation history, the age and the
star formation rate of each system individually. For this purpose, it is first imperative to obtain 
comprehensive sets of elemental abundances. 

By combining our UVES-VLT spectra of a sample of four DLAs in the redshift interval $z_{\rm abs} = 
1.7-2.5$ toward the quasars Q0100+13, Q1331+17, Q2231$-$00 and Q2343+12 with the existing HIRES-Keck 
spectra, we covered the total optical spectral range from 3150 to 10\,000 \AA\ for the four quasars. 
Thanks to this large wavelength coverage and the high quality of the spectra, we succeeded in 
measuring the column densities of up to 21 ions, namely of 15 elements $-$ N, O, Mg, Al, Si, P, S, 
Cl, Ar, Ti, Cr, Mn, Fe, Ni, Zn. With the detections of adjacent ions of the same element, such
as Al$^{+}$/Al$^{++}$, Fe$^{+}$/Fe$^{++}$ and N$^0$/N$^+$, and of the Ar/Si,S ratios, we 
constrained the photoionization effects, which may affect the DLA gas-phase abundances. Our analysis 
revealed that the DLA toward Q2343+12 requires important ionization corrections, while in the three 
other DLAs the ionization corrections are negligible. With the detection of both refractory and 
mildly/non-refractory elements of the same nucleosynthetic origin, we evaluated the dust depletion 
effects and computed the dust corrections using the method of \citet{vladilo02a,vladilo02b}. The 
dust corrections are particularly important in the DLA toward Q1331+17. This system with [Zn/Fe] 
$= +0.75\pm 0.05$ exhibits one of the largest dust depletion level of any DLA. The constraint of 
both the photoionization and dust depletion effects allowed us to determine relatively robust {\it 
intrinsic} chemical abundance patterns of three out of the four DLAs studied.

The intrinsic chemical abundance patterns are the signature of the star formation history of 
galaxies, which one needs to correctly interpret. For this purpose, we called on the chemical 
evolution models for spiral and dwarf irregular galaxies \citep[see][]{calura03a}. By comparing 
these models with the distributions of the abundance ratios of two elements produced on different 
timescales as a function of the metallicity we can constrain the star formation history of a DLA 
galaxy, and with the distributions of the same abundance ratios considered as a function of the 
redshift we can constrain the age of a galaxy. The access to several abundance ratios, [S/Zn], 
[S/Fe], [Si/Fe], [O/Zn], [Mg/Fe], [Ni/Fe], [N/S], [N/Si] and [N/O], in the same DLA allowed us for 
the first time to determine the star formation history and the age of three DLA systems. We were 
also able to estimate the star formation rates per unit area of the DLAs, since the SFR is a direct 
output of the chemical evolution models, once they are constrained. Our results show that the galaxy 
associated with the DLA toward Q0100+13 is either a spiral galaxy showing similar characteristics 
as those observed in the solar neighborhood ($R\sim 8$ kpc) or a dwarf irregular galaxy with a 
single burst of SF with propertiessimilar to the ones observed in the BCGs. It is a young galaxy 
with an age of $250\pm 200$ Myr or between $50-200$ Myr, respectively. The galaxy associated with 
the DLA toward Q1331+17 is very likely a spiral galaxy observed in the outer regions of its disk 
($R\sim 16$ kpc) and $1.5\pm 0.4$ Gyr old. But, it can also well be a dwarf irregular galaxy with 
an inefficient continuous SFH and an age $\gtrsim 3.5$ Gyr. Finally, the galaxy associated with the 
DLA toward Q2231$-$00 is a dwarf irregular galaxy with a single intense burst of SF and an age 
between $50-150$ Myr. The estimated star formation rates per unit area of these three objects are 
between $-2.1 < \log \psi < -1.5$ M$_{\odot}$ yr$^{-1}$ kpc$^{-2}$. They are in agreement with the 
values obtained by \citet{wolfe03a,wolfe03b} in DLAs using a different technique, and similar to 
the SFR per unit area measured in the Milky Way disk.

This work shows that the DLAs studied may either be associated with the outer regions of disks of 
spiral galaxies ($R\geq 8$ kpc) or with dwarf irregular galaxies with bursting or continuous star 
formation histories. Observed at redshifts $z_{\rm abs} = 1.7-2.5$, they may be very young galaxies 
with ages between $50-250$ Myr, but also galaxies with ages larger than $> 1.5$ Gyr. Although 
this work needs to be extended to a larger sample of DLA systems to be representative of the overall 
population of DLAs at high redshift, we can already derive a number of conclusions. Our limited 
sample is centered at $z\sim 2$, a redshift region where spectroscopic data on light-emitting 
galaxies just start to become available. We do not find any association of DLAs with episodes of 
massive star formation, the derived star formation rates per unit area are moderate. Clearly this 
is not the standard scenario, and we cannot exclude this possibility for all objects in the DLA 
sample. Actually there is an indication of a recent massive burst of star formation in at least the 
DLA at $z_{\rm abs} = 3.025$ toward Q0347$-$3819 \citep{levshakov02,dessauges02b}. The DLAs at 
$z\sim 2$ do not appear to be associated with a homogeneous population of galaxies, neither in 
terms of star formation history (morphological type) and age. This result is in agreement with the 
studies of DLAs at low redshifts ($z<1$) where different types of galaxies have been shown to be 
the optical counterparts \citep[e.g.][]{lebrun97,nestor02}. Our results confirm once again that 
simple plots of DLA abundances versus redshift cannot be used to estimate the overall metal 
evolution of the Universe, because the local parameters (as determined by the star formation history 
of the associated galaxy) would dominate the scatter in the measurements. We plan to extend our 
comparison of the DLA abundance patterns with chemical evolution models in 2004 to additional 5 
objects for which the data have already been acquired.
%

\begin{acknowledgements}

The authors wish to extend special thanks to all people working at ESO/Paranal for the high quality 
of the UVES spectra obtained in service mode. We are grateful to G.~Meynet for useful discussions
and to A.~M. Wolfe for interesting comments. We particularly thank C.~Chiappini for many interesting 
advises and for her help to clarify important points. M.D.-Z. is supported by the Swiss National 
Funds. This work has partly been done during M.D.-Z. ESO studentship and has benefited of the 
support from the European Commission Research and Training Network ``The Physics of the 
Intergalactic Medium''.

\end{acknowledgements}

\end{document}